\documentclass[12pt]{article}

\usepackage{epsf,epsfig}
\usepackage{color,colortbl}
\usepackage{graphics}
\usepackage{axodraw}

\renewcommand{\arraystretch}{1.2}

\def\slash#1{#1 \hskip-0.45em /}
\def\Slash#1{#1 \hskip-0.59em /}

\def\simgt{\rlap{\lower 3.5 pt \hbox{$\mathchar \sim$}} \raise 1pt \hbox {$>$}}
\def\simlt{\rlap{\lower 3.5 pt \hbox{$\mathchar \sim$}} \raise 1pt \hbox {$<$}}

\def\be{\begin{equation}}
\def\ee{\end{equation}}
\def\beq{\begin{eqnarray}}
\def\eeq{\end{eqnarray}}

\def\np{n_+}
\def\nm{n_-}

\newcommand{\SCETI}{\mbox{SCET${}_{\rm I}$}}
\newcommand{\SCETII}{\mbox{SCET${}_{\rm II}$}}

\begin{document}

\begin{titlepage}

\begin{flushright}
{\small
PITHA~05/10\\
hep-ph/0508250\\[0.2cm]
August 16, 2005}
\end{flushright}

\vspace{0.7cm}
\begin{center}
\Large\bf\boldmath
Heavy-to-light $B$ meson form factors 
at large recoil energy -- spectator-scattering 
corrections 
\unboldmath
\end{center}

\vspace{0.8cm}
\begin{center}
{\sc M.~Beneke$^a$ and D.~Yang$^b$}\\
\vspace{0.7cm}
{\sl ${}^a$Institut f\"ur Theoretische Physik E, RWTH Aachen\\
D--52056 Aachen, Germany\\[0.3cm]
${}^b$ Department of Physics, Nagoya University\\
Nagoya 464-8602, Japan}
\end{center}

\vspace{1.0cm}
\begin{abstract}
\vspace{0.2cm}\noindent
We complete the investigation of loop corrections to 
hard spectator-scattering in exclusive $B$ meson to light 
meson transitions by computing the short-distance 
coefficient (jet-function) from the hard-collinear scale. 
Adding together the two coefficients from 
matching QCD $\to$ \SCETI{} $\to$ \SCETII{}, we 
investigate the size of loop effects on the ratios of 
heavy-to-light meson form factors at large recoil. 
We find the corrections from the hard and hard-collinear 
scales to be of approximately the same size, and significant, 
but the perturbative expansions appear to be well-behaved. 
Our calculation provides a non-trivial verification of the 
factorization arguments. We observe considerable differences 
between the predictions based on factorization in the heavy-quark 
limit and current QCD sum rule calculations of the form factors. 
We also include the hard-collinear correction in the 
$B\to \pi\pi$ tree amplitudes, and find an enhancement of 
the colour-suppressed amplitude relative to the colour-allowed 
amplitude.
\end{abstract}

\vfil
\end{titlepage}

\section{Introduction}
\label{sec:intro}

The matrix elements of flavour-changing currents 
$\bar q\hspace*{0.05cm}\Gamma_i b$ 
are important strong interaction parameters in low-energy 
weak-interaction processes. The strong interaction dynamics of 
semi-leptonic $B$ decays is encoded in these form factors. They are 
also inputs to the factorization formulae for
hadronic two-body $B$ decays~\cite{Beneke:1999br} 
and radiative decays~\cite{Beneke:2001at}. A better understanding
of such quantities improves the accuracy of the extraction of
the CKM matrix parameters from experimental data, and of searches 
for new phenomena in flavour-changing processes. Thus efforts 
are being made to compute the form factors with  different methods 
including QCD lattice simulations \cite{lattice}, light-cone 
QCD sum rules~\cite{sumrules} and quark models \cite{quarkmodel}. 

It is also interesting to investigate these form factors in the 
heavy-quark expansion. It is well-known that all $B \to D^{(*)}$ 
form factors reduce to a single (Isgur-Wise) function \cite{Isgur:1990ed} 
up to calculable short-distance corrections at leading order in 
this expansion. In this paper we consider transitions of $B$ mesons to light 
mesons in the large-recoil regime, where the light meson momentum 
is parametrically of order of the heavy-quark mass. In this regime 
a similar simplification applies to heavy-to-light form 
factors \cite{Charles:1998dr}: the three (seven) independent 
$B\to$ pseudoscalar (vector) meson form factors reduce to one (two) 
function(s) up to corrections that can be calculated in the
hard-scattering formalism at leading order in the heavy-quark 
expansion \cite{Beneke:2000wa}. The different form factors 
can therefore be related in a systematic way. The factorization 
formula that summarizes these statements reads \cite{Beneke:2000wa}
\begin{eqnarray}
\label{eq:factorization} F_i^{B\to M} (E)\,=\,C_i(E)
\,\xi_a(E)+\int_0^\infty\,\frac{d\omega}{\omega}\,\int_0^1dv\,T_i(E;\ln
\omega,v)\phi_{B+}(\omega)\phi_M(v)
\end{eqnarray}
with $E$ the energy of the light meson $M$, 
$\xi_a(E)$ the single non-perturbative form factor (one of the two
form factors when $M$ is a vector meson), and
$\phi_X$ the light-cone distribution amplitudes of the $B$ meson 
and the light meson. The short-distance coefficients 
$C_i$ and the hard-scattering kernel $T_i$ can be calculated in 
perturbation theory. The heavy-to-light form factors 
are more complicated than both, the $B\to D^{(*)}$ form factors 
and light-light meson transition form factors at large momentum transfer. 
Contrary to the case of $B\to D^{(*)}$, a spectator-scattering correction, 
the second term on the right hand side of (\ref{eq:factorization}),  
appears. On the other hand, the form factor cannot be expressed in 
terms of a convolution of light-cone distribution amplitudes alone, 
because the corresponding convolution integrals are dominated by 
endpoint singularities \cite{Szczepaniak:1990dt}. 
In (\ref{eq:factorization}) these contributions are factored into 
the the function $\xi_a(E)$.\footnote{The statement that the endpoint 
contributions are not calculable is challenged in the PQCD 
approach \cite{Keum:2000wi}, which assumes that Sudakov 
resummation renders them perturbative. This point is critically 
examined in \cite{Descotes-Genon:2001hm}. We also note here that 
our notation $\xi_a(E)$ does not show the dependence of the form 
factor on the nature of the meson $M$.} The factorization 
formula (\ref{eq:factorization}) has been shown to be valid to 
all orders in perturbation theory~\cite{Beneke:2003pa} (see 
also \cite{Lange:2003pk,Bauer:2002aj}) 
in the framework of soft-collinear effective 
theory (SCET)~\cite{Bauer:2000yr,BCDF,Hill:2002vw}. In particular, 
since the two relevant short-distance scales  $m_b$ and $(m_b 
\Lambda)^{1/2}$ ($\Lambda$ is the characteristic scale of QCD) 
can be separated in SCET, the short-distance coefficients $T_i$ 
pertaining to spectator-scattering are 
represented as convolutions $C^{(B1)}_i\!\star J$ with the two factors 
associated with the two different scales. 

In the limit that not only power corrections in $\Lambda/m_b$ 
but also radiative corrections in the strong coupling $\alpha_s$ 
are neglected, the second term on the right hand side 
of (\ref{eq:factorization}) is absent, and 
parameter-free relations between ratios of form 
factors follow \cite{Charles:1998dr}. 
The $\alpha_s$ contributions to (\ref{eq:factorization}) have been 
computed in \cite{Beneke:2000wa}, and the spectator-scattering 
term $T_i$ has been found to dominate the correction. This motivates 
an investigation of 
the subsequent term in the perturbative expansion of $T_i$. 
Since the leading $\alpha_s$ term is due to a tree diagram 
with gluon exchange between the current quarks and the spectator 
anti-quark, this amounts to the computation of the 1-loop correction 
to spectator-scattering. Since $T_i = C^{(B1)}_i\!\star J_a$, 
the calculation splits into two parts. In a previous paper 
\cite{Beneke:2004rc} (see also \cite{Becher:2004kk}) we reported  
the first part of the calculation which consisted of the 1-loop 
correction to the coefficients $C^{(B1)}_i$ originating from the 
hard scale $m_b$. In this paper we complete the calculation 
with the 1-loop computation of the ``jet-functions'' $J_a$ 
originating from the ``hard-collinear'' scale  $(m_b 
\Lambda)^{1/2}$. The jet-functions have also been computed by 
Hill {\em et al.} \cite{Becher:2004kk,Hill:2004if}. Nevertheless, 
an independent calculation is useful, since the computation 
is quite involved and a comparison showed that the result 
of \cite{Hill:2004if} was originally not given in a scheme 
consistent with the $\overline{\rm MS}$ definition of the 
light-meson light-cone distribution amplitude (see the 
discussion in \cite{Becher:2004kk,Hill:2004if}). Furthermore, 
the numerical impact of these calculations on the relation between 
form factors and other observables in $B$ decays 
has not yet been discussed in any detail in the literature. 

The organization of the calculation of the short-distance 
coefficients $C_i$ and $T_i$ follows closely the derivation of 
the factorization formula in~\cite{Beneke:2003pa}. In a first 
step, the effects from the hard scale $m_b$ are computed 
and QCD is matched to an intermediate effective theory,
called \SCETI. In \SCETI{} the term $\xi_a$ and the hard-scattering term
are naturally defined by the matrix elements of two distinct 
operator structures, the so-called $A$-type
and $B$-type operators. At this step, the form factors 
can be represented as
\begin{eqnarray}
\label{ff2} 
F_i^{B\to M} (E)\,=\,C_i(E)\,\xi_a(E)+\int d\tau\,C^{(B1)}_i(E,\tau)\,
\Xi_a(\tau,E).
\end{eqnarray}
The point to note here is that the three (seven) 
form factors of a $B\to P$ ($B\to V$) transition can be expressed 
in terms of one (two) form factor(s) $\xi_a(E)$ and one (two) 
non-local form factor(s) $\Xi_a(\tau,E)$. A 
number of relations between form factors emerge already at this stage.
In Section~\ref{sect2} we define the \SCETI{} operator basis, 
and express the QCD heavy-to-light form factors in terms of the 
SCET$_{\rm I}$ hadronic matrix elements, which leads to 
(\ref{ff2}). All the required short-distance 
coefficients of the \SCETI{} operators 
can be inferred from \cite{Beneke:2004rc,Becher:2004kk}.

Eq.~(\ref{ff2}) is useful only to a limited extent, because it introduces 
the form factors $\Xi_a(\tau,E)$, which depend on two variables. However, 
it has been shown that, contrary to the $\xi_a(E)$, the 
$\Xi_a(\tau,E)$ can be factorized further into a convolution of 
light-cone distribution amplitudes with a hard-scattering kernel 
(jet-function)~\cite{Beneke:2003pa}. This amounts  
to performing a second matching to $\SCETII$, in which the effects at 
the hard-collinear scale $(m_b\Lambda)^{1/2}$ 
are computed. This is done in Section~\ref{sec:jetfunc}. Here 
we discuss in detail the 1-loop calculation and renormalization of the
jet-functions $J_a$ that follow from representing the \SCETI{} matrix
element of the B-type operators in the form 
\be
\Xi_a(\tau,E) = \frac{1}{4}\,\int_0^\infty\,d\omega\,\int_0^1\,dv\,
J_a(\tau;v,\omega)\,\hat{f}_B\phi_{B+}(\omega)f_M\phi_M(v).
\label{b1match}
\ee
Combining this with (\ref{ff2}) we obtain the spectator-scattering term 
in (\ref{eq:factorization}). The calculation is done in dimensional 
regularization which requires dealing with evanescent 
Dirac structures specific to $d$ dimensions. As will be discussed, 
a subtlety arises due to the fact that the factorization properties 
of $\SCETII$ require a specific choice of reduction scheme. 
Together with $J_a$ we also determine the anomalous dimensions 
of the B-type operators confirming the results of~\cite{Hill:2004if}. 

The detailed numerical analysis of the corrections from the two 
matching steps is contained in
Sections~\ref{sec:numerics} and~\ref{sec:symmbr}. 
In addition to the next-to-leading order correction we also include 
the summation of formally large logarithms from the ratio of the hard 
and hard-collinear scale by deriving a renormalization group improved 
expression for the coefficient functions $C^{(B1)}_i$. From the size 
of the 1-loop correction we conclude that the perturbative calculation 
of spectator-scattering is under reasonable control 
despite the comparatively low scale of order $(m_b\Lambda)^{1/2} 
\sim 1.5\,$GeV. The combined hard and hard-collinear 1-loop correction 
is about $(50-70)\%$ depending on the observable. 
This is also of interest in the context of QCD
factorization calculations of hadronic $B$ decays, since the 
same jet-function enters the spectator-scattering contributions 
to two-body decays~\cite{Bauer:2004tj}. 
Section~\ref{sec:symmbr} is devoted to a discussion of the 
symmetry-breaking effects on the form factor ratios and a comparison 
of these ratios to QCD sum rule calculations. We then consider the 
tensor-to-(axial-)vector form factor ratios that appear in 
electromagnetic and electroweak penguin decays, and the numerical 
impact of our jet-function calculation on 
hadronic decays to two pions. Here we find that the new contribution  
increases the ratio of the colour-suppressed to the colour-allowed 
tree decay amplitude, which leads to a better description of the 
branching fraction data. We conclude in Section~\ref{sec:conclusion}. 

In Appendix~\ref{app:Wilson} we summarize the short-distance 
coefficients $C^{(B1)}_i(E,\tau)$ relevant to (\ref{ff2}). 
Some of the convolution integrals of the jet-functions and the 
coefficients $C^{(B1)}_i(E,\tau)$ needed for the numerical analysis of the
spectator-scattering term are collected in 
Appendix~\ref{app:convolution}.


\section{Heavy-to-light form factors in SCET$_{\rm I}$}
\label{sect2}

Our first task is to express the QCD form factors in terms of matrix 
elements of \SCETI{} currents and the corresponding short-distance 
coefficients. We use the position-space SCET formalism and the notation 
of \cite{Beneke:2003pa,BCDF,Beneke:2004rc} to which we refer for further 
details.  The ``collinear'' fields $\xi$ and $A_c$ that appear in this 
section describe both, hard-collinear (virtuality 
$m_b\Lambda$) {\em and} collinear  (virtuality $\Lambda^2$) 
modes. The reference vectors $v$, $n_\mp$ are defined such that 
$v^2=1$, $n_-^2=n_+^2=0$, $n_- n_+=2$. Except for 
Section~\ref{subsec} we adopt a frame of reference 
where $n_- v=1$ and $v=(n_-+n_+)/2$. In scalar products of 
$n_-$, $n_+$ with other vectors we omit the scalar-product ``dot''.

\subsection{Operator basis}
\label{subsec}

The relevant terms in the \SCETI{} expansion of a heavy-to-light 
current $\psi\hspace*{0.03cm}\Gamma_i Q$ 
read \cite{Beneke:2003pa,Bauer:2002aj,Beneke:2004rc}
\begin{eqnarray}
\label{match}
    (\bar{\psi}\hspace*{0.03cm}\Gamma_i \hspace*{0.03cm}Q)(0)&=&
     \int d\hat{s}
    \sum\limits_{j} \widetilde{C}^{(A0)}_{ij}(\hat{s})\,
    O_{j}^{(A0)}(s;0) \nonumber \\
   &&+\int d\hat{s}
    \sum\limits_{j} \widetilde{C}^{(A1)}_{ij\mu} (\hat{s})\,
    O_{j}^{(A1)\mu}(s;0) \nonumber \\
    &&+\int d\hat{s}_1 d\hat{s}_2
    \sum\limits_{j} \widetilde{C}^{(B1)}_{ij\mu}(\hat{s}_1,\hat{s}_2)\,
    O_{j}^{(B1)\mu}(s_1,s_2;0)+\cdots,
\label{currentrep}
\end{eqnarray}
where 
\begin{eqnarray}
   O_j^{(A0)}(s;x) &\equiv& (\bar{\xi} W_c)(x+sn_+) \Gamma_j^{\prime}
    h_v(x_-) \equiv (\bar{\xi} W_c)_s \Gamma_j^{\prime} h_v,
   \nonumber\\[0.2cm]
   O_{j\mu}^{(A1)}(s;x) &\equiv&
    (\bar{\xi}i\overleftarrow{D}_{\perp c\mu}
    (in_-v n_+\overleftarrow{D}_c)^{-1}W_c)_s \Gamma_j^{\prime} h_v,
    \nonumber\\[0.1cm]
   O_{j\mu}^{(B1)}(s_1,s_2;x) &\equiv& \frac{1}{m_b} \,
    (\bar{\xi}W_c)_{s_1} (W_c^\dag i
    D_{\perp c\mu}W_c)_{s_2}\Gamma_j^{\prime} h_v,
\label{opbasis}
\end{eqnarray}
and $\hat{s}_i\equiv s_i m_b/n_- v$.
Since the collinear fields $\xi$ and $A_c$ describe modes of 
different virtuality, no simple $\Lambda/m_b$-scaling rules apply to 
these fields. The power-counting argument that shows that the three 
types of operators contribute to the form factors at leading power in 
the heavy-quark expansion has been given in \cite{Beneke:2003pa}. 
The main difference between 
the two types of operators is their dependence on position arguments. 
The B-type operators are tri-local, and for this reason are sometimes 
also referred to as ``three-body'' operators. The 1-loop corrections to 
the coefficient functions of the A-type currents have been calculated 
in \cite{Bauer:2000yr,Beneke:2004rc}, to those of the B-type currents 
in \cite{Beneke:2004rc,Becher:2004kk}. 

The basis (\ref{opbasis}) is motivated by the simple expressions 
of the tree-level matching coefficients in this 
basis~\cite {BCDF}. However, the analysis of \cite{Beneke:2003pa} shows 
that $O_{j\mu}^{(A1)}$ and $O_{j\mu}^{(B1)}$ are operators relevant 
at leading power in the $1/m_b$-expansion 
only because of the transverse collinear gluon field in the covariant 
derivative $D_{\perp c}$. It is therefore advantageous to perform a 
basis redefinition such that the transverse collinear gluon field appears 
only in $O_{j\mu}^{(B1)}$. This can be done by replacing 
$O_{j\mu}^{(A1)}$ by 
\begin{eqnarray}
    (\bar{\xi}W_c)_s \,i\overleftarrow{\partial}_{\!\perp \mu}
    (in_-v n_+\overleftarrow{\partial})^{-1} \Gamma_j^{\prime} h_v,
\label{newa1}
\end{eqnarray}
which is the choice that has been 
adopted in \cite{Becher:2004kk,Hill:2004if}. The redefinition 
involves the identity 
\begin{eqnarray}
&& 
(\bar{\xi}i\overleftarrow{D}_{\perp c\mu}
(in_-v n_+\overleftarrow{D}_c)^{-1}W_c)_s \Gamma_j^{\prime} h_v 
= (\bar{\xi}i\overleftarrow{D}_{\perp c\mu} W_c)_s \,
\frac{1}{in_-v n_+\overleftarrow{\partial}} \,\Gamma_j^{\prime} h_v 
\nonumber\\
&&\hspace*{1cm}
= (\bar{\xi}W_c)_s \,\frac{i\overleftarrow{\partial}_{\perp \mu}}
{in_-v n_+\overleftarrow{\partial}}\,\Gamma_j^{\prime} h_v
- (\bar{\xi}W_c)_s (W_c^\dag i D_{\perp c\mu}W_c)_{s}\,
\frac{1}{in_-v n_+\overleftarrow{\partial}}\,\Gamma_j^{\prime} h_v
\\
&&\hspace*{1cm}
= (\bar{\xi}W_c)_s \,\frac{i\overleftarrow{\partial}_{\perp \mu}}
{in_-v n_+\overleftarrow{\partial}}\, \Gamma_j^{\prime} h_v 
- i \int_{-\infty}^\infty  d\hat{r} \,\frac{\theta(\hat r-\hat s)}{m_b}\,
(\bar{\xi}W_c)_r (W_c^\dag i D_{\perp c\mu}W_c)_{r} \Gamma_j^{\prime} h_v.
\nonumber
\end{eqnarray}
The second term modifies 
$\widetilde{C}^{(B1)}_{ij\mu}(\hat{s_1},\hat{s_2})$ in the new basis 
by an amount proportional to $\widetilde{C}^{(A1)}_{ij\mu}
(\hat{s})$.\footnote{In momentum space the generic modification 
is $C^{(B1)}_{\rm new}(E,\tau) = C^{(B1)}_{\rm old}(E,\tau)+
m_b/(2 E)\, C^{(A1)}_{\rm old}(E)$, see also 
Appendix~\ref{app:Wilson}.} In the new basis only the A0- and B-type operators 
contribute to the form factors at leading power in the
$1/m_b$-expansion. Our new basis of operators for a given Dirac structure
$\Gamma_i$ is:

\begin{itemize}

\item Scalar current $J=\bar{\psi} Q$:
\begin{eqnarray}
\label{scalar:newbasis}
    J^{(A0)}&=&(\bar{\xi}W_c)
    \bigg(1-\frac{i\overleftarrow{\slash{\partial}}_{\!\!\perp}}
         {i n_+\overleftarrow{\partial}}\,\frac{\slash{n}_+}{2}\bigg)
         h_v\nonumber \\
    J^{(B1)} &=& \frac{1}{m_b} (\bar{\xi}W_c) [W_c^\dag i\not\!\!
    D_{\perp c}W_c] h_v
\end{eqnarray}

\item Vector current $J_\mu=\bar{\psi}\gamma_\mu Q$:
\begin{eqnarray}
\label{vector:newbasis}
    J^{(A0)1-2}_\mu&=&(\bar{\xi}W_c)
    \bigg(1-\frac{i\overleftarrow{\slash{\partial}}_{\!\!\perp}}
         {i n_+\overleftarrow{\partial}}\,\frac{\slash{n}_+}{2}\bigg)
   \{\gamma_{\mu},v_\mu\}  h_v \nonumber\\
    J^{(A0)3}_\mu&=&(\bar{\xi}W_c)
    \bigg(1-\frac{i\overleftarrow{\slash{\partial}}_{\!\!\perp}}
         {i n_+\overleftarrow{\partial}}\,\frac{\slash{n}_+}{2}\bigg)
     \frac{n_{-\mu}}{n_-v} h_v  +
     \frac{2}{n_- v} \,(\bar{\xi}W_c)
     \frac{i\overleftarrow{\partial}_{\!\mu_\perp}}
     {i n_+\overleftarrow{\partial}}\,h_v \nonumber\\
    J^{(B1)1-3}_\mu &=& \frac{1}{m_b} (\bar{\xi}W_c) [W_c^\dag i\not\!\!
    D_{\perp c}W_c] \{v_\mu, \frac{n_{-\mu}}{n_-v},\gamma_{\mu_\perp} \}
    h_v\nonumber\\
    J^{(B1)4}_\mu &=& \frac{1}{m_b} (\bar{\xi}W_c) \gamma_{\mu_\perp}
    [W_c^\dag i\not\!\! D_{\perp c}W_c]h_v
\end{eqnarray}

\item Tensor current $J_{\mu \nu}=\bar{\psi} i \sigma_{\mu\nu} Q$:
\begin{eqnarray}
\label{tensor:newbasis}
    J^{(A0)1-2}_{\mu\nu}&=& (\bar{\xi}W_c)
    \bigg(1-\frac{i\overleftarrow{\slash{\partial}}_{\!\!\perp}}
         {i n_+\overleftarrow{\partial}}\,\frac{\slash{n}_+}{2}\bigg)
   \{\gamma_{[\mu}\gamma_{\nu]},v_{[\mu} \gamma_{\nu]}\}
    h_v\nonumber \\
    J^{(A0)3-4}_{\mu\nu}&=& (\bar{\xi}W_c)
    \bigg(1-\frac{i\overleftarrow{\slash{\partial}}_{\!\!\perp}}
         {i n_+\overleftarrow{\partial}}\,\frac{\slash{n}_+}{2}\bigg)
   \Big\{\frac{n_{-[\mu}\gamma_{\nu]}}{n_-v},
   \frac{n_{-[\mu} v_{\nu]}}{n_-v}\Big\} h_v\nonumber \\
   &&+\,\frac{2}{n_- v}\,(\bar{\xi}W_c)
     \frac{i\overleftarrow{\partial}_{\![\mu_\perp}}
     {i n_+\overleftarrow{\partial}}\,\{\gamma_{\nu]},v_{\nu]}\} h_v
   \nonumber\\
   J^{(B1)1-2}_{\mu\nu}&=& \frac{1}{m_b} (\bar{\xi}W_c) [W_c^\dag i\not\!\!
    D_{\perp c}W_c]  \Big\{v_{[\mu} \gamma_{\nu_\perp]},
    \frac{n_{-[\mu}\gamma_{\nu\perp]}}{n_-v}\Big\}h_v \nonumber\\
   J^{(B1)3-4}_{\mu\nu}&=& \frac{1}{m_b} (\bar{\xi}W_c)
    \Big\{v_{[\mu} \gamma_{\nu_\perp]},
    \frac{n_{-[\mu}\gamma_{\nu\perp]}}{n_-v}\Big\} [W_c^\dag i\not\!\!
    D_{\perp c}W_c]  h_v \nonumber\\
   J^{(B1)5-6}_{\mu\nu}&=& \frac{1}{m_b} (\bar{\xi}W_c) [W_c^\dag i\not\!\!
    D_{\perp c}W_c]  \Big\{\frac{n_{-[\mu} v_{\nu]}}{n_-v},
    \gamma_{[\mu_\perp} \gamma_{\nu_\perp]}\Big\} h_v \nonumber\\
   J^{(B1)7}_{\mu\nu}&=& \frac{1}{m_b} (\bar{\xi}W_c)
   \gamma_{[\mu_\perp} \gamma_{\nu_\perp]}[W_c^\dag i\not\!\!
    D_{\perp c}W_c] \,
    h_v + J^{(B1)6}_{\mu\nu}
\end{eqnarray}
Here $a_{[\mu} b_{\nu]}=a_\mu b_\nu-a_\nu b_\mu$. The operator
$J_{\mu\nu}^{(B1)7}$ vanishes in four dimensions, but must be kept
since we regularize dimensionally.
\end{itemize}

Here we dropped the position indices $s_{1,2}$ which should be 
clear from (\ref{opbasis}).  We also dropped the operators 
involving explicit factors of position $x^\mu$, which come from the
multipole expansion (see \cite{BCDF}), since we can always 
work with the QCD currents at $x=0$. 
The choice of the A0 operators is identical to that
of \cite{Hill:2004if}, but the basis of B-operators is slightly
different. As in \cite{Hill:2004if} we combined the A1-operators 
with the A0-operators using that their coefficient functions 
are related \cite{Beneke:2004rc,Pirjol:2002km}. Since the 
A1-operators without the transverse hard-collinear gluon field do 
not contribute to the form factors at leading power~\cite{Beneke:2003pa}, 
these extra terms to the $J^{(A0)}$ will not be considered in the 
following. The SCET representation of the QCD current $J_X(0)$ is then 
\begin{equation}
J_X(0) = \sum_i \widetilde{C}_X^{(A0)i} \star J_X^{(A0)i}
  + \sum_k
  \widetilde{C}_X^{(B1)k} \star J_X^{(B1)k} +\ldots,
\label{Jx}
\end{equation}
which defines the coefficient functions for the scalar ($X=S$),
pseudoscalar ($P$), vector ($V$), axial-vector ($A$) and tensor ($T$)
currents. The star-product of coefficient function and operator in
position space is a convolution over the arguments
$\hat{s}_i$ as in (\ref{match}). 

The basis of the pseudoscalar (axial-vector) operators can be inferred 
from the scalar (vector) basis by the replacement 
$(\bar{\xi}W_c)\to (\bar{\xi}W_c)\gamma_5$. In a renormalization
with anti-commuting $\gamma_5$ (as adopted in 
 \cite{Beneke:2004rc} and in this paper), the short-distance coefficients 
$\widetilde{C}_X^{(A0)i}$, $\widetilde{C}_X^{(B1)k}$ of the 
scalar and pseudoscalar current are then equal, as are those of the 
vector and the axial-vector 
current $J_{\mu 5}=\bar{\psi} \gamma_5 \gamma_\mu Q$
(note the order of $\gamma_5$ and $\gamma_\mu$).  
In Appendix~\ref{app:Wilson} we give the transformation  
of the momentum-space coefficient functions calculated 
in \cite{Beneke:2004rc} to the new basis.
For the remainder of the paper we adopt the frame where $n_- v=1$ 
and $v=(n_-+n_+)/2$.

\subsection{Definition of the QCD form factors}

The matrix elements of the QCD currents are decomposed into
Lorentz-invariant form factors. Following the conventions of 
\cite{Beneke:2000wa} the independent form factors are 
\begin{eqnarray}
 \langle P (p^\prime)\vert \bar q \gamma^\mu b\vert \bar
B(p)\rangle &=& f_+(q^2) \bigg[p^\mu+p^{\prime
\mu}-\frac{m_B^2-m_P^2}{q^2}q^\mu\bigg]+f_0(q^2)\frac{m_B^2-m_P^2}{q^2}
q^\mu,\nonumber \\
\langle P (p^\prime)\vert \bar q \sigma^{\mu\nu}q_\nu b\vert \bar
B(p)\rangle &=&
\frac{if_T(q^2)}{m_B+m_P}\Big[q^2(p^\mu+p^{\prime\mu})-(m_B^2-m_P^2)
q^\mu\Big]
\end{eqnarray}
for pseudoscalar mesons, and 
\begin{eqnarray}
\langle V(p^\prime,\epsilon^*)\vert \bar q \gamma^\mu b\vert \bar
B(p)\rangle &=& \frac{2
iV(q^2)}{m_B+m_V}\epsilon^{\mu\nu\rho\sigma}\epsilon^*_\nu
p^\prime_\rho p_\sigma,\nonumber \\
\langle V(p^\prime,\epsilon^*)\vert \bar q \gamma^\mu \gamma_5
b\vert \bar B(p)\rangle &=& 2 m_V A_0(q^2)\frac{\epsilon^*\cdot
q}{q^2} q^\mu+ (m_B+m_V) A_1(q^2)\bigg [\epsilon^{*\mu}-
\frac{\epsilon^*\cdot q}{q^2} q^\mu\bigg ]\nonumber \\
&&-A_2(q^2)\frac{\epsilon^*\cdot q}{m_B+m_V}\bigg
[p^\mu+p^{\prime\mu}-\frac{m_B^2-m_V^2}{q^2}q^\mu\bigg ],\nonumber
\\
\langle V(p^\prime,\epsilon^*)\vert \bar q i \sigma^{\mu\nu} q_\nu
b\vert \bar B(p)\rangle &=&2i T_1
(q^2)\epsilon^{\mu\nu\rho\sigma}\epsilon^*_\nu
p^\prime_\rho p_\sigma,\nonumber \\
\langle V(p^\prime,\epsilon^*)\vert \bar q i\sigma^{\mu\nu}\gamma_5
q_\nu b\vert \bar B(p)\rangle &=&T_2(q^2)\Big [(m_B^2-m_V^2)
\epsilon^{*\mu}-(\epsilon^*\cdot q)(p^\mu+p^{\prime \mu})\Big
]\nonumber \\
&&+T_3(q^2)(\epsilon^*\cdot
q)\bigg[q^\mu-\frac{q^2}{m_B^2-m_V^2}(p^\mu+p^{\prime \mu})\bigg]
\end{eqnarray}
for vector mesons. We define $q=p-p^\prime$ and use 
the convention $\epsilon^{0123}=-1$. From now on we neglect terms
quadratic in the light meson masses $m_{P,V}$, 
but keep linear terms. In this
approximation we can put $p^\prime = E n_-$ with 
$E=n_+p^\prime/2 = (m_B^2-q^2)/(2 m_B)$, and 
$\epsilon^*\cdot n_-=0$. 

\subsection{Definition of the SCET$_{\rm I}$ form factors}

Taking into account the quantum numbers of the mesons, it is
straightforward to relate the matrix elements of the 
operators defined in (\ref{scalar:newbasis}) to
(\ref{tensor:newbasis}) and the corresponding pseudoscalar 
and axial-vector operators to a few non-vanishing \SCETI{} matrix
elements. We first note that one of the $\hat{s}_i$-integrations 
in (\ref{currentrep}) can be done explicitly using collinear 
momentum conservation~\cite{Beneke:2003pa}. This allows us to focus 
on  matrix elements of A0-operators with $s=0$ and of B-type 
operators with $s_1=0$. To see this for the case of the B-type
operators, we represent the position-space coefficient functions 
in terms of 
\be
\tilde{C}^{(B1)}_{ij\mu}(\hat{s}_1,\hat{s}_2) = 
\int\frac{dx_1}{2\pi}\frac{d x_2}{2\pi}\,
e^{-i(x_1\hat{s}_1+x_2\hat{s}_2)}\,C^{(B1)}_{ij\mu}(x_1,x_2),
\ee
where the arguments $x_i$ of the momentum-space coefficient 
functions correspond to the momentum fractions $x_i=n_+
p^\prime_i/m_b$ of the collinear building blocks $(\bar \xi
W_c)_{s_1}$ and $(W_c^\dagger i D_{\perp c}^\mu W_c)_{s_2}$ 
of the current operator. Then with (\ref{currentrep}) and  
(\ref{opbasis}) we obtain 
\begin{eqnarray}
&&\langle M(p^\prime)| \int d\hat{s}_1 d\hat{s}_2\,
  \widetilde{C}^{(B1)}_{ij\mu}(\hat{s}_1,\hat{s}_2)\,
  O_{j}^{(B1)\mu}(s_1,s_2;0)|\bar B_v\rangle 
= \frac{1}{m_b} \int d\tau \,
  C^{(B1)}_{ij\mu}\!\left(\frac{2 E\bar \tau}{m_b},\frac{2 E\tau}{m_b}\right) 
\nonumber\\
&& \hspace*{1cm}\times  \,(2 E)\int\frac{dr}{2\pi}\,e^{-i\,2E\tau r}\,
  \langle M(p^\prime)|(\bar \xi W_c)(0) 
  (W_c^\dagger i D_{\perp c}^\mu W_c)(r n_+) \Gamma^\prime_j h_v(0)
  |\bar B_v\rangle
\label{ft2}
\end{eqnarray}
with $\bar\tau=1-\tau$. Abusing notation we will write the
coefficient functions simply as $C^{(B1)}_{ij\mu}(E,\tau)$ in the 
following. Next, the $J^{(B1)}$-operators (except 
for the tensor operators with two transverse indices, which 
we do not need in the following) can be written as (Fourier 
transforms of) linear combinations of 
\begin{equation}
{\cal J}_k(\tau)=2 E \int \frac{dr}{2\pi}\,e^{-i\,2 E \tau r}\, 
(\bar\xi W_c)(0)(W_c^\dagger i D_{\perp c}^\mu
W_c)(r n_+) \Gamma_k h_v(0),
\label{odef}
\end{equation}
where for $k=1,2,3$ the Dirac matrix $\Gamma_k$ can take one of 
the three expressions  
\be
\Gamma_k = \{(\gamma_5) \gamma^{\mu_\perp}, 
(\gamma_5)\gamma_{\nu_\perp}\gamma^{\mu_\perp}, 
(\gamma_5)\gamma^{\mu_\perp}\gamma_{\nu_\perp} \},
\label{gammaset}
\ee
and $\hat r=r m_b$. Here the $\gamma_5$ in brackets means that this 
factor may be added. This notation is convenient, because many of 
the results below do not depend on the extra factor of $\gamma_5$. 
In the following definitions we leave out the position argument of 
a field, when it is $x=0$. Eq.~(\ref{ft2}) suggests defining the 
B-type form factors as the matrix elements 
of the operators ${\cal J}_k(\tau)$. 

We therefore define the two leading-power 
\SCETI{} form factors for pseudoscalar mesons through   
\begin{eqnarray}
&& \langle P(p^\prime)| (\bar\xi W_c)h_v|\bar B_v\rangle
 = 2 E\,\xi_P(E),
\nonumber \\[0.2cm]
&& \langle P(p^\prime)| (\bar\xi W_c)(W_c^\dagger i
\Slash{D}_{c\perp} W_c)(rn_+) h_v|\bar B_v\rangle
 = 2 m_b E \int d\tau \,e^{i\,2E\tau r}\,\Xi_P(\tau,E).
\label{XiP}
\end{eqnarray}
Here $|\bar B_v\rangle$ denotes the $\bar B$ meson state in the 
static limit (see the Lagrangian (\ref{lagrangian}) below) normalized 
to $2 m_B$ (rather than 1 as is conventional in heavy quark 
effective theory). The second definition is such that 
\be 
\Xi_P(\tau,E) =
(2 m_b E)^{-1}\,\langle P(p^\prime)|{\cal J}_1(\tau)|\bar B_v\rangle
\label{normXi}
\ee 
(no $\gamma_5$ in ${\cal J}_1(\tau)$).
Similarly, for vector mesons 
\begin{eqnarray}
&& \langle V(p^\prime)| (\bar\xi W_c)\gamma_5 h_v|\bar B_v\rangle
 = -2 E \,\epsilon^*\cdot v \,\xi_\parallel(E),
\nonumber \\[0.4cm]
&& \langle V(p^\prime)| (\bar\xi W_c)\gamma_5 \gamma_{\mu_\perp}
h_v|\bar B_v\rangle
 = -2 E \,(\epsilon_\mu-\epsilon\cdot v \,n_{-\mu})\,\xi_\perp(E),
\nonumber \\[0.4cm]
&& \langle V(p^\prime)| (\bar\xi W_c)\gamma_5 (W_c^\dagger i
\Slash{D}_{c\perp} W_c)(rn_+) h_v|\bar B_v\rangle
\nonumber \\
&&\hspace*{2cm} = -2 m_b E \,\epsilon^*\cdot v \int d\tau
\,e^{i\,2E\tau r}\,\Xi_\parallel(\tau,E),
\nonumber\\[0.2cm]
&& \langle V(p^\prime)| (\bar\xi W_c)\gamma_5
\gamma_{\mu\perp}(W_c^\dagger i \Slash{D}_{c\perp}
W_c)(rn_+)h_v|\bar B_v\rangle
\nonumber \\
&&\hspace*{2cm}  = -2 m_b E \,(\epsilon^*_\mu-\epsilon^*\cdot v
\,n_{-\mu}) \int d\tau \,e^{i\,2E\tau r}\,\Xi_\perp(\tau,E),
\nonumber \\[0.2cm]
&& \langle V(p^\prime)| (\bar\xi W_c)\gamma_5 (W_c^\dagger i
\Slash{D}_{c\perp} W_c)(rn_+) \gamma_{\mu_\perp} h_v|\bar
B_v\rangle
\nonumber \\
&& \hspace*{2cm} = -2 m_b E \,(\epsilon^*_\mu-\epsilon^*\cdot v
\,n_{-\mu}) \int d\tau \,e^{i\,2E\tau r}\,\tilde
\Xi_\perp(\tau,E).
\label{XiPerp}
\end{eqnarray}
The tensor operators $J_{\mu\nu}^{(B1)6}$, 
$J_{\mu\nu}^{(B1)7}$ must have vanishing matrix elements 
between a pseudoscalar $B$ meson and a pseudoscalar or 
vector meson, so the set of B-type operators ${\cal J}_{1-3}(\tau)$ 
(now including $\gamma_5$) is complete. We shall find in 
Section~\ref{sec:jetfunc} that the matrix element that defines 
$\tilde \Xi_\perp(\tau,E)$ (corresponding to ${\cal J}_3(\tau)$) 
vanishes at leading order in the 
$1/m_b$-expansion, hence we set $\tilde \Xi_\perp(\tau,E)=0$ in 
the remainder of this section. The dependence on the polarization 
vector $\epsilon^*$ shows that the form factors with subscript 
$a=\,\parallel$ ($a=\,\perp$) refer to longitudinally (transversely) 
polarized vector mesons.

\subsection{Form factor expressions}

Taking the matrix element of (\ref{Jx}), using (\ref{ft2}), 
and inserting the definitions 
of the QCD and the \SCETI{} form factors we derive  
expressions for the QCD form factors, in which the effects at the scale 
$m_b$ are explicitly factorized. 
For the pseudoscalar meson form factors, we find 
\begin{eqnarray}
\label{QCDinSCET} f_+(E) &=& C^{(A0)}_{f_+}(E) \,\xi_P(E)+\int d\tau
\,C^{(B1)}_{f_+}(E,\tau) \,\Xi_P(\tau,E)\,,\nonumber\\
 \frac{m_B}{2 E} \,f_0(E) &=& C^{(A0)}_{f_0}(E) \,\xi_P(E)+\int d\tau
\,C^{(B1)}_{f_0}(E,\tau) \,\Xi_P(\tau,E)\,,\nonumber\\
\frac{m_B}{m_B+m_P} \,f_T(E) &=&C^{(A0)}_{f_T}(E) \,\xi_P(E)+\int
d\tau \,C^{(B1)}_{f_T}(E,\tau) \,\Xi_P(\tau,E).
\label{pscet1}
\end{eqnarray}
Similarly, for the form factors of vector mesons
\begin{eqnarray}
\frac{m_B}{m_B+m_V} \,V(E) &=& C^{(A0)}_{V}(E) \,\xi_\perp(E)+\int
d\tau
\,C^{(B1)}_{V}(E,\tau) \,\Xi_\perp(\tau,E)\,,\nonumber\\
\frac{m_V}{E}\,A_0(E) &=&C^{(A0)}_{f_0}(E) \,\xi_\parallel(E)+\int
d\tau
\,C^{(B1)}_{f_0}(E,\tau) \,\Xi_\parallel(\tau,E)\,,\nonumber\\
\frac{m_B+m_V}{2 E} \,A_1(E) &=&C^{(A0)}_{V}(E) \,\xi_\perp(E)+\int
d\tau
\,C^{(B1)}_{V}(E,\tau) \,\Xi_\perp(\tau,E)\,,\nonumber\\
\frac{m_B+m_V}{2 E} \,A_1(E) &-&\frac{m_B-m_V}{m_B}
\,A_2(E)\nonumber\\
&=&C^{(A0)}_{f_+}(E) \,\xi_\parallel(E)+\int d\tau
\,C^{(B1)}_{f_+}(E,\tau) \,\Xi_\parallel(\tau,E)\,,\nonumber\\
T_1(E) &=&C^{(A0)}_{T_1}(E) \,\xi_\perp(E)+\int d\tau
\,C^{(B1)}_{T_1}(E,\tau) \,\Xi_\perp(\tau,E)\,,\nonumber \\
\frac{m_B}{2 E} \,T_2(E) &=&C^{(A0)}_{T_1}(E) \,\xi_\perp(E)+\int
d\tau
\,C^{(B1)}_{T_1}(E,\tau) \,\Xi_\perp(\tau,E)\,,\nonumber\\
\frac{m_B}{2 E} \,T_2(E) - T_3(E)
&=&C^{(A0)}_{f_T}(E) \,\xi_\parallel(E)+\int d\tau
\,C^{(B1)}_{f_T}(E,\tau) \,\Xi_\parallel(\tau,E).
\label{vscet1} 
\end{eqnarray}
We recall that $E$ denotes the energy of the light meson. 
The coefficient functions $C^{(A0)}_F$ and $C^{(B1)}_F$ 
are defined as linear combinations of momentum-space 
coefficients functions 
of A0- and B-type operators. Remarkably, the ten different 
form factors involve only five independent linear combinations 
of the A0- and B-type coefficients as a consequence of helicity 
conservation of the strong interactions at leading power in the 
$1/m_b$-expansion. This implies 
\be 
\frac{m_B}{m_B+m_V} \,V(E) = \frac{m_B+m_V}{2 E} \,A_1(E), 
\qquad 
T_1(E) = \frac{m_B}{2 E} \,T_2(E) 
\label{exactrel}
\ee
up to power corrections \cite{Burdman:2000ku}, and the equality 
of the coefficients pertaining to pseudoscalar mesons and longitudinally 
polarized vector mesons. The five pairs 
$(C^{(A0)}_F,C^{(B1)}_F)$ constitute the main result of the 
first matching step from QCD to \SCETI. The 1-loop expressions can 
be inferred from \cite{Beneke:2004rc,Becher:2004kk}. They are  
collected in Appendix~\ref{app:Wilson}.

\subsection{Physical form factor scheme}

Since the \SCETI{} form factors $\xi_a(E)$ are not known, it has been 
suggested in \cite{Beneke:2000wa} to define them in terms of three 
QCD (or ``physical'') form factors. Let 
\be
\xi_P^{\rm FF} \equiv f_+, 
\qquad 
\xi_\perp^{\rm FF} \equiv \frac{m_B}{m_B+m_V} \,V,
\qquad 
\xi_{\parallel}^{\rm FF} \equiv 
\frac{m_B+m_V}{2 E} \,A_1 -\frac{m_B-m_V}{m_B}
\,A_2,
\label{ffscheme}
\ee
which corresponds to \cite{Beneke:2000wa} except for 
the longitudinal form factor $\xi_{\parallel}^{\rm FF}$. The 
above definition, which has already been adopted in \cite{Beneke:2004dp}, 
is preferred, since it preserves the equality of the 
short-distance coefficients for the 
pseudoscalar meson and longitudinal vector meson form factors. 

In the physical form factor scheme we have again the two 
identities (\ref{exactrel}), and the five remaining form factors 
read
\beq
\frac{m_B}{2 E} \,f_0 &=&    
R_0 \, \xi_P^{\rm FF} + 
\left(C^{(B1)}_{f_0}-C^{(B1)}_{f_+} \,R_0\right)
\star \,\Xi_P,  
\nonumber \\[0.2cm]
\frac{m_B}{m_B+m_P} \,f_T &=&  
R_T\, \xi_P^{\rm FF} + 
\left(C^{(B1)}_{f_T}-C^{(B1)}_{f_+}\, R_T\right)
\star \,\Xi_P,
\nonumber\\[0.2cm]
T_1 &=&    
R_\perp \, \xi_\perp^{\rm FF} + 
\left(C^{(B1)}_{T_1}-C^{(B1)}_{V} \,R_\perp\right)
\star \,\Xi_\perp, 
\nonumber \\[0.2cm]
\frac{m_V}{E}\,A_0  &=& 
R_0 \, \xi_\parallel^{\rm FF} + 
\left(C^{(B1)}_{f_0}-C^{(B1)}_{f_+} \,R_0\right)
\star \,\Xi_\parallel,  
\nonumber \\[0.2cm]
\frac{m_B}{2 E} \,T_2 - T_3 &=& 
R_T\, \xi_\parallel^{\rm FF} + 
\left(C^{(B1)}_{f_T}-C^{(B1)}_{f_+}\, R_T\right)
\star \,\Xi_\parallel.
\label{physscet1}
\eeq
Here the asterisk is a shorthand for the convolution integral over 
$\tau$. The ratios $R$ of A0-coefficients take much simpler expressions 
than the individual coefficients given in Appendix~\ref{app:Wilson}. 
Up to one loop \cite{Beneke:2000wa}
\beq
R_0 &\equiv& \frac{C^{(A0)}_{f_0}}{C^{(A0)}_{f_+}} 
= 1+\frac{\alpha_s C_F}{4\pi} \Big[2-2 \ell\Big],
\nonumber\\
R_T &\equiv& \frac{C^{(A0)}_{f_T}}{C^{(A0)}_{f_+}}
= 1+\frac{\alpha_s C_F}{4\pi} \Big[\ln\frac{m_b^2}{\nu^2}+ 2 \ell\Big],
\nonumber\\
R_\perp  &\equiv& \frac{C^{(A0)}_{T_1}}{C^{(A0)}_{V}}
= 1+\frac{\alpha_s C_F}{4\pi} \Big[\ln\frac{m_b^2}{\nu^2}- \ell\Big]
\label{rfactors}
\eeq 
with 
\be 
\ell\equiv -\frac{2 E}{m_b-2 E}\ln\frac{2 E}{m_b},
\ee
$C_F=4/3$, and $\nu$ the renormalization scale of the 
QCD tensor current. In 
the physical form factor scheme there are only three non-trivial 
ratios $R$ and three non-trivial combinations of B-coefficients.


\section{Jet-functions}
\label{sec:jetfunc}

In this section we turn to the main part of this paper, the
calculation of the \SCETI{} form factors $\Xi_a(\tau,E)$. Technically,
this amounts to matching the B-type \SCETI{} currents, 
${\cal J}_k(\tau)$, whose matrix
elements define the $\Xi_a(\tau,E)$, to four-fermion 
operators in \SCETII{}. These four-fermion operators factorize into a
product of two light-cone distribution amplitudes resulting in the
desired expression (\ref{b1match}). That this can be done 
follows from \cite{Beneke:2003pa}, where it has been shown that 
at leading power in the heavy quark expansion the B-type currents 
match only to four-fermion operators with convergent 
convolution integrals. In terms of operators, we derive the matching relation 
\begin{eqnarray}
&& {\cal J}_k(\tau) = 
2 E \int \frac{d r}{2\pi}\,e^{-i\,2 E \tau r}\, (\bar\xi
W_c)(0) (W_c^\dagger i {D}_{\perp c\mu} W_c)(r n_+)\Gamma_k h_v(0)
\nonumber  \\
&& = \, \int d\omega dv \,J_{k}(\tau;v,
\omega) \Big[(\bar\xi W_c)(s n_+)\frac{\slash{n}_+}{2}\Gamma^c_k
(W_c^\dagger\xi)(0)\Big]_{\rm FT} \Big[(\bar q_s Y_s)(t
n_-)\frac{\slash{n}_-}{2}\gamma_5 (Y_s^\dagger h_v)(0)\Big]_{\rm
FT}\nonumber
 \\
&& \hspace*{0.5cm}+\,\ldots
\label{matchrel}
\end{eqnarray}
to the 1-loop order, 
where the ellipses denote terms that have vanishing matrix elements
between $\bar B$ mesons and pseudoscalar or vector mesons, or are 
power-suppressed in $1/m_b$, $\Gamma_k^c$ will be defined after 
(\ref{fierz3}), and the
subscript ``FT'' denotes a Fourier transformation with respect to the
light-cone variables $s,t$ that will be made more precise
later (see (\ref{Pdef})). The functions $J_{k}$ are the short-distance 
coefficients of the \SCETII{} operators, which contain the
hard-collinear effects from the scale $(m_b\Lambda)^{1/2}$ 
integrated out in passing to \SCETII. 
These short-distance coefficients will be called ``jet-functions''.

\subsection{Set-up of the calculation}

The jet-functions $J_{k}(\tau;v,\omega)$ are extracted by computing 
both sides of (\ref{matchrel}) between appropriate quark states. We 
therefore consider the four-quark matrix element of the 
operators ${\cal J}_k(\tau)$,
\begin{equation}
{\cal A}_k(\tau;v,\omega)=\langle q(p_1^\prime)\bar q(p_2^\prime)| 
{\cal J}_k(\tau)|\bar q(l) b(m_b v)\rangle,
\label{adef}
\end{equation}
where $k=1,2,3$, and the momenta $l$,
$p_i^\prime$ ($i=1,2$) are soft and collinear, respectively. $q$ 
denotes a light quark, $b$ the $b$-quark. The quark and anti-quark 
in the final state may be of different flavour, but the flavours of 
the initial and final state anti-quark are identical. The quark-antiquark 
initial and final states are assumed to be in a colour-singlet state. 
Since the operators on the right-hand side of (\ref{matchrel}) do not 
contain derivatives, the transverse momenta of the collinear states 
can be set to zero, and we can define $p_1^\prime = v p^\prime = 
v E n_-$, $p_2^\prime =\bar v p^\prime$ with $\bar v\equiv 1-v$. Likewise 
for the soft states the momenta can be taken to be $m_b v$ for the 
heavy quark and $l=\omega n_+/2$ for the light anti-quark. The four 
functions $\Xi_a(\tau,E)$ defined in (\ref{XiP}), (\ref{XiPerp})
correspond to setting $\Gamma_k = (\gamma_5)\gamma^{\mu_\perp}$ 
$(\Xi_P,\Xi_\parallel)$, $\Gamma_k = \gamma_5\gamma_{\nu_\perp}
\gamma^{\mu_\perp}$ ($\Xi_\perp$), and $\Gamma_k=
\gamma_5\gamma^{\mu_\perp}\gamma_{\nu_\perp}$ ($\tilde\Xi_\perp$). 

The operators ${\cal J}_k(\tau)$ generate 
momentum-space vertices with an arbitrary number of $n_+ A_c$ gluons 
due to the Wilson lines $W_c$. Of these only the one- and two-gluon 
vertices are needed in the 1-loop calculation. The corresponding Feynman 
rules read (collinear quark lines dashed, all gluon momenta are out-going)
{\small 
\begin{eqnarray}
&&\nonumber\\[-0.5cm]
&&
  \begin{picture}(100,30)(0,30)
    \SetScale{1} 
    \SetWidth{1.2} 
    \ArrowLine(0,30)(50,30)
    \SetWidth{0.8} 
    \DashArrowLine(50,30)(100,30){3}
    \SetWidth{0.5} 
    \GCirc(50,30){3}{0}
    \DashLine(50,30)(50,60){3}
    \Gluon(50,30)(50,-5){3}{4}
    \Text(55,35)[lb]{\small $\Gamma_k,\,\mu$}
    \Text(50,-10)[ct]{\small $k,\,\rho,\,A$}
    \Text(100,35)[lb]{\small $p_1^\prime$}
   \end{picture}
\hspace*{1.5cm}
g_s\,\delta\bigg(\tau -\frac{n_+ k}{n_+ p^\prime}\bigg)
\left(g_\perp^{\mu\rho}-\frac{n_+^\rho k_\perp^\mu}{n_+ k}\right)
T^A\,\Gamma_k, 
\label{opvertex1}\\[1.8cm]
&& 
  \begin{picture}(100,30)(0,30)
    \SetScale{1} 
    \SetWidth{1.2} 
    \ArrowLine(0,30)(50,30)
    \SetWidth{0.8} 
    \DashArrowLine(50,30)(100,30){3}
    \SetWidth{0.5} 
    \GCirc(50,30){3}{0}
    \DashLine(50,30)(50,60){3}
    \Gluon(50,30)(30,-5){3}{4}
    \Gluon(50,30)(70,-5){3}{4}
    \Text(55,35)[lb]{\small $\Gamma_k,\,\mu$}
    \Text(20,-10)[ct]{\small $k_1,\,\rho,\,A$}
    \Text(80,-10)[ct]{\small $k_2,\,\sigma,\,B$}
    \Text(100,35)[lb]{\small $p_1^\prime$}
   \end{picture}
\vspace*{2mm}
\hspace*{1.5cm}
g_s^2 \,\Bigg\{\left(\delta\bigg(\tau-\frac{n_+ k_2}{n_+ p^\prime}\bigg)
-\delta\bigg(\tau-\frac{n_+ (k_1+k_2)}{n_+ p^\prime}\bigg)\right)
\nonumber\\
&&\hspace*{6cm}
\times \frac{n_+^\rho}{n_+ k_1}
\left(g_\perp^{\mu\sigma}-\frac{n_+^\sigma k_{2\perp}^\mu}{n_+ k_2}\right)
+\delta\bigg(\tau-\frac{n_+ (k_1+k_2)}{n_+ p^\prime}\bigg)
\nonumber\\
&&\hspace*{6cm}
\times \frac{n_+^\sigma}{n_+ k_2}
\left(g_\perp^{\mu\rho}-\frac{n_+^\rho (k_{1}+k_{2})_\perp^\mu}
{n_+ (k_1+k_2)}\right)
\Bigg\}\,T^A T^B\,\Gamma_k 
\nonumber\\
&&\hspace*{5cm}
+ \,(k_1\leftrightarrow k_2, \,A\leftrightarrow B, 
\,\rho\leftrightarrow\sigma).
\label{opvertex2}
\end{eqnarray}
}
\noindent {}\hspace*{-2.5mm} In light-cone gauge 
$n_+ A_c=0$ the variable $\tau$ corresponds to the fraction of total 
collinear longitudinal momentum $n_+ p^\prime$ carried 
by the transverse hard-collinear gluon.

In addition the calculation requires the collinear interactions from 
the leading-power SCET Lagrangian, 
\begin{eqnarray}
{\cal L} &=&\bar{\xi}\left(i \nm D +i\Slash{D}_{\perp c}\frac{1}{i\np D_c}
i\Slash{D}_{\perp c}\right)\frac{\slash{n}_+}{2}\xi 
 -\frac{1}{2}\,\mbox{tr}\left(F_c^{\mu\nu}
  F_{\mu\nu c}\right) 
\nonumber\\
&&+\, \bar h_v i v D_s h_v + \bar q_s i\Slash{D}_{s} q_s 
 - \frac{1}{2}\,\mbox{tr}\left(F_s^{\mu\nu} F_{\mu\nu s}\right), 
\label{lagrangian}
\end{eqnarray}
as well as the sub-leading interaction \cite{BCDF}
\begin{eqnarray}
{\cal L}^{(1)}_{\xi q}&=&\bar q_s W_c^\dagger i\Slash{D}_{\perp c}\xi - 
\bar \xi i\overleftarrow{\Slash D}_{\perp c} W_c q_s 
\label{eq:Lqxi}
\end{eqnarray}
that converts the soft spectator anti-quark in the $\bar B$ meson 
into an energetic, collinear anti-quark. The Feynman rules 
for the collinear interactions can be found in \cite{Bauer:2000yr}, 
while (\ref{eq:Lqxi}) implies the vertices (collinear quark line dashed,
soft quark line solid, gluon momenta outgoing) 
{\small 
\begin{eqnarray}
&&\nonumber\\[0.3cm]
&&
  \begin{picture}(100,30)(0,30)
    \SetScale{1} 
    \SetWidth{0.8} 
    \DashArrowLine(0,30)(50,30){3}
    \SetWidth{0.5} 
    \ArrowLine(50,30)(100,30)
    \GCirc(50,30){1}{0}
    \Gluon(50,30)(50,65){3}{4}
    \Text(50,70)[cb]{\small $k,\,\rho,\,A$}
    \Text(0,35)[rb]{\small $p$}
   \end{picture}
\hspace*{1.5cm}
i g_s T^A 
\left(\gamma_\perp^{\rho}-\frac{n_+^\rho \slash{p}_\perp}{n_+ p}\right),
\\
&&
\nonumber\\
&&
\nonumber\\
&&
  \begin{picture}(100,30)(0,30)
    \SetScale{1} 
    \SetWidth{0.8} 
    \DashArrowLine(0,30)(50,30){3}
    \SetWidth{0.5} 
    \ArrowLine(50,30)(100,30)
    \GCirc(50,30){1}{0}
    \Gluon(50,30)(30,65){3}{4}
    \Gluon(50,30)(70,65){3}{4}
    \Text(20,70)[cb]{\small $k_1,\,\rho,\,A$}
    \Text(80,70)[cb]{\small $k_2,\,\sigma,\,B$}
    \Text(0,35)[rb]{\small $p$}
   \end{picture}
\hspace*{1.5cm}
 -i g_s^2 T^A T^B \,\frac{n_+^\rho}{n_+ k_1}
\left(\gamma_\perp^{\sigma}-\frac{n_+^\sigma \slash{p}_\perp}{n_+ p}\right)
\nonumber\\
&&\hspace*{5.5cm} 
+ \,(k_1\leftrightarrow k_2, \,A\leftrightarrow B, 
\,\rho\leftrightarrow\sigma).
\end{eqnarray}}
\noindent\hspace*{-1.5mm}{}We note 
that $n_+ p=n_+ (k_1+k_2)$ and $p_\perp = (k_1+k_2)_\perp$, since 
the corresponding soft momentum components are neglected in collinear-soft 
interaction vertices (multipole expansion). 

\subsection{Unrenormalized matrix element}

\subsubsection{Tree}

\begin{figure}[b]
\vspace*{-3.5cm}
\hspace*{1.5cm}
\epsfig{file=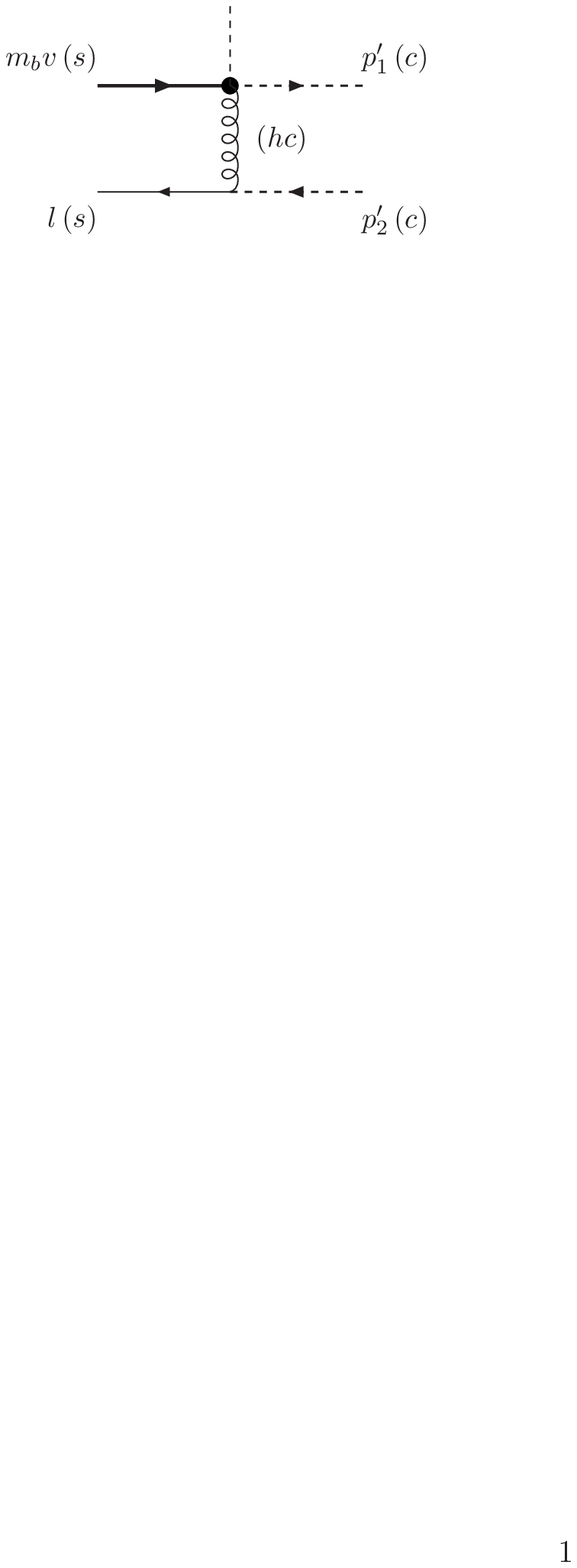,angle=0}
\vspace*{-22.7cm}
\caption{Tree diagram.}
\label{fig:tree}
\end{figure}

The tree contribution to (\ref{adef}) is shown in Figure~\ref{fig:tree}. 
The gluon momentum (outgoing from the operator vertex) is given by 
$k=p_2^\prime -l$, and with $k^2=-\bar v n_+ p^\prime n_- l = 
-2 E \omega \bar v$, we obtain 
\be 
{\cal A}_k^{(0)}(\tau;v,\omega) = 
- \frac{g_s^2 C_F}{N_c}\frac{1}{2 E \omega \bar v}\,
\delta (\tau-\bar v) \,\Gamma_k \otimes \gamma_\perp^\mu\,,
\label{treeJ}
\end{equation}
where $\Gamma_1 \otimes \Gamma_2$ means
$\bar{u}_c(p_1^\prime) \Gamma_1 u_h(m_b v) \,\bar{v}(l)\Gamma_2
v_c(p_2^\prime)$.
The heavy quark spinor satisfies $\slash{v}u_h(m_b v)=u_h(m_b v)$, 
and for the collinear spinors $\slash{n}_- v_c(p_2^\prime)=
\bar{u}_c(p_1^\prime)\slash{n}_-  = 0$.

\subsubsection{One loop}

\begin{figure}[t]
\vspace*{0cm}
\centerline{\epsfig{file=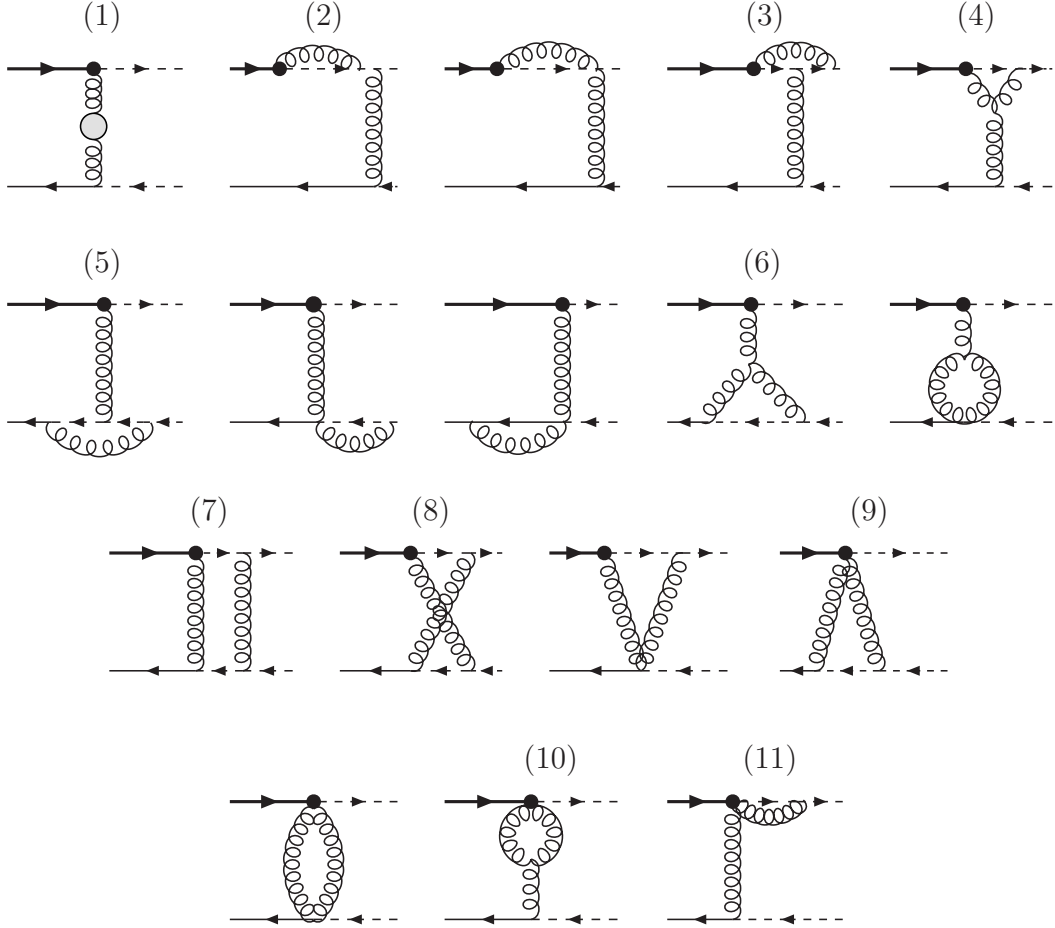,width=14cm,angle=0}}
\vskip0.5cm
\caption{1-loop diagrams. The first diagram summarizes all gluon 
propagator corrections. Diagrams without numbers are omitted 
when the calculation is done by expansion of QCD diagrams.}
\label{fig:oneloop}
\end{figure}

A generic 1-loop diagram in \SCETI{} contains contributions 
from the hard-collinear, collinear and soft momentum region. 
For our external momentum configuration soft and collinear 
loop integrals are scaleless, so the diagram computation extracts 
the hard-collinear contribution. The \SCETI{} diagrams are shown in 
Figure~\ref{fig:oneloop}, omitting diagrams that are obviously 
scaleless. We use dimensional regularization ($d=4-2\epsilon$) 
for both ultraviolet and infrared singularities, hence scaleless 
integrals vanish. 

We computed the 1-loop diagrams in two different ways. First, 
we used the \SCETI{} Feynman rules as given above and computed the 
diagrams as shown in the figure. Second, we computed the 
matrix element (\ref{adef}) with full QCD Feynman rules, but 
expanded the Feynman integrand under the assumption that 
the loop momentum is hard-collinear, and the external momenta 
collinear and soft, respectively. This method, also known as 
the ``strategy of expanding by 
regions''~\cite{Beneke:1998zp,Smirnov:1998vk}, gives precisely 
the same result as the first computation, but it turns out 
to be algebraically simpler, because it avoids having to use 
the more complicated vertex 
Feynman rules generated by the SCET Lagrangian. There are 
also fewer diagrams to compute. There are no 
two-gluon $q\bar q gg$ vertices in full QCD, and the 
unnumbered diagrams in Figure~\ref{fig:oneloop} are absent. In both 
computations we write 
\be
\frac{d^d k}{(2\pi)^d} = \frac{1}{8\pi^2}\,d n_+ k d n_-k \,
\frac{d^{d-2}k_\perp}{(2\pi)^{d-2}},
\ee
and first perform the $n_- k$-integral by contour integration. 
The $k_\perp$-integral reduces to a conventional Feynman integral;  
the $n_+ k$-integration is eliminated by the $\delta$-function 
in (\ref{opvertex1}) or (\ref{opvertex2}) 
with the exception of diagrams such as (1), (5) or (6).

It is convenient to perform the calculation without specifying 
the Dirac matrix $\Gamma_k$ of the \SCETI{} operator.
The unrenormalized matrix element (\ref{adef}) can be written 
as
\begin{eqnarray} 
{\cal A}_k^{(ur)} &=&
\Big[A^{(0)}\!+A^{(1)}\Big]\,\Gamma_k\otimes\gamma^{\mu_\perp} +
B^{(1)}\gamma^{\rho_\perp}\gamma^{\mu_\perp}\Gamma_k\otimes
\gamma_{\rho_\perp}
+ C^{(1)}\gamma^{\rho_\perp}\gamma^{\lambda_\perp}\Gamma_k\otimes
\gamma^{\mu_\perp}\gamma_{\lambda_\perp}\gamma_{\rho_\perp}
\hspace*{0.7cm}
\label{aur}
\end{eqnarray}
in terms of Dirac structures that cannot be reduced further in 
$d$ dimensions. The notation is such 
that for any quantity the superscript (0) denotes
the tree and (1) the 1-loop contribution. The coefficients of 
all three structures are infrared divergent, but only 
$A^{(1)}$ and $B^{(1)}$ are ultraviolet divergent. It follows 
that all B-type currents can be renormalized with only two
independent renormalization constants. Specifically, inserting 
the three Dirac structures (\ref{gammaset}), we obtain 
\begin{eqnarray}
{\cal A}_1^{(ur)} &=& \left[A+(d-2) B\right]\,(\gamma_5)
\gamma_{\mu_\perp}\otimes\gamma^{\mu_\perp} + C \,(\gamma_5)
\gamma^{\rho_\perp}\gamma^{\lambda_\perp}\gamma^{\mu_\perp}\otimes
\gamma_{\mu_\perp}\gamma_{\lambda_\perp}\gamma_{\rho_\perp},
\nonumber\\[0.2cm]
{\cal A}_2^{(ur)} &=& A\, (\gamma_5)\gamma_{\nu_\perp}\gamma_{\mu_\perp}\otimes
\gamma^{\mu_\perp} + (4-d) B\,(\gamma_5)
\gamma_{\mu_\perp}\gamma_{\nu_\perp} \otimes\gamma^{\mu_\perp} 
\nonumber\\[0.2cm]
&&\hspace*{1cm}+\,C \,(\gamma_5)
\gamma^{\rho_\perp}\gamma^{\lambda_\perp}\gamma_{\nu_\perp}\gamma^{\mu_\perp}
\otimes\gamma_{\mu_\perp}\gamma_{\lambda_\perp}\gamma_{\rho_\perp},
\nonumber\\[0.2cm]
{\cal A}_3^{(ur)} &=&\left[A+(d-2) B\right]\,(\gamma_5)  \gamma_{\mu_\perp}
\gamma_{\nu_\perp} \otimes \gamma^{\mu_\perp} + C \,(\gamma_5)
\gamma^{\rho_\perp}\gamma^{\lambda_\perp} \gamma^{\mu_\perp}
\gamma_{\nu_\perp}\otimes
\gamma_{\mu_\perp}\gamma_{\lambda_\perp}\gamma_{\rho_\perp}.
\phantom{--}
\label{bare3}
\end{eqnarray}
Here and in the following we use anti-commuting $\gamma_5$, and the 
bracket around $\gamma_5$ refers to the two cases, where $\gamma_5$
is or is not included in $\Gamma_k$. From 
this it can be seen that the UV divergent parts have the same Dirac
structure as the original operator. Hence, the 
operators ${\cal J}_k(\tau)$ and all the B-type current operators 
do not mix under renormalization (in the basis adopted here). 
The renormalization constant for the operator ${\cal J}_k(\tau)$ 
with $\Gamma_{1,3}$ is related to the
divergent part of $A^{(1)}+ 2 B^{(1)}$, the one for ${\cal J}_k(\tau)$ 
with $\Gamma_2$ to $A^{(1)}$.

\subsection{Ultraviolet counterterms}

The counterterm diagrams are obtained from the tree diagram by 
insertions of a counterterm vertex into the gluon propagator, 
the quark gluon vertex and the operator vertex. Finally, 
the on-shell matrix element must be multiplied by the propagator residue 
factors $R_{h_v}^{1/2} R_{q_s}^{1/2} R_\xi$. 
They are related to the field renormalization constants by 
\begin{eqnarray}
R_X=\frac{Z_X^{\rm OS}}{Z_X}, 
\end{eqnarray}
where $Z_X^{\rm OS}$ is the renormalization constant of field $X$ in the
on-shell scheme (by definition $R_X^{\rm OS}=1$) and $Z_X$ is the
field renormalization constant in the $\overline{\rm MS}$ scheme.

The Lagrangian counterterms are standard. As regards the operator 
counterterms, we note that the three operators ${\cal J}_k(\tau)$ 
do not mix (see above), 
hence the renormalized operator is related to the bare operator 
(expressed in terms of bare fields) by 
\begin{eqnarray}
{\cal J}_{1,3}(\tau) &=& \int d\tau^\prime 
\,Z_\parallel(\tau,\tau^\prime) \,{\cal J}_{1,3}^{\rm bare}(\tau^\prime),
\nonumber\\
{\cal J}_{2}(\tau) &=& \int d\tau^\prime 
\,Z_\perp(\tau,\tau^\prime) \,{\cal J}_{2}^{\rm bare}(\tau^\prime),
\label{zfact2}
\end{eqnarray}
which defines the operator renormalization kernels. 
Here we used that ${\cal J}_1(\tau)$ and ${\cal J}_3(\tau)$ renormalize 
identically.

Putting together all the renormalization constants, the on-shell 
ultraviolet-renor\-ma\-lized matrix elements of ${\cal J}_1$ and 
${\cal J}_3$ follow from (\ref{bare3}) by the replacement
\begin{eqnarray}
A+(d-2) B &\to& A+(d-2) B + Z_\parallel^{(1)}\star A^{(0)}
\nonumber\\
&&+\,\Big(2 Z_g^{(1)}+\frac{1}{2}\, \Big(Z_{h_v}^{{\rm
OS}(1)}+Z_{q_s}^{{\rm OS}(1)}+ 2 Z_\xi^{{\rm OS}(1)}\Big)\Big)\,
A^{(0)},
\label{ren13}
\end{eqnarray}
while the renormalized matrix element of $J_2$ is the 
second line of (\ref{bare3}) with 
\begin{eqnarray}
A &\to& A  + Z_\perp^{(1)}\star A^{(0)}
\nonumber\\
&&+\,\Big(2 Z_g^{(1)}+\frac{1}{2}\, \Big(Z_{h_v}^{{\rm
OS}(1)}+Z_{q_s}^{{\rm OS}(1)}+ 2 Z_\xi^{{\rm OS}(1)}\Big)\Big)\,
A^{(0)}.
\label{ren2}
\end{eqnarray}
The asterisk denotes convolution in the variable $\tau^\prime$ 
as in the definition of the operator renormalization kernels. 
The on-shell field renormalization factors are all equal to 1 in
dimensional regularization, so $Z_X^{{\rm OS}(1)}=0$. $Z_g$ is 
the standard $\overline{\rm MS}$ strong coupling renormalization 
factor, $g_s^{\rm bare}=Z_g g_s$ with ($C_A=3$, $T_f=1/2$)
\be
Z_g=1-\frac{\alpha_s\beta_0}{8\pi\epsilon},\qquad 
\beta_0=\frac{11 C_A}{3}-\frac{4}{3}n_f T_f.
\label{beta0}
\ee
The matrix elements (\ref{bare3}) with the substitutions 
(\ref{ren13}), (\ref{ren2}) are now ultraviolet finite, 
but infrared divergent. The infrared divergences are reproduced 
by the \SCETII{} computation of the matrix element, resulting in 
a finite jet-function. This can be used to {\em determine} the
operator renormalization kernels $Z_{\parallel}$ and $Z_\perp$ alternative 
to the direct computation of the operators' ultraviolet 
singularities performed in \cite{Hill:2004if}. That is, after 
extracting the jet-function, we shall obtain the renormalization
factors by requiring that they render the result finite.

\subsection{Matching to SCET${}_{\rm II}$ and extraction of the jet
function}

\subsubsection{\SCETII{} matrix element}

To obtain the jet-function, we need the \SCETII{} matrix elements 
on the right hand side of the matching equation (\ref{matchrel}) to
the 1-loop order. In the absence of collinear-soft interactions 
in \SCETII{} the four-quark operator factorizes into a collinear and a
soft bilinear. We define
\begin{eqnarray}
Q[\Gamma_k^c](v) &=& \Big[(\bar\xi W_c)(s n_+)\frac{\slash{n}_+}{2}\Gamma^c_k
(W_c^\dagger\xi)(0)\Big]_{\rm FT} 
\nonumber\\
&=&  \frac{n_+
p^\prime}{2\pi}\int ds\,e^{-is v n_+ p^\prime}\, (\bar\xi W_c)(s
n_+)\frac{\slash{n}_+}{2}\Gamma^c_k(W_c^\dagger\xi)(0),
\nonumber\\
P(\omega) &=& 
\Big[(\bar q_s Y_s)(t
n_-)\frac{\slash{n}_-}{2}\gamma_5 (Y_s^\dagger h_v)(0)\Big]_{\rm FT}
\nonumber\\
&=& \frac{1}{2\pi}\int dt\,e^{it \omega}\,
(\bar q_s Y_s)(t n_-)\frac{\slash{n}_-}{2}\gamma_5(Y_s^\dagger h_v)(0),
\label{Pdef}
\end{eqnarray} 
and the quark matrix elements
\begin{eqnarray} 
\Phi_{q\bar q}(v^\prime,v) &=& \langle
q(p_1^\prime)\bar q(p_2^\prime)|Q[\Gamma_k^c](v^\prime)|0\rangle,
\nonumber\\[0.0cm]
\Phi_{b\bar q}(\omega^\prime,\omega) &=& 
\langle 0|P(\omega^\prime)|\bar q(l) b_v(0)\rangle.
\label{quarkP}
\end{eqnarray} 
The Dirac matrices $\Gamma_k^c$ will be determined later by the 
Fierz transformation of the first Dirac matrix product on the 
right-hand sides of (\ref{bare3}). 
Note that the hadronic matrix elements of $Q$ and $P$ are precisely 
the meson light-cone distribution amplitudes. Collinear-soft 
factorization\footnote{We recall here that in general 
the apparent factorization of collinear and soft degrees of freedom in
\SCETII{} is not valid \cite{Beneke:2003pa,Becher:2003qh} due to the 
non-existence of a regulator that preserves factorization. The
unregulated integrals correspond to endpoint-divergent convolution 
integrals. However, it has been shown that for the particular case 
of the matrix elements of the B-type currents, the 
convolution integrals must be convergent~\cite{Beneke:2003pa}, 
so collinear-soft factorization 
{\em is} valid here.} in \SCETII{} means that the matrix element of 
$Q[\Gamma_k^c](v)P(\omega)$, which appears in (\ref{matchrel}),  
factorizes into 
\begin{equation}
\langle M(p^\prime)| Q[\Gamma_k^c](v)P(\omega)|\bar B_v\rangle 
= \langle M(p^\prime)| Q[\Gamma_k^c](v)|0 \rangle 
\langle 0| P(\omega)|\bar B_v\rangle,
\label{scet2fact}
\end{equation} 
which is a product of light-cone distribution amplitudes.

With these definitions the tree quark matrix element of the 
\SCETII{} four-quark operator is given by 
\be
\langle q(p_1^\prime)\bar q(p_2^\prime)|
Q[\Gamma_k^c](v^\prime)P(\omega^\prime)|\bar q(l) b(m_b v)\rangle^{(0)}
= \delta(v-v^\prime)\delta(\omega-\omega^\prime)\,
\frac{\slash{n}_+}{2}\Gamma^c_k\,\tilde\otimes\,
\frac{\slash{n}_-}{2}\gamma_5,
\ee
where now $\Gamma_1 \,\tilde \otimes \,\Gamma_2$ means
$\bar{u}_c(p_1^\prime) \Gamma_1 v_c(p_2^\prime)\,\bar{v}(l)\Gamma_2
u_h(m_b v)$. The ``tensor products'' $\otimes$ and $\tilde\otimes$ 
are related by a Fierz transformation as discussed below.  

The \SCETII{} computation of the matrix elements (\ref{quarkP}) 
is very simple, since in the collinear sector \SCETII{} is 
equivalent to full QCD, and in the soft sector to heavy quark effective 
theory \cite{Beneke:2003pa}. The external momenta do not allow to 
form a non-vanishing kinematic invariant, hence all loop integrals 
are scaleless and vanish. The matrix elements are given by tree
diagrams including tree diagrams with counterterm insertions. We can
therefore write
\begin{eqnarray} 
\Phi_{q\bar q}(v^\prime,v) &=& Z_Q(v^\prime,v)\,
\frac{\slash{n}_+}{2}\Gamma^c_k,
\nonumber\\[-0.2cm]
\Phi_{b\bar q}(\omega^\prime,\omega) &=& 
Z_P(\omega^\prime,\omega) \,\frac{\slash{n}_-}{2}\gamma_5,
\end{eqnarray} 
with $Z_Q(v^\prime,v)$ and $Z_P(\omega^\prime,\omega)$ 
the renormalization kernels that relate the renormalized operator to
the bare operator expressed in terms of the bare fields, 
\begin{eqnarray}
Q[\Gamma_k^c](v) &=& \int_0^1 dw\, Z_Q(v,w)\,Q[\Gamma_k^c]^{\rm bare}(w),
\nonumber\\
P(\omega) &=& \int_0^1
d\omega^{\prime}\,Z_P(\omega,\omega^{\prime})\,
P^{\rm bare}(\omega^{\prime}).
\end{eqnarray}

The required 1-loop renormalization factors have been computed in 
other contexts. For the collinear operator $Z_Q(v,w)$ is the 
Brodsky-Lepage kernel \cite{Lepage:1980fj}
\begin{eqnarray} Z_Q(v,w)&=& \delta(v-w) + Z_Q^{(1)}(v,w) +
\ldots,
\nonumber\\[0.2cm]
Z_Q^{(1)}(v,w)&=& 
-\frac{\alpha_s C_F}{2\pi\epsilon}\,
\Bigg\{\frac{1}{w\bar w} \left[v\bar w\frac{\theta(w-v)}{w-v}+
\bar v w \frac{\theta(v-w)}{v-w} \right]_+
\nonumber\\
&& -\,\frac{1}{2}\,\delta(v-w)
+ \Delta \left(\frac{v}{w}\theta(w-v)+\frac{\bar v}{\bar w}
\theta(v-w)\right)\Bigg\}, 
\label{blkernel}
\end{eqnarray} 
where $\Delta=1$ applies to $\Gamma_k^c=(\gamma_5) 1$ (pseudoscalar meson,
longitudinally polarized vector meson distribution amplitudes) and 
$\Delta=0$ to $\Gamma_k^c=\gamma_\perp^\nu$ (transversely
polarized vector meson distribution amplitude), and the plus-distribution 
is defined for symmetric kernels $f$ as  
 \begin{eqnarray}
\int dw [f(v,w)]_+ \,g(w) = \int dw f(v,w)\,\left(g(w)-g(v)\right). 
\end{eqnarray}
Similarly, the renormalization of $P(\omega)$ has been
worked out in \cite{Lange:2003ff} with the result
\begin{eqnarray} 
Z_P(\omega,\omega^\prime)&=&
\delta(\omega-\omega^\prime) + Z_P^{(1)}(\omega,\omega^\prime) +
\ldots,
\nonumber \\[0.2cm]
Z_P^{(1)}(\omega,\omega^\prime)& =& \frac{\alpha_s C_F}{4\pi}\,
\Bigg[\left(\frac{1}{\epsilon^2}+\frac{2}{\epsilon}\, 
\ln\frac{\mu}{\omega}-\frac{5}{2\epsilon}\right)
\delta(\omega-\omega^\prime)
\nonumber\\[0.2cm]
&&-\, \frac{2\omega}{\epsilon}\left(\frac{1}{\omega^\prime}\,
\frac{\theta(\omega^\prime-\omega)}{\omega^\prime-\omega}+
\frac{1}{\omega}\,
\frac{\theta(\omega-\omega^\prime)}{\omega-\omega^\prime}\right)_+
\Bigg]. 
\label{lnkernel}
\end{eqnarray}

\subsubsection{Extraction of the jet-function}

The jet-function is extracted from the quark matrix element of 
(\ref{matchrel}), 
\begin{eqnarray}
{\cal A}_k(\tau;v,\omega) = \int_0^\infty d\omega^\prime
\int_0^1 dv^\prime\,J_k(\tau;\omega^\prime,v^\prime)\, \Phi_{b\bar
q}(\omega^\prime,\omega)\,\Phi_{q\bar q}(v^\prime,v).
\end{eqnarray}
This gives
\be
J_k^{(0)}(\tau;\omega,v)\,\frac{\slash{n}_+}{2}\tilde\Gamma_k^c\tilde \otimes
\,\frac{\slash{n}_-}{2}\gamma_5
= {\cal A}^{(0)}(\tau;v,\omega),
\label{exttree}
\ee
at tree level, and 
\begin{eqnarray}
J_k^{(1)}(\tau;\omega,v)\,
\frac{\slash{n}_+}{2}\tilde\Gamma_k^c\tilde \otimes
\,\frac{\slash{n}_-}{2}\gamma_5
&=& {\cal A}_k^{(1)}(\tau;v,\omega) - \bigg[\int_0^\infty d\omega^\prime
\, Z_P^{(1)}(\omega^\prime,\omega)\, J_k^{(0)}(\tau;\omega^\prime,v)
\nonumber\\[0.2cm]
&&\hspace*{-2cm} +\, \int_0^1 dv^\prime\,Z_Q^{(1)}(v^\prime,v)\,
J_k^{(0)}(\tau;\omega,v^\prime) \bigg]\,\frac{\slash{n}_+}{2}
\tilde\Gamma_k^c\tilde
\otimes\,\frac{\slash{n}_-}{2}\gamma_5
\label{subtraction}
\end{eqnarray} 
at the 1-loop order. ${\cal A}_k^{(1)}(\tau;v,\omega)$ 
is the ultraviolet-renormalized but infrared divergent 
1-loop matrix element of ${\cal J}_k(\tau)$. The subtraction 
on the right-hand side precisely cancels the infrared divergences, 
so that the short-distance jet-function is finite as it should be. 
There is however a difficulty in executing the subtraction 
in the form of (\ref{subtraction}), since ${\cal A}_k^{(1)}(\tau;v,\omega)$ 
is expressed in terms of the spinor product $\otimes$, corresponding
to \SCETII{} operators with spinor indices contracted in $[\bar\xi h_v] \,
[\bar q_s \xi]$, while the left-hand side  and the subtraction term 
involve the product $\tilde \otimes$, corresponding
to operators with spinor indices contracted in $[\bar\xi \xi] \,
[\bar q_s h_v]$. The standard Fierz identities that relate these 
structures are valid only in four dimensions. We shall discuss 
this issue together with the reduction of evanescent Dirac structures 
appearing in (\ref{bare3}) in the following subsection. 

None of this applies to the tree-level equation (\ref{exttree}), where
all quantities are finite. We can therefore apply the four-dimensional
Fierz identities to the tree-level terms in 
(\ref{bare3}), so that for 
\begin{eqnarray}
&k=1:\qquad&(\gamma_5)\gamma_{\mu_\perp}\otimes \gamma^{\mu_\perp} 
 = (\pm1)\,\frac{\Slash{n}_+}{2}\gamma_5(\gamma_5)\,\tilde\otimes\,
\frac{\Slash{n}_-}{2}\gamma_5 +\ldots,
\nonumber\\
&k=2:\qquad&(\gamma_5)\gamma_{\nu_\perp}
\gamma_{\mu_\perp}\otimes \gamma^{\mu_\perp} 
 = -\frac{\Slash{n}_+}{2}\gamma_{\nu_\perp}
\gamma_5(\gamma_5)\,\tilde\otimes\,
\frac{\Slash{n}_-}{2}\gamma_5 +\ldots,
\nonumber\\[0.2cm]
&k=3:\qquad&(\gamma_5)\gamma_{\mu_\perp}\gamma_{\nu_\perp}
\otimes \gamma^{\mu_\perp} 
 = \ldots.
\label{fierz3}
\end{eqnarray}
The Fierz transformation produces a number of 4-fermion operators 
with different Dirac structures. 
In the following we discuss only those operators  
(jet-functions) that contribute to the decay of a pseudoscalar $\bar
B$ meson. The ellipses denote terms involving $\Slash{n}_-/2$ and 
$\Slash{n}_-/2\,\gamma_{\nu_\perp}(\gamma_5)$ to the right of
$\tilde\otimes$, which vanish when the matrix 
element $\langle 0|\bar q_s [\ldots]h_v|\bar B\rangle$ 
with the pseudoscalar $\bar B$ meson state is evaluated. 
Hence, these terms will not be considered further. 
The upper (lower) sign refers to the operators on the left-hand side 
without (with) the $\gamma_5$ factor. Here and below the Fierz
identities are given for four-fermion {\em operators} and the extra
minus sign from the field permutation 
relative to the identities for matrix products is already 
included. Comparison of (\ref{fierz3}) with the definition
(\ref{Pdef}) of $Q[\Gamma_k^c]$ determines the Dirac matrix 
$\Gamma_k^c$ in the collinear \SCETII{} operator. For 
$k=1$, we have $\Gamma_k=(\gamma_5)\gamma_{\mu_\perp}$ and 
the corresponding $\tilde \Gamma_k^c=(\pm 1) \gamma_5(\gamma_5)$; for $k=2$, 
$\Gamma_k=(\gamma_5)\gamma_{\nu_\perp}\gamma_{\mu_\perp}$ 
and $\tilde \Gamma_k^c=-\gamma_{\nu_\perp}\gamma_5(\gamma_5)$; 
for $k=3$, $\Gamma_k=(\gamma_5)\gamma_{\mu_\perp}\gamma_{\nu_\perp}$ 
and $\tilde \Gamma_k^c=0$, i.e. there is no contribution for 
pseudoscalar $B$ mesons. Hence, comparing (\ref{exttree}),
(\ref{fierz3}) to (\ref{treeJ}) leads to the tree-level jet
function 
\be 
J_k^{(0)}(\tau;v,\omega) = 
- \frac{g_s^2 C_F}{N_c}\frac{1}{2 E \omega \bar v}\,
\delta (\tau-\bar v)
\label{treeJmatch}
\end{equation}
for $k=1,2$, and 0 for $k=3$. 

Using the 1-loop expressions for the \SCETII{} renormalization kernels,
we can evaluate the subtraction term in square brackets in 
(\ref{subtraction}) with the result
\begin{eqnarray} -\frac{g_s^2 C_F}{N_c}\,\frac{1}{2 E \omega \bar
v} \, \left\{\frac{\alpha_s C_F}{4\pi}\, \delta(\tau-\bar
v)\left(\frac{1}{\epsilon^2}+ \frac{2}{\epsilon}
\ln\frac{\mu}{\omega}- \frac{5}{2\epsilon} \right) +
\frac{\bar v}{\tau} Z_Q^{(1)}(\bar\tau,v)
\right\}. 
\label{explicitsubtract}
\end{eqnarray}

\subsubsection{Evanescent operators and Fierz transformation}

We now discuss the reduction of the Dirac structure and the Fierz
transformation. We consider in detail the case 
$\Gamma_k=\gamma^{\mu_\perp}$. 
According to the first equation of (\ref{bare3}) 
the UV renormalized matrix element of the corresponding 
SCET${}_{\rm I}$ current ${\cal J}_1(\tau)$ is
\begin{eqnarray}
{\cal A}_1(\tau;v,\omega) = 
\hat A\, \gamma_{\mu_\perp}\otimes\gamma^{\mu_\perp} + C \,
\gamma^{\rho_\perp}\gamma^{\lambda_\perp}\gamma^{\mu_\perp}\otimes
\gamma_{\mu_\perp}\gamma_{\lambda_\perp}\gamma_{\rho_\perp},
\end{eqnarray}
where $\hat A$ stands for the right hand side of (\ref{ren13}). 
The second Dirac structure reduces to a multiple of the first one in 
four dimensions, and the first one satisfies the Fierz relation
(\ref{fierz3}) in four dimensions. In the following we discuss
separately the reduction of the Dirac structure and the Fierz
transformation in $d$ dimensions, although it is possible to perform
both reductions in a single step. 

Indicating only the field content and Dirac structure we define 
the operators $O_1=(\bar\xi \gamma_{\mu_\perp} h_v)
(\bar q_s \gamma^{\mu_\perp} \xi)$ and 
$O_1^\prime=(\bar\xi \gamma^{\rho_\perp}\gamma^{\lambda_\perp}
\gamma^{\mu_\perp} h_v)
(\bar q_s \gamma_{\mu_\perp}\gamma_{\lambda_\perp}\gamma_{\rho_\perp}
\xi)$, and rewrite the previous equation as 
\begin{eqnarray}
{\cal A}_1(\tau;v,\omega) =
\left[\hat A^{(0)}+\hat A^{(1)}\right] \langle O_1\rangle^{(0)} + 
C^{(1)} \langle O_1^\prime\rangle^{(0)}, 
\end{eqnarray}
where the tree-level matrix elements of $O_1^{(\prime)}$ just
reproduce the Dirac
structures, i.e.  $\langle O_1\rangle^{(0)} = \bar u_c(p_1^\prime) 
\gamma^{\mu_\perp} u_h(m_b v) \,\bar v(l)\gamma_{\mu_\perp} 
v_c(p_2^\prime)$ etc. In $d=4$, $O_1^\prime=4 O_1$, so we define the evanescent
operator $E=O_1^\prime-f(\epsilon) O_1$, where $f(0)=4$, but otherwise
$f(\epsilon)$ is arbitrary. Hence, 
\begin{eqnarray}
{\cal A}_1(\tau;v,\omega) =
\left[\hat A^{(0)}+\hat A^{(1)}+f(\epsilon) C^{(1)}\right]
\langle O_1\rangle^{(0)} + C^{(1)} \langle E\rangle^{(0)}.
\end{eqnarray}
Since $C^{(1)}$ has a $1/\epsilon$ infrared divergence, 
the coefficient of the ``physical'' operator $O_1$ depends on the 
up to now arbitrary prescription $f(\epsilon)$ in the definition 
of the evanescent operator. The jet-functions are obtained 
by expressing this equation in terms of the renormalized 
operator matrix elements. To the 1-loop order it is sufficient 
to use 
\begin{eqnarray}
\langle O_1\rangle =\left(1+M_{O_1}^{(1)}\right)
\langle O_1\rangle^{(0)} +M_{O_1 E}^{(1)}\langle E\rangle^{(0)},
\qquad 
\langle E\rangle=\langle E^{(0)}\rangle,
\label{1lopme}
\end{eqnarray}
which results in 
\begin{eqnarray}
{\cal A}_1(\tau;v,\omega) &=&
\left[\hat A^{(0)}+\hat A^{(1)}+f(\epsilon) C^{(1)}-
\hat A^{(0)}M_{O_1}^{(1)}\right]
\langle O_1\rangle + \left[C^{(1)}-
\hat A^{(0)}M_{O_1E}^{(1)}\right] \langle E\rangle
\nonumber\\
&\stackrel{d\to 4}\to& 
\left[J_{O_1}^{(0)}+J_{O_1}^{(1)}\right] \langle O_1\rangle.
\label{wrongfierz}
\end{eqnarray}
The coefficients of the operator matrix elements in the first line 
of this equation are now infrared-finite short-distance quantities, 
and the equation is interpreted 
as an operator matching relation valid also when the matrix elements 
are taken between hadronic final states. Hence we can take the limit 
$d\to 4$ in which $\langle E\rangle=0$, since the tree-level matrix 
element vanishes in four dimensions.\footnote{In higher orders one usually 
uses a non-minimal subtraction scheme to ensure that $\langle
E\rangle=0$, but this is not relevant to the present calculation.} 
The 1-loop jet-function $J_{O_1}^{(1)}$ can be read off from 
(\ref{wrongfierz}). It depends on the choice
of the evanescent operator, i.e. the order-$\epsilon$ term of
$f(\epsilon)$, but so does $ \langle O_1\rangle$ as can be 
seen from (\ref{1lopme}), and
the physical amplitude is scheme-independent.

The jet-function $J_{O_1}$ is not the desired result, because instead 
of $O_1$ we must use the Fierz-transformed operator 
\begin{equation}
P_1 = \Big(\bar\xi \frac{\Slash{n}_+}{2}\gamma_5\,\xi\Big) 
\,\Big(\bar q_s \frac{\Slash{n}_-}{2}\gamma_5 h_v\Big) +\ldots,
\end{equation}
where the ellipses denote terms that are irrelevant for us, as in 
(\ref{fierz3}). This Fierz-ordering is uniquely singled out 
by collinear-soft factorization in SCET${}_{\rm II}$. Radiative 
corrections to $P_1$ occur only within the collinear factor or 
within the soft factor, hence the Dirac structures in arbitrary 
loop diagrams can be reduced to the tree structure $P_1$. It follows 
that whatever evanescent operator one may write down in this 
Fierz-ordering decouples from $P_1$, and can simply 
be ignored. Hence, even in $d$ dimensions we have 
\begin{eqnarray}
{\cal A}_1(\tau;v,\omega)=
\left[J_{P_1}^{(0)}+J_{P_1}^{(1)}\right] \langle P_1\rangle,
\label{rightfierz} 
\end{eqnarray} 
with $J_{P_1}^{(1)}$ the desired 1-loop jet-function. Comparing this equation
to (\ref{wrongfierz}) we have $J_{P_1}=J_{O_1}$, if the 
renormalized matrix elements $\langle P_1\rangle$ and 
$\langle O_1\rangle$ are equal. In general this is not the case,
because the Fierz identity that relates $O_1$ and $P_1$ in 
four dimensions is not valid for $d\not=4$. Now, the 
renormalization scheme for $P_1$ is fixed by the requirement 
that the collinear and soft matrix elements coincide with the 
standard $\overline{\rm MS}$ definition of the light meson 
and heavy meson light-cone distribution 
amplitudes, respectively, so $\langle P_1\rangle$ is completely
defined. However, by choosing $f(\epsilon)$, 
we can adjust the definition of $O_1$, such that the infrared-finite 
matrix elements are equal in the limit $d\to 4$. 

Since the jet-functions are independent of the infrared 
regularization, any one can be chosen for the calculation, 
and it is convenient to 
compare the matrix elements of $O_1$ and $P_1$ by assuming 
an off-shell infrared regularization. The ultraviolet-renormalized 
1-loop matrix elements are 
\begin{eqnarray}
\langle O_1\rangle &=&\left(1+M_{O_1}^{\rm off}\right)
\langle O_1\rangle^{(0)} +M_{O_1 E}^{\rm off}\langle E\rangle^{(0)},
\nonumber \\
\langle P_1\rangle &=&\left(1+M_{P_1}^{\rm
off}\right) \langle P_1\rangle^{(0)}.
\end{eqnarray}
There are no $1/\epsilon$ poles in these equations due to the use
of off-shell IR regularization, so the evanescent term drops out
for $d\to 4$. Furthermore, $\langle O_1\rangle^{(0)} =
\langle P_1\rangle^{(0)}$ by the four-dimensional Fierz-equivalence 
of $O_1$ and $P_1$, hence we need to define $O_1$ such that 
the difference $M_{O_1}^{\rm off}-
M_{P_1}^{\rm off}=0$. At the 1-loop order one finds 
that the self-energy corrections and the vertex correction, where 
the gluon is exchanged between the soft light quark and the heavy
quark, drop out, because the coupling to the heavy quark 
is proportional to $v^\mu$ and does not introduce a new Dirac
structure. Only the gluon exchange between the collinear 
$\xi$-fields can give a non-zero difference. In the case of $O_1$ 
the part of this diagram that does not cancel in the difference 
$M_{O_1}^{\rm off}-M_{P_1}^{\rm off}$ involves 
$\gamma^{\rho_\perp}\gamma^{\lambda_\perp}\gamma^{\mu_\perp}\otimes
\gamma_{\mu_\perp}\gamma_{\lambda_\perp}\gamma_{\rho_\perp}$, 
and its contribution to $M_{O_1}$ is therefore proportional to 
$f(\epsilon)$. In the case of $P_1$, the Dirac structure is 
\begin{eqnarray}
\gamma^{\alpha_\perp}\gamma^{\beta_\perp} \frac{\Slash{n}_+}{2}
\gamma_{\beta_\perp}\gamma_{\alpha_\perp}\tilde\otimes
\frac{\Slash{n}_-}{2}  =
(d-2)^2 \langle  P_1\rangle^{(0)}.
\end{eqnarray}
Including the overall coefficient, we find 
\begin{equation}
M_{O_1}^{\rm off}-M_{P_1}^{\rm off} = 
-[C^{(1)}]_{\rm div} \left(f(\epsilon)-(d-2)^2\right) \langle
P_1^{(0)}\rangle 
\end{equation}
where $[C^{(1)}]_{\rm div}$ is the divergent part of $C^{(1)}$. 
In order for this to vanish, we must choose $f(\epsilon) =(d-2)^2$. 
In other words, we must define 
\begin{eqnarray}
\gamma^{\rho_\perp}\gamma^{\lambda_\perp}\gamma^{\mu_\perp}\otimes
\gamma_{\mu_\perp}\gamma_{\lambda_\perp}\gamma_{\rho_\perp}
 = (d-2)^2 \,\gamma^{\mu_\perp}\otimes
\gamma_{\mu_\perp}. 
\end{eqnarray}
With this $J_{P_1}=J_{O_1}$, so returning to (\ref{wrongfierz}) 
we have 
\begin{equation}
J_{P_1}^{(1)} = \lim_{d\to 4} \left(
\hat A^{(1)}+(d-2)^2 C^{(1)}-
\hat A^{(0)} M_{P_1}^{(1)}\right).
\end{equation}
The term $\hat A^{(0)} M_{P_1}^{(1)}$ is nothing but the subtraction 
term (\ref{explicitsubtract}) that appears in (\ref{subtraction}), 
while $\hat A^{(1)}$ and $C^{(1)}$ can be read off 
from the 1-loop calculation that leads to (\ref{aur}) 
and (\ref{ren13}), so the previous equation gives the final result 
for the correctly renormalized and subtracted jet-function. 
The above argument can be repeated for
$\Gamma_k=\gamma_5\gamma^{\mu_\perp}$, and one finds that the
jet-function is identical to the one for $\Gamma_k=\gamma^{\mu_\perp}$
as was expected.

The case of B-type currents ${\cal J}_{2,3}(\tau)$ with one uncontracted 
transverse index is slightly more complicated
than the scalar case, because there are two physical and two
evanescent operators. Corresponding to 
$\Gamma_k = (\gamma_5)\gamma^{\nu_\perp}\gamma^{\mu_\perp}$ 
and $(\gamma_5)\gamma^{\mu_\perp}\gamma^{\nu_\perp}$ 
we define the two physical operators  
$O_2=(\bar\xi (\gamma_5)\gamma_{\nu_\perp}\gamma_{\mu_\perp}  h_v)
(\bar q_s \gamma^{\mu_\perp} \xi)$ 
and $O_3=(\bar\xi (\gamma_5)\gamma_{\mu_\perp}\gamma_{\nu_\perp}  h_v)
(\bar q_s \gamma^{\mu_\perp} \xi)$. Requiring that the renormalized 
matrix elements of $O_{2,3}$ equal the renormalized matrix elements 
of the corresponding operators $P_{2,3}$ in the other Fierz-ordering 
fixes uniquely the prescription for reducing the Dirac structures 
multiplying $C$ in (\ref{bare3}) and ensures that 
the \SCETII{} operators are 
correctly minimally subtracted. A short calculation 
analogous to the one discussed above gives 
\begin{eqnarray}
&&(\gamma_5) \gamma^{\rho_\perp}\gamma^{\lambda_\perp}
\gamma_{\nu_\perp}\gamma^{\mu_\perp} \otimes
\gamma_{\mu_\perp}\gamma_{\lambda_\perp}\gamma_{\rho_\perp}  =
(d-4)^2\,(\gamma_5)\gamma_{\nu_\perp}\gamma_{\mu_\perp}
\otimes\gamma^{\mu_\perp},
\nonumber\\[0,2cm]
&& (\gamma_5)\gamma^{\rho_\perp}\gamma^{\lambda_\perp}
\gamma^{\mu_\perp} \gamma_{\nu_\perp}\otimes
\gamma_{\mu_\perp}\gamma_{\lambda_\perp}\gamma_{\rho_\perp} =
(d-2)^2\,(\gamma_5) \gamma_{\mu_\perp}\gamma_{\nu_\perp}
\otimes\gamma^{\mu_\perp}.
\label{gam2}\end{eqnarray}
Referring to (\ref{fierz3}) we see that the matrix element of 
$O_3$ vanishes for a pseudoscalar $\bar B$ meson. Hence 
the B-type operator ${\cal J}_3(\tau)$ with 
$\Gamma_k=(\gamma_5)\gamma_{\mu_\perp}\gamma_{\nu_\perp}$ has
a vanishing jet-function just as at tree level. On the other hand, 
inserting the first equation of (\ref{gam2}) into 
the second of (\ref{bare3}), we find that 
the B-type operator ${\cal J}_2(\tau)$ matches to $J_{P_2}P_2 $ with 
\begin{equation}
J_{P_2}^{(1)} = \lim_{d\to 4} \left(
A^{(1)}+(d-4)^2 C^{(1)}-
A^{(0)} M_{P_2}^{(1)}\right).
\end{equation}
Here $A^{(1)}$ and $C^{(1)}$ are defined by (\ref{bare3}) and 
(\ref{ren2}) and the subtraction term is given by 
(\ref{subtraction}) except that now $\Delta=0$ must be used in the 
Brodsky-Lepage kernel. Since $C^{(1)}$ has only a single $1/\epsilon$ pole, 
the term $(d-4)^2 C^{(1)}$ does not contribute to the final 
result for the jet-function. 

\subsection{Final results for the jet-functions and B-type 
current re\-nor\-ma\-lization kernels}

We shall now denote the jet-function $J_{P_1}$ ($J_{P_2}$) that 
arises in the matching of ${\cal J}_1(\tau)$ (${\cal J}_2(\tau)$) 
as $J_\parallel$ ($J_\perp$). 
To the 1-loop order we write
 \begin{eqnarray} 
J_a &=& -\frac{g_s^2 C_F}{N_c} \,
\frac{1}{2 E\omega\bar v} \left(\delta(\tau-\bar v) +
\frac{\alpha_s}{4\pi} \,j_a(\tau;v, \omega)\right)
\end{eqnarray}
with $a=\,\parallel,\perp$. Applying the subtraction procedure 
described in the previous subsections we obtain the ultraviolet and 
infrared finite 1-loop corrections 
$j_a(\tau;v, \omega)$. The resulting expressions still depend 
on the yet undetermined renormalization factors $Z_a^{(1)}$ 
of the B-type currents through (\ref{ren13}), (\ref{ren2}). They can now be 
determined by the condition that $j_a(\tau;v, \omega)$ 
must not contain $1/\epsilon$ poles. We choose the 
$\overline{\rm MS}$ scheme to be consistent with the definition 
of the B-type short-distance coefficients in \cite{Beneke:2004rc}.

\subsubsection{Renormalization kernels}

We expand the $Z$-factors in (\ref{zfact2}) as
\be
Z_a(\tau,\tau^\prime)=\delta(\tau-\tau^\prime)+
\frac{\alpha_s}{4\pi} \,z_a^{(1)}(\tau,\tau^\prime),
\ee
and obtain
\begin{eqnarray}
\label{zfact}
z_\parallel^{(1)}(\tau,\tau^\prime) &=& (-C_F)
\left(\frac{1}{\epsilon^2}+\frac{2}{\epsilon}\ln\frac{\mu}{2 E}\right)
\delta(\tau-\tau^\prime)-\frac{1}{\epsilon}\,\Big[
z_1(\tau,\tau^\prime)+z_2(\tau,\tau^\prime)\Big],
\nonumber\\[0.0cm]
z_\perp^{(1)}(\tau,\tau^\prime) &=&  (-C_F)
\left(\frac{1}{\epsilon^2}+\frac{2}{\epsilon}\ln\frac{\mu}{2 E}\right)
\delta(\tau-\tau^\prime)-\frac{z_1(\tau,\tau^\prime)}{\epsilon},
\end{eqnarray}
with 
\begin{eqnarray}
z_1(\tau,\tau^\prime) &=& \delta(\tau-\tau^\prime)\left\{
C_F \left[-2 \ln \bar \tau+ \frac{5}{2}\right ]+C_A
\ln\frac{\bar \tau}{\tau}\right\}
\nonumber\\
&&+\,C_A \left[\frac{\theta(\tau-\tau^\prime)}{\tau-\tau^\prime}+
\frac{\theta(\tau^\prime-\tau)}{\tau^\prime-\tau}\right]_+
+ \left(C_F-\frac{C_A}{2}\right)\,\frac{2\bar \tau}{\tau 
\tau^\prime}\theta(\tau-\bar\tau^\prime)
\nonumber\\
&&-\,C_A\left(
\frac{1}{\tau\bar \tau^\prime}\theta(\tau-\tau^\prime)+
\frac{1}{\tau^\prime}\theta(\tau^\prime-\tau)\right),
\nonumber\\
z_2(\tau,\tau^\prime) &=& (-2) \left(C_F-\frac{C_A}{2}\right) 
\left(\frac{\tau
    \tau^\prime}{\bar\tau^\prime}\theta(\bar\tau^\prime-\tau) 
+ \bar\tau\left(1+\frac{1}{\tau}+\frac{1}{\tau^\prime}\right)
\theta(\tau-\bar\tau^\prime)\right)
\nonumber\\
&&+\,C_A\left(\frac{\bar\tau\tau^\prime}{\bar \tau^\prime}
\left(1+\frac{1}{\tau}\right)\theta(\tau-\tau^\prime)+
\tau \left(1+\frac{1}{\tau^\prime}\right)
\theta(\tau^\prime-\tau)\right).
\label{re-kernel} 
\end{eqnarray}
The anomalous dimensions of the B-type operators are derived from 
the renormalization kernels in the standard way (see (\ref{andim}) 
below). Our
result is in complete agreement with \cite{Hill:2004if}, where the 
anomalous dimension has been obtained by extracting directly the
ultraviolet divergent parts of the \SCETI{} 
diagrams.\footnote{The variable $u$ in \cite{Hill:2004if} 
corresponds to our $\bar\tau=1-\tau$, their $v$ to our $\bar\tau^\prime$.}

\subsubsection{Jet-functions}

The 1-loop jet-functions read
\beq
 j_\parallel(\tau;v, \omega) &=& A \,\delta(\tau-\bar v) + 
\bigg (C_F-\frac{C_A}{2}\bigg ) \,\big[2 B\big]_+ 
\nonumber\\
&&+\,C_F\bigg[\theta(\bar v-\tau)\,\frac{2\bar\tau}{v \bar
  v}\bigg(L+\ln\frac{(\bar v-\tau)\tau}{\bar v}+
  \frac{v(\bar v-\tau)}{\tau\bar\tau}\bigg)-
  \frac{2\bar v \bar \tau}{v}\bigg(L+\ln\tau\bar\tau+
  \frac{v\bar\tau}{\bar v\tau}\bigg)\bigg]
\nonumber\\
&&-\,\bigg (C_F-\frac{C_A}{2}\bigg)\Bigg[
\theta(\tau-v)\,\frac{2(v-\tau)^2}{v\bar v\tau}
\bigg(L+\ln\frac{(\tau-v)\bar\tau}{\bar v}-\frac{v\bar \tau}{(v-\tau)^2}\bigg)
\nonumber\\
&&+\,\theta(\bar v-\tau)\,\frac{2 (v-\tau)}{v\bar v}
\bigg(L+\ln(\bar v-\tau)+\frac{\tau\bar \tau}{(\bar v-\tau) (v-\tau)}
\ln\frac{\bar v}{\tau}-\frac{v}{v-\tau}\bigg)
\nonumber\\
&&+\,\theta(\tau-\bar v)\,\frac{2}{\tau}
\bigg(L+\ln(\tau-\bar v)+\frac{\bar v}{\tau-\bar v}
\ln\frac{v}{\bar\tau}-\frac{\bar\tau}{v}\bigg)\Bigg],
\label{jparfinal}\\ 
 j_\perp(\tau;v, \omega) &=& A \,\delta(\tau-\bar v) + 
\bigg (C_F-\frac{C_A}{2}\bigg ) \,\big[2 B\big]_+ 
\nonumber\\
&&+\,C_F\bigg[\theta(\bar v-\tau)\,\frac{2}{\bar
  v}\bigg(L+\ln\frac{(\bar v-\tau)\tau}{\bar v}-1\bigg)-\theta(\tau-\bar
  v)\,\frac{2\bar \tau}{v\tau}\bigg]
\nonumber\\
&&-\,\bigg (C_F-\frac{C_A}{2}\bigg)\Bigg[
\theta(\tau-v)\,\frac{2\bar\tau}{\bar  v\tau}
\bigg(L+\ln\frac{(\tau-v)\bar\tau}{\bar v}-\frac{v-\tau}{v\bar\tau}\bigg)
\nonumber\\
&&+\,\theta(\bar v-\tau)\,\frac{2}{\bar v}
\bigg(L+\ln(\bar v-\tau)+\frac{\tau}{\bar v-\tau}\ln\frac{\bar v}{\tau}-
\frac{1}{v}\bigg)
\nonumber\\
&&+\,\theta(\tau-\bar v)\,\frac{2}{v\tau}
\bigg(L+\ln(\tau-\bar v)+\frac{\bar v\bar \tau}{\tau-\bar v}
\ln\frac{v}{\bar\tau}-1\bigg)\Bigg]
\label{jperpfinal}
\eeq
with 
\begin{eqnarray}
A &=& C_F\bigg [\Big(L+\ln \bar v\Big )^2-\frac{13}{3}\Big(L+\ln\bar
v\Big)-\frac{\pi^2}{6}+\frac{80}{9}\bigg]\nonumber \\
&&+\,\bigg (C_F-\frac{C_A}{2}\bigg )\bigg[\Big(L+\ln v\Big)^2-
\Big(L+\ln\bar v\Big)^2+
\frac{22}{3}\Big(L+\ln \bar v\Big)+
\frac{2 \pi^2}{3}-\frac{152}{9}\bigg ] \nonumber \\ 
&&+\,n_f T_f \bigg[\frac{4}{3} \Big(L+\ln \bar v\Big)-\frac{20}{9}\bigg],
\nonumber \\[0.2cm]
B &=& \frac{\theta(\tau-\bar v)}{\tau-\bar v}\bigg (L+\ln(\tau-\bar
v)\bigg )+\frac{\theta(\bar v-\tau)}{\bar v-\tau}\bigg (L+\ln(\bar
v-\tau)\bigg)
\end{eqnarray}
and $L=\ln(n_+ p^\prime n_- l/\mu^2)=\ln(2 E\omega/\mu^2)$. 
The SU(3) group factors are $C_F=4/3$, $C_A=3$, $T_f=1/2$, and 
$n_f$ denotes the number of light quark flavours.
Once again we find agreement with \cite{Becher:2004kk,Hill:2004if}.

\subsubsection{Hard-scattering form factors}

Here we express the hard-scattering (\SCETI) form factors $\Xi(\tau,E)$ 
defined in (\ref{XiP}), (\ref{XiPerp}) as
convolutions of the above jet-functions and light-cone distribution
amplitudes. 

The light-cone distribution amplitudes of the light mesons follow from
the matrix element of $Q[\Gamma^c_k](v)$ defined in (\ref{Pdef}), 
\beq 
\langle M(p^\prime)|Q[\Gamma^c_k](v)|0\rangle =
\left\{\begin{array}{ll}
-if_P E\,\phi_P(v) & \quad \Gamma^c_k = 
{\displaystyle\frac{\Slash{n}_+}{2}}\gamma_5 
\\[0.2cm]
-if_{V\parallel} E\,{\displaystyle\frac{m_V\epsilon^*\cdot v}{E}} 
\,\phi_{V\parallel}(v) & 
 \quad \Gamma^c_k = {\displaystyle\frac{\Slash{n}_+}{2}}
\\[0.2cm]
-if_{V\perp} E\,(\epsilon^{*\alpha}-\epsilon^*\cdot v\,n_-^\alpha)\,
\phi_{V\perp}(v) & 
 \quad \Gamma^c_k = {\displaystyle\frac{\Slash{n}_+}{2}\gamma^\alpha_\perp}
\end{array}
\right.
\eeq
The three cases correspond to $M$ being a pseudoscalar meson, longitudinally
polarized or transversely polarized vector meson,
respectively. Similarly, the $B$ meson distribution amplitude is 
related to the matrix element of $P(\omega)$ in (\ref{Pdef}) such that
\be 
\langle 0|P(\omega)|\bar B_v\rangle = 
 \frac{i \hat{f}_B m_B}{2}\,\phi_{B+}(\omega)
\label{blcda}
\ee
with $\hat{f}_B$ the HQET $B$ meson decay constant (but defined
such that is has mass dimension 1)\footnote{With the conventions 
of heavy-quark effective theory our $|\bar B_v\rangle$ 
corresponds to   $\sqrt{m_B}|\bar B_v\rangle$, and 
the right-hand side of (\ref{blcda}) is $m_B$-independent when 
expressed in terms of $F_{\rm stat}$.}, that is
\begin{equation}
f_B = K(\mu)\,\hat{f}_B(\mu) = \frac{K(\mu) F_{\rm
stat}(\mu)}{\sqrt{m_B}}
\end{equation}
with $ F_{\rm stat}(\mu)$ the $m_B$-independent decay constant in
the static limit (HQET) and 
\begin{equation}
K(\mu) = 1+\frac{\alpha_s C_F}{4\pi} 
\left(3\ln\frac{m_b}{\mu}-2\right)
\end{equation}
a short-distance coefficient \cite{Ji:1991pr}. 

We can now evaluate
\beq
\langle P(p^\prime)|{\cal J}_1(\tau)|\bar B_v\rangle  
&=&  2 E \int \frac{dr}{2\pi}\,e^{-i\,2E\tau r}\, \langle
P(p^\prime)| (\bar\xi W_c)(0) (W_c^\dagger i \Slash{D}_{\perp c}
W_c)(r n_+) h_v(0) |\bar B_v\rangle
\nonumber \\
&=& \int d\omega dv \,J_{\parallel}(\tau;v, \omega) \langle
P(p^\prime)| Q\Big[\frac{\Slash{n}_+}{2}\gamma_5\Big](v)P(\omega) 
|\bar B_v\rangle
\nonumber \\
&=& \frac{m_B E}{2} \int d\omega dv
\,J_{\parallel}(\tau;v, \omega)
\,\hat{f}_B\phi_{B+}(\omega)\,f_P\phi_P(v),
\label{matr1}
\eeq
where we have used (\ref{matchrel}) and (\ref{scet2fact}).
Comparing this to the definition (\ref{XiP}) and (\ref{normXi}) 
gives the expression for
$\Xi_P$. Proceeding in the same way for the other form factors we
obtain 
\begin{eqnarray}
\Xi_P(\tau,E)&=& \frac{m_B}{4 m_b}\,\hat{f}_B\phi_{B+}\star
f_P\phi_P\star J_\parallel,
\nonumber \\
\frac{E}{m_V}\,\Xi_\parallel(\tau,E)&=& \frac{m_B}{4
m_b}\,\hat{f}_B\phi_{B+}\star f_{V\parallel}\phi_{V\parallel}\star
J_\parallel,
\nonumber \\
\Xi_\perp(\tau,E)&=& \frac{m_B}{4 m_b}\,\hat{f}_B\phi_{B+}\star
f_{V\perp}\phi_{V\perp}\star J_\perp,
\nonumber\\
\tilde\Xi_\perp(\tau,E)&=&0.
\label{Xiperpres}
\end{eqnarray}
The asterisk stands for the convolutions in $\omega$ and $v$ as in 
(\ref{matr1}). This together with the explicit expressions for the
1-loop jet-functions is the main technical result of this paper. 
Using this result in (\ref{physscet1}) allows us to investigate
numerically the corrections to the symmetry relations for
heavy-to-light form factors at the 1-loop order. The subsequent
sections are devoted to this numerical investigation.


\section{Numerical analysis}
\label{sec:numerics}

\subsection{Jet-functions}

It will be seen below that the jet-function appears in the form
factors in the form of the integral
\begin{equation}
I_a \equiv  \frac{\lambda_B}{\langle \bar v^{-1}\rangle_M} 
\int_0^1\frac{dv}{\bar v}\,\phi_M(v)\int_0^\infty\frac{d\omega}{\omega}\,
\phi_{B+}(\omega)\int_0^1 d\tau \left(\delta(\tau-\bar v) +
\frac{\alpha_s}{4\pi} \,j_a(\tau;v, \omega)\right),
\label{jetintterm}
\end{equation}
which is normalized to 1 in the absence of the $\alpha_s$-correction. 
We now evaluate the 1-loop correction. 

\subsubsection{Light-cone distribution amplitudes} 

The light-cone distribution amplitude (LCDA) of the light meson, 
$\phi_M(v)$, is conventionally expanded into the eigenfunctions of 
the 1-loop renormalization kernel, 
\begin{equation}
\phi_M(v)\,=\, 6 v\bar v \left[1+\sum\limits_{n=1}^{\infty}
a_n^M C^{(3/2)}_n(2 v-1)\right]\,,
\end{equation}
where $a_n^M$ and $C^{(3/2)}_n(v)$ are the Gegenbauer
moments and polynomials, respectively. We define the quantity
\begin{equation}
\langle \bar v^{-1}\rangle_M \equiv \int_0^1\frac{dv}{\bar v}\,\phi_M(v) 
= 3\left(1 +\sum_{n=1}^\infty a_n^M\right).
\end{equation}
In practice the expansion will be truncated after the second term, 
since it is believed that the higher Gegenbauer moments are negligible, 
or accounted for approximately 
by phenomenological determinations of the first 
two moments. The LCDAs and Gegenbauer moments are scale- and 
scheme-dependent. Our computation of the jet-function corresponds 
to the modified minimal subtraction ($\overline{\rm MS}$) 
scheme and the renormalization 
scale $\mu$ of the LCDA (not indicated by its arguments) is equal 
to the scale $\mu$ that appears in the expressions 
(\ref{jparfinal}), (\ref{jperpfinal}) for the jet-functions. 
In particular the scale-dependence of $\langle \bar v^{-1}\rangle_M$  
is given by
\begin{equation}
\mu\frac{d}{d\mu} \langle \bar v^{-1}\rangle_M 
= \frac{\alpha_s C_F}{\pi} \int_0^1\frac{dv}{\bar v}\,\phi_M(v) 
\left\{\frac{4-\Delta}{2}+\frac{\ln\bar v}{v}\Big[1-\bar v \Delta
\Big]\right\},
\label{mrun}
\end{equation}
which follows from (\ref{blkernel}).\footnote{Note that (\ref{blkernel}) 
describes the scale dependence of $f_M\phi_M(v)$, and 
$$\mu\frac{d}{d\mu} f_M = \frac{\alpha_s C_F}{\pi}
\frac{\Delta-1}{2}\,f_M.$$}
We recall that 
$\Delta =0$ for transversely 
polarized vector mesons $M$, and $\Delta=1$ for pseudoscalar mesons or 
longitudinally polarized vector mesons. 

The first inverse moment of the LCDA of the $B$ meson, 
\begin{equation}
\frac{1}{\lambda_B} \equiv \int_0^\infty
\frac{d\omega}{\omega}\,\phi_{B+}(\omega)
\end{equation}
is a key quantity in exclusive $B$ decays \cite{Beneke:1999br}. 
We define the averages
\begin{equation}
\langle f(\omega)\rangle \equiv \lambda_B \int_0^\infty
\frac{d\omega}{\omega}\,\phi_{B+}(\omega) f(\omega).
\end{equation}
The LCDA and $\lambda_B$ are scale-dependent. 
Our computation of the jet-function corresponds 
to the modified minimal subtraction scheme and the renormalization 
scale $\mu$ is equal to the scale $\mu$ that appears in the expressions 
for the jet-functions. The scale-dependence of $1/\lambda_B$  
is given by
\begin{equation}
\mu\frac{d}{d\mu} \left(\frac{\hat{f}_B}{\lambda_B}\right) 
= \frac{\alpha_s C_F}{\pi} \frac{\hat{f}_B}{\lambda_B} \left\{
\frac{3}{4}+\frac{1}{2}-
\left\langle \ln\frac{\mu}{\omega}\right\rangle\right\}
\label{brun}
\end{equation}
which follows from (\ref{lnkernel}). The first term in the bracket 
comes from the scale-dependence of the static decay constant 
$\hat{f}_B$.

Since $\omega$ is of order $\Lambda$, only logarithmic modifications 
of the first inverse moment appear at leading order in the 
$1/m_b$-expansion. It can be seen from (\ref{jparfinal}) and 
(\ref{jperpfinal}) that the 1-loop calculation involves the two logarithmic 
moments $\langle L\rangle$,  $\langle L^2\rangle$ with 
$L=\ln(2 E\omega/\mu^2)$. The entire energy and scale-dependence of 
the 1-loop jet-functions is contained in these two quantities. We adopt 
a simple one-parameter model for the shape of the distribution 
amplitude \cite{Grozin:1996pq}, 
\begin{equation}
\phi_{B+}(\omega) \,=\, \frac{\omega}{\lambda_B^2}
e^{-\omega/\lambda_B}, 
\label{bmodel}
\end{equation}
which relates the two logarithmic moments to the parameter 
$\lambda_B$, 
\begin{equation}
\langle L\rangle = \ln\frac{2 E \lambda_B e^{-\gamma_E}}{\mu^2}, 
\qquad 
\langle L^2\,\rangle = \ln^2\frac{2 E \lambda_B e^{-\gamma_E}}{\mu^2} 
+\frac{\pi^2}{6}
\label{bmoms}
\end{equation} 
with $\gamma_E = 0.577216\ldots$. 

The functional form (\ref{bmodel}) and the Gegenbauer moments  
$a_n^M$ are assumed to be given at some reference scale $\mu_0$ of order 
$(m_b\Lambda)^{1/2}$. This avoids having to evolve the meson 
parameters from the hadronic scale $\Lambda$ to the hard-collinear 
scale with the \SCETII{} renormalization group equations.\footnote{
The corresponding anomalous dimensions are given by (\ref{blkernel}), 
(\ref{lnkernel}).} 

\subsubsection{Integrated jet-functions}

We proceed to the evaluation of $I_a$. The integrals 
$\int_0^1 d\tau\,j_a(\tau;v, \omega)$ are given in analytic 
form in Appendix~\ref{app:convolution}. Integration over 
$\omega$ introduces the logarithmic moments 
$\langle L\rangle$,  $\langle L^2\rangle$  defined above. 
The final $v$-integration can be done numerically, or 
term by term in the Gegenbauer expansion of the light meson 
light-cone distribution amplitude. Up to the second moment, 
we obtain 
\begin{eqnarray}
I_\parallel &=& 1+\frac{\alpha_s(\mu)}{4\pi} \,
\frac{3}{\langle \bar v^{-1}\rangle_M} 
\Bigg(\frac{4}{3} \Big[1+a_1^M+a_2^M\Big]\langle L^2\rangle - 
\Big[ 5.24 + 8.93 a_1^M +10.86 a_2^M\Big] \langle L\rangle 
\nonumber\\
&& +\, \Big[3.99 +8.67 a_1^M + 13.47 a_2^M\Big] \Bigg),
\nonumber\\
I_\perp &=& 1+\frac{\alpha_s(\mu)}{4\pi} \,
\frac{3}{\langle \bar v^{-1}\rangle_M} 
\Bigg(\frac{4}{3} \Big[1+a_1^M+a_2^M\Big]\langle L^2\rangle - 
\Big[4.90 + 8.93 a_1^M +10.81 a_2^M\Big] \langle L\rangle 
\nonumber\\
&& +\, \Big[0.73 +6.78 a_1^M + 11.48 a_2^M\Big] \Bigg),
\label{iperp}
\end{eqnarray}
with\footnote{We assume four massless quark flavours throughout this 
numerical analysis for simplicity. This is not a good approximation for 
the charm quark, whose mass is of the same order as the hard-collinear 
scale. A more precise treatment would keep the charm quark mass in 
the fermion loop correction to the jet-function. This could be 
done without difficulty if such precision were required.} 
$n_f=4$, $T_f=1/2$, $C_F=4/3$ and $C_A=3$. The analytic
expressions of these convolutions are also given in 
Appendix~\ref{app:convolution}.

These results do not contain large logarithms, when $\mu$ is of 
order of the hard-collinear scale 
$(m_b\Lambda)^{1/2} \approx 1.5\,\mbox{GeV}$, i.e. 
$\langle L\rangle$,  $\langle L^2\rangle$ are of order 1 
for $E\sim m_b$ and $\mu\sim (m_b\Lambda)^{1/2}$. 
Since $\alpha_s(1.5\,\mbox{GeV})/(4 \pi)$ 
is approximately 0.029, $\langle L^2\rangle\approx 2.5$ and 
$\langle L\rangle\approx -1$ (for typical parameters), 
the perturbative corrections to the jet-functions 
are about (20-50)\%, depending on $a=\parallel,\perp$ and the precise 
values of the Gegenbauer moments. We may therefore conclude that 
perturbative corrections to hard spectator-scattering are 
non-negligible. At the same time there is no sign that the 
series expansion is not well-behaved despite the comparatively 
low scale, lending support to the possibility of performing 
perturbative factorization at the hard-collinear scale. This 
is an important result, already mentioned 
in \cite{Hill:2004if,Beneke:2004bn}, since theoretical 
calculations of exclusive $B$ decays in general rely on 
this possibility. The present calculation and the one in 
\cite{Becher:2004kk,Hill:2004if} are the first computations 
of quantum corrections to spectator-scattering. 

\subsection{Renormalization group improvement of the $C^{(B1)}$ 
coefficients}

The complete hard-scattering term involves $\int_0^1 d\tau\,[C^{B1}\star J]$, 
so we now turn to the evaluation of the $C^{B1}$. With 
$\mu$ of order $(m_b\Lambda)^{1/2}$, the jet-function is free 
from large logarithms, but $C^{B1}$ involves up to two 
powers of logarithms $\ln (m_b/\Lambda)$ per loop from the 
ratio of the hard to the hard-collinear scale. In the following 
we derive an expression that sums the leading-logarithmic and 
double-logarithmic terms to all orders in perturbation theory. We shall 
refer to this simply as the leading-logarithmic approximation 
(LL).\footnote{In the literature on Sudakov resummation the analogous 
approximation is usually called ``next-to-leading-logarithmic 
approximation''. We prefer the term ``leading-logarithmic'', since, as 
in other applications of renormalization-group improved perturbation 
theory, the approximation requires only the 1-loop anomalous 
dimension of a generic operator. The complication from
double-logarithms is reflected by the presence of the so-called 
cusp anomalous dimension, for which the 2-loop coefficient 
is needed already in the leading-logarithmic approximation.}

\subsubsection{Solution to the renormalization group equation}

The renormalization-group formalism that accomplishes this summation 
is standard and has been applied to the B-type \SCETI{} currents  
in \cite{Hill:2004if}. We shall recapitulate the relevant expressions 
to define the notation. In the following we drop the superscript 
``B1'' on $C^{(B1)}$ and denote by $C(E,\tau;\mu)$ a generic 
short-distance coefficient. 
With the renormalization kernels (\ref{zfact}),  
(\ref{re-kernel}) the renormalization group equation is derived from 
the requirement that $C(E,\tau;\mu) \,J^{(B1)}(\mu)$ is independent 
of the QCD/\SCETI{} factorization scale. This implies 
\begin{equation}
\mu\frac{d}{d\mu} C(E,\tau;\mu) = 
-\Gamma_{\rm cusp}(\alpha_s) \ln
\frac{\mu}{2 E} \,C(E,\tau;\mu) + \int_0^1 d\tau^\prime
\gamma_a(\tau^\prime,\tau)\,C(E,\tau^\prime;\mu)
\label{rge}
\end{equation}
with 
\begin{equation}
\Gamma_{\rm
cusp}(\alpha_s)=\sum\limits_{n=0}^{\infty} \Gamma_n \left
(\frac{\alpha_s}{4\pi}\right)^{n+1}
\end{equation}
the so-called universal cusp anomalous dimension. Comparison with 
(\ref{zfact}), (\ref{re-kernel}) gives $\Gamma_0=4 C_F$ and 
\begin{eqnarray}
\gamma_a(\tau,\tau^\prime)
\,=\,-\frac{\alpha_s(\mu)}{2\pi}\left
[z_1(\tau,\tau^\prime)+\Delta_a \,z_2(\tau,\tau^\prime)\right ],
\label{andim}
\end{eqnarray}
where $\Delta_a =1$ for $a=\,\parallel$ and $\Delta_a=0$ for $a=\,\perp$. 
Due to the presence of double logarithms we also need the two-loop 
cusp anomalous dimension \cite{Kodaira:1981nh}
\begin{equation}
\Gamma_1=4
C_F\left(\bigg[\frac{67}{9}-\frac{\pi^2}{3}\bigg] C_A-
\frac{20}{9} n_f T_f\right).
\end{equation}
The short-distance coefficients of the B-type operators 
(\ref{scalar:newbasis}) to (\ref{tensor:newbasis}) and the 
coefficients appearing in (\ref{pscet1}), (\ref{vscet1}) 
evolve with the anomalous dimensions 
$\gamma_a(\tau^\prime,\tau)$ as follows:
\begin{eqnarray}
&a=\,\parallel&\qquad C_{S,P}^{(B1)}, \,C_{V,A}^{(B1)1-3}, 
\,C_T^{(B1)1,2,5-7}, \,C_{f_+}^{(B1)},\,C_{f_0}^{(B1)},\,C_{f_T}^{(B1)},
\nonumber \\
&a=\,\perp&\qquad C_{V,A}^{(B1)4}, \,C_T^{(B1)3,4},\,C_{V}^{(B1)},
\,C_{T_1}^{(B1)}.
\end{eqnarray} 

The general 
solution to the renormalization group equation (\ref{rge}) reads 
\begin{eqnarray}
\label{eq:RGEsolution} 
C(E,\tau;\mu) = 
e^{-S(E;\mu_h,\mu)} \int_0^1 d\tau^\prime\,
U_a(\tau,\tau^\prime;\mu_h,\mu)\,
C(E,\tau^\prime;\mu_h),
\end{eqnarray}
where
\begin{eqnarray}
S(E;\mu_h,\mu)&=&\int_{\alpha_s(\mu_h)}^{\alpha_s(\mu)}d\alpha_s\,
\frac{\Gamma_{\rm cusp}(\alpha_s)}{\beta(\alpha_s)}
\int_{\alpha_s(2
E)}^{\alpha_s}\frac{d\alpha_s^\prime}{\beta(\alpha_s^\prime)} 
\nonumber\\ 
&& \hspace*{-2cm} = \int_{\alpha_s(\mu_h)}^{\alpha_s(\mu)}d\alpha_s\,
\frac{\Gamma_{\rm cusp}(\alpha_s)}{\beta(\alpha_s)}
\int_{\alpha_s(\mu_h)}^{\alpha_s}
\frac{d\alpha_s^\prime}{\beta(\alpha_s^\prime)} 
+\ln\frac{\mu_h}{2 E} \,\int_{\alpha_s(\mu_h)}^{\alpha_s(\mu)}d\alpha_s\,
\frac{\Gamma_{\rm cusp}(\alpha_s)}{\beta(\alpha_s)}
\label{sudakov}
\end{eqnarray}
with the QCD $\beta$-function
\begin{eqnarray}
 \beta(\alpha_s) &=& \mu\frac{d\alpha_s}{d\mu} 
=-2 \alpha_s\sum\limits_{n=0}^{\infty} \beta_n
\left(\frac{\alpha_s}{4\pi}\right)^{n+1},
\nonumber\\
&& \hspace*{-1.5cm} \beta_0=\frac{11}{3}C_A-\frac{4}{3} n_f T_f, 
\quad \beta_1=
\frac{34}{3} C_A^2-\left(\frac{20}{3} C_A+4 C_F\right) n_f T_f.
\end{eqnarray}
The evolution kernel $U_a(\tau,\tau^\prime;\mu_h,\mu)$ satisfies the 
integro-differential equation
\begin{eqnarray}
\mu \frac{d}{d\mu}U_a(\tau,\tau^\prime;\mu_h,\mu)=
\int_0^1 d\tau^{\prime\prime} \gamma_a(\tau^{\prime\prime},\tau)
\,U_a(\tau^{\prime\prime},\tau^\prime;\mu_h,\mu)\,,
\end{eqnarray}
with initial condition $U_a(\tau,\tau^\prime;\mu_h,\mu_h)=
\delta(\tau-\tau^\prime)$. To sum the large logarithms the initial 
scale $\mu_h$ should be of order $m_b$, and the evolution ends at
$\mu$ of order $(m_b\Lambda)^{1/2}$.

\begin{figure}
\epsfig{file=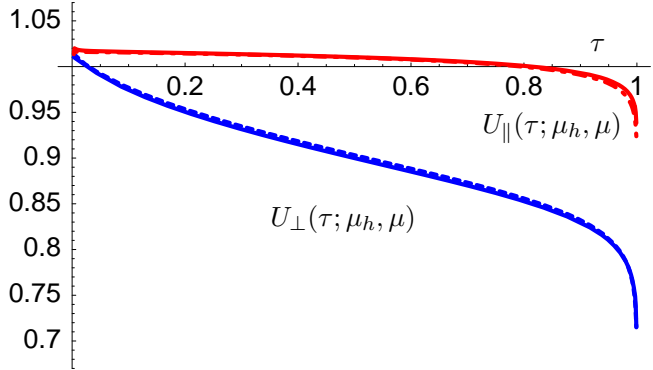,width=12cm,angle=0}
\caption{$U_a(\tau,\mu_h,\mu)$ for $\mu_h=m_b=4.8$ GeV
and $\mu=1.5$ GeV. The upper curves refer to $U_\parallel$, the lower
ones to $U_\perp$. Solid lines: exact numerical integration. 
Dashed lines: approximate solutions.}
\label{fig:UGamma}
\end{figure}

Several simplifications occur in the leading-logarithmic approximation. 
Eq.~(\ref{sudakov}) can be integrated to 
\begin{eqnarray}
S(E;\mu_h,\mu)&=&- \frac{\Gamma_0}{2\beta_0}\ln r\ln\frac{\mu_h}{2
E} + \frac{\Gamma_0}{4\beta_0^2}\Bigg(\frac{4\pi}{\alpha_s(\mu_h)}
\left[\ln r-1+\frac{1}{r}\right]-
\frac{\beta_1}{2 \beta_0} \ln^2 r
\nonumber\\
&& +\,\left(\frac{\Gamma_1}{\Gamma_0}-
\frac{\beta_1}{\beta_0}\right)
\left[r-1-\ln r\right]\Bigg),
\end{eqnarray}
with $r=\alpha_s(\mu)/\alpha_s(\mu_h)>1$. Furthermore, the initial 
condition is given by the tree-level expression for $C(E,\tau;\mu_h)$, 
which is independent of $\tau$ (and $\mu_h$), hence the $\tau^\prime$ 
integration in (\ref{eq:RGEsolution}) can be done. We then have 
\begin{eqnarray}
\label{eq:RGEsolution2} 
C^{(LL)}(E,\tau;\mu)\,=\,
e^{-S(E;\mu_h,\mu)} \,U_a(\tau;\mu_h,\mu)\,
C^{(0)}(E)\,,
\end{eqnarray}
where $C^{(0)}(E)$ is the tree coefficient, and 
$U_a(\tau;\mu_h,\mu)=\int_0^1 d\tau^\prime\,U_a(\tau,\tau^\prime;\mu_h,\mu)$ 
satisfies 
\begin{eqnarray}
\mu \frac{d}{d\mu}U_a(\tau;\mu_h,\mu)=
\int_0^1 d\tau^{\prime} \gamma_a(\tau^{\prime},\tau)
\,U_a(\tau^{\prime};\mu_h,\mu)\,,
\end{eqnarray}
with initial condition $U_a(\tau;\mu_h,\mu_h)=1$. This equation 
must be solved numerically. We always use two-loop running of 
$\alpha_s(\mu)$, and put $\mu_h=m_b=4.8\,\mbox{GeV}$. As input 
we take $\alpha_s(4.8\,\mbox{GeV})=0.215$, which gives 
 $\alpha_s(1.5\,\mbox{GeV})=0.359$ (four massless flavours). 
The result of this integration is shown in 
Figure~\ref{fig:UGamma} for $\mu=1.5\,$GeV. We have found 
that the solution to 
\begin{eqnarray}
\mu \frac{d}{d\mu}U_a^{\rm app}(\tau,\mu_h,\mu)=\left
[\int_0^1 d\tau^\prime \gamma_a(\tau^\prime,\tau)\right
]U_a^{\rm app}(\tau,\mu_h,\mu),
\end{eqnarray}
given by 
\begin{equation}
U_a^{\rm app}(\tau,\mu_h,\mu) = \left(\frac{\alpha_s(\mu)}{\alpha_s(\mu_h)}
\right)^{-\gamma_a(\tau)/(2\beta_0)}
\end{equation}
with $\frac{\alpha_s}{4\pi}\gamma_a(\tau) = 
\int_0^1 d\tau^\prime \gamma_a(\tau^\prime,\tau)$ and 
\begin{eqnarray}
\gamma_\parallel(\tau) &=& - C_F +
4 \left(C_F-\frac{C_A}{2}\right) \frac{\ln\bar\tau}{\tau}, 
\nonumber \\
\gamma_\perp(\tau) &=& 
- C_F\left(\frac{4 \tau\ln\tau}{\bar
\tau}+1\right) + 4 \left(C_F-\frac{C_A}{2}\right )\left
(\frac{1+\tau}{\tau}\ln\bar \tau+ \frac{\tau\ln\tau}{\bar
\tau}\right)
\label{gam0}
\end{eqnarray}
provides a very good approximation (better than $1\%$) to the 
exact solution, provided one uses 1-loop running 
for $\alpha_s$ with $\alpha_s(4.8\,\mbox{GeV})=0.215$ 
in the approximate solution. The approximate 
expressions are also shown in Figure~\ref{fig:UGamma}. 

\subsubsection{NLO+LL approximation}

We are now in the position to give expressions for the B-type 
short-distance coefficients $C(E,\tau;\mu)$, which include the 
complete 1-loop correction as well as the leading-logarithmic 
terms. The formula is 
\begin{eqnarray}
C(E,\tau;\mu) &=& C^{(0)}(E) + C^{(1)}(E,\tau;\mu) - 
C^{(0)}(E) \left[e^{-S} \,U_a\right]_{\alpha_s}\!(E,\tau;\mu)
\nonumber\\ 
&&  + \,
C^{(0)}(E) \left[e^{-S} \,U_a-1\right](E,\tau;\mu).
\label{master1}
\end{eqnarray}
The meaning of the four terms on the right-hand side is as follows: 
the first and second terms are the tree and 1-loop coefficients, 
respectively. Together they constitute the next-to-leading order
(NLO) approximation to $C(E,\tau;\mu)$. 
The fourth term is the sum of leading-logarithmic terms 
to all orders minus the tree. Finally, the third term subtracts 
the logarithmic terms already included in the full 1-loop correction 
$C^{(1)}(E,\tau;\mu)$. The subtraction is given by 
\begin{eqnarray}
\left[e^{-S} \,U_a\right]_{\alpha_s}\!(E,\tau;\mu) &=& 
\frac{\alpha_s(\mu)}{4\pi}\Bigg\{-\frac{\Gamma_0}{2}
\left [\ln^2\left (\frac{\mu}{\mu_h}\right)+2 \ln\left
(\frac{\mu_h}{2 E}\right )\ln\left(\frac{\mu}{\mu_h}\right )\right ]
\nonumber\\
&& +\,\ln \left
(\frac{\mu}{\mu_h}\right ) \gamma_a(\tau)\Bigg\}
\label{subtraction2}
\end{eqnarray}

\begin{figure}[p]
\vspace*{-3.9cm}
\hspace*{-2cm}
\epsfig{file=fig_cb1.eps,width=18.5cm}
\vskip-3.3cm
\caption{The short-distance coefficients $C^{(B1)}_X(E,\tau,\mu)$ relevant 
to the form factors $X$ at two representative energy values 
($x=1$ (left) and $x=0.6$ (right)) at $\mu=1.5\,$GeV normalized to 
the tree approximation. Dash-dotted: logarithmic terms at order 
$\alpha_s$; dashed: full NLO approximation; solid: NLO plus 
logarithmic summation.
\label{fig:b1type}}
\end{figure}

To analyze the structure of the correction, we display in 
Figure~\ref{fig:b1type} the following approximations, all normalized 
to the tree coefficient: (i) tree plus the logarithmic terms 
at 1-loop (dash-dotted); (ii) previous approximation plus the non-logarithmic 
term, i.e. the complete next-to-leading order result (dashed); (iii) 
previous approximation plus the sum of leading-logarithmic terms 
at order $\alpha_s^2$ and beyond (solid). In the numerical implementation 
we set $\mu_h=m_b=4.8\,$GeV, $E=m_b x/2$, and regard the coefficients 
as functions of energy fraction $x$ and the convolution variable 
$\tau$. Since $E$ must be of order $m_b$, $x$ cannot be chosen 
too small. We take $x=1$ and $x=0.6$ as representative examples 
of large (maximal) and small energy of the light meson, and 
evolve to $\mu=1.5\,$GeV. We also fix the scale of the QCD tensor 
current operator to $\nu=m_b=4.8\,$GeV. 
Figure~\ref{fig:b1type} shows four of the five combinations, 
$C^{(B1)}_X$, $X\in \{f_+,f_0,f_T,T_1,V\}$,  
which appear in the hard-scattering  
contribution to the vector and tensor current form factors
(\ref{pscet1}), (\ref{vscet1}). Only 
$C^{(B1)}_V$ is not shown, because its tree coefficient vanishes, 
hence only $C^{(1)}(E,\tau;\mu)$ in (\ref{master1}) is non-zero. 

The following observations can be made from the figure: a) the logarithmic 
term at order $\alpha_s$ (dash-dotted lines) is a very poor 
approximation to the full $\alpha_s$ coefficient (dashed lines), 
especially for $C^{(B1)}_{T_1}$, which is the only one of the 
four coefficients shown involving the transverse anomalous 
dimension $\gamma_\perp$. b) Except near the endpoints $\tau=0,1$, where the 
relative correction diverges, the typical next-to-leading order 
correction from the hard scale $m_b$ is of order 30\%. The endpoint 
singularities are logarithmic and disappear when the correction is 
folded with the light-meson distribution amplitude (integration over 
$\tau$). c) The effect of the logarithmically enhanced terms 
beyond the order $\alpha_s$ is negligible (difference between the 
solid and dashed lines). It is largest for  $C^{(B1)}_{T_1}$ 
towards larger $\tau$ since here $U_\perp(\tau;\mu_h,\mu)$ is significantly 
different from 1, see Figure~\ref{fig:UGamma}.

\subsection{Spectator-scattering correction}

According to (\ref{pscet1}), (\ref{vscet1}), (\ref{physscet1}) 
and (\ref{Xiperpres})  
the form factors $F_X(E)$ are given by  
\begin{equation}
F_X(E) = C^{(A0)}_X \, \xi_a(E) + H_X(E) 
\end{equation}
with the spectator-scattering term 
\begin{eqnarray}
H_X(E) &=& \frac{m_B}{4 m_b}\,\hat{f}_B f_M 
\int_0^1 dv \,\phi_M(v)\int_0^\infty d\omega \,\phi_{B+}(\omega) 
\nonumber\\ 
&&   \times \int_0^1 d\tau \, C^{(B1)}_X(E,\tau)\,J_a(\tau;v,\omega).
\label{nextf}
\end{eqnarray}
Here $a=\,\parallel$ (or $P$ in case of $\xi_a$) for 
$X=\{f_+,f_0,f_T\}$ and $a=\,\perp$ for $X=\{V,T_1\}$. We have now 
assembled all the pieces required for the evaluation of $H_X(E)$ 
at order $\alpha_s^2$ (1-loop). From now on we set $m_B/m_b$ in 
(\ref{nextf}) to 1, since the difference between the meson and the 
quark mass is a power correction beyond the accuracy of the 
present calculation. 

At the leading order we insert the tree expressions for the 
B-type coefficient function (hereafter we again drop the superscript 
``B1'') and the jet-function and obtain
\begin{equation} 
H_X^{(0)}(E) = -\frac{\pi \alpha_s(\mu) C_F}{N_c}
\frac{\hat f_B f_M \langle \bar v^{-1}\rangle_M}{2 E \lambda_B} 
\,C^{(0)}_{X}(E).
\label{htree}
\end{equation}
Here as before the superscript ``(0)'' refers to the tree 
approximation. This agrees with 
the results of \cite{Beneke:2000wa}. 

To obtain the next-to-leading order result including the 
renormalization-group summation, we insert (\ref{master1}) and the  
jet-function into (\ref{nextf}), 
and neglect cross terms of order $\alpha_s^3$ in 
the product $C_X \star J_a$. The result is 
\begin{eqnarray}
H_X(E) &=& H_X^{(0)}(E) 
\times\, \Bigg\{1 + 
\frac{1}{\langle \bar v^{-1} \rangle_M}\int \frac{dv}{\bar v}
\,\phi_M(v)\,\frac{C^{(1)}(E,\bar v)}{C^{(0)}(E)}
+ \Big[I_a-1\Big]
\nonumber\\ 
&& -\, \frac{1}{\langle
\bar v^{-1} \rangle_M}\int \frac{dv}{\bar v}\,\phi_M(v)\left [e^{-S}\,
U_a\right ]_{\alpha_s}(E,\bar v) 
\nonumber\\ 
&& + \frac{1}{\langle \bar v^{-1} \rangle_M}\int
\frac{dv}{\bar v}\,\phi_M(v)\left [e^{-S} \,U_a-1\right
](E,\bar v)\Bigg\}.
\label{master2}
\end{eqnarray}
The second term in the bracket is the 1-loop hard correction; the third 
comes from the jet-function and is defined in  
(\ref{iperp}); the fourth and fifth are related to the renormalization 
group summation as in (\ref{master1}). The integration of the second 
term can be done analytically, but the expressions are lengthy. They 
are given for selected short-distance coefficients and for 
the integration with 
the asymptotic distribution amplitude in Appendix~\ref{app:convolution}. 
It is as straightforward to perform the integration over $\phi_M(v)$ 
numerically. The integration of the subtraction term is elementary and
given by (\ref{subtraction2}) together with (\ref{gam0}). The integration of 
the last term can only be done numerically using the numerical 
solution of the integro-differential equation for $U_a$. Since $S$ is 
independent of $v$, one needs the integrals 
\begin{eqnarray}
\int_0^1 \frac{dv}{\bar v} \,\phi_M(v,\mu)\, U_\parallel(\bar
v,\mu_h,\mu)&=& 3.037+3.058 \,a_1^M+3.051 \,a_2^M+...\,,
\nonumber\\
\int_0^1 \frac{dv}{\bar v}\,\phi_M(v,\mu)\, U_\perp(\bar
v,\mu_h,\mu)&=& 2.795+2.980 \,a_1^M+3.003\,a_2^M+...\,.
\end{eqnarray}
The numerical values are given for $\mu_h=4.8\,$GeV and $\mu=1.5\,$GeV.

\begin{figure}[t]
\vspace*{0.5cm}
\hspace*{-2.8cm}
\epsfig{file=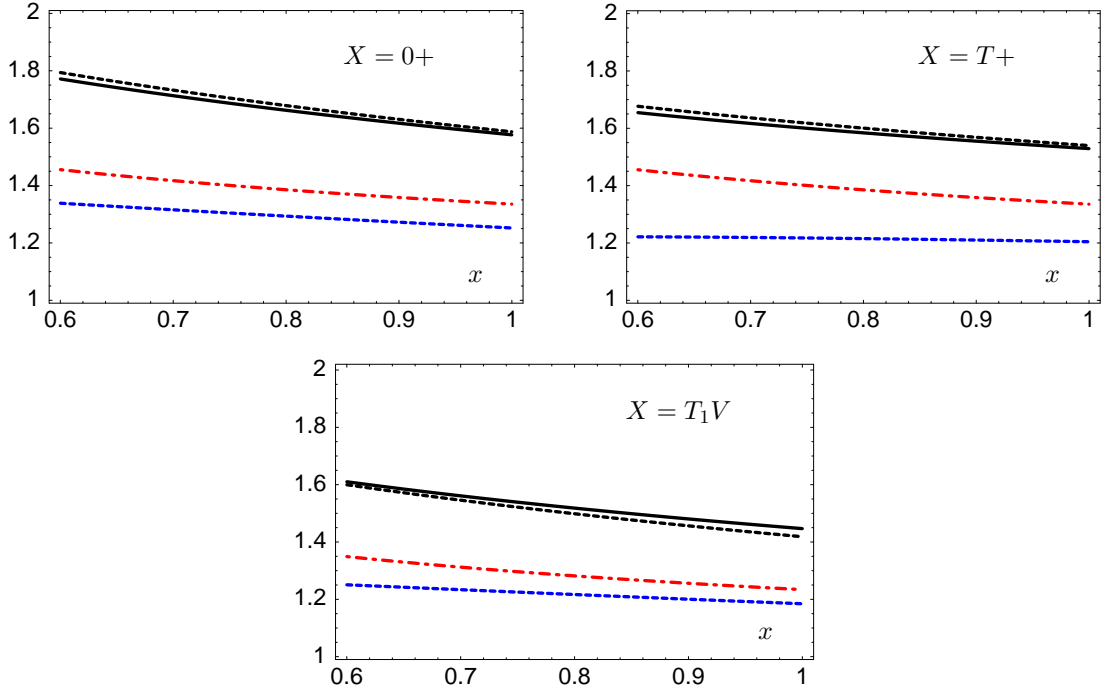,width=18cm}
\vskip0.3cm
\caption{$H_X(E)/H_X^{(0)}(E)$ for $X=0+,T+,T_1V$. The two upper 
curves represent the complete next-to-leading order result 
with (solid) and without (dashed) the renormalization group 
summation. The middle (dash-dotted) line shows the jet-function 
(hard-collinear) correction alone, the lower (dashed) line the 
hard correction alone.
\label{fig:full}}
\end{figure}

To illustrate these results we consider the three coefficients 
relevant to the form factors in the physical form factor scheme 
defined in (\ref{physscet1}). Let
\begin{eqnarray}
C_{0+}^{(B1)}(\tau,E)&=&C_{f_0}^{(B1)}(\tau,E)-
C_{f_+}^{(B1)}(\tau,E)\,R_0(E), 
\nonumber \\
C_{T+}^{(B1)}(\tau,E)&=&C_{f_T}^{(B1)}(\tau,E)-
C_{f_+}^{(B1)}(\tau,E)\,R_T(E), 
\nonumber \\
C_{T_1V}^{(B1)}(\tau,E)&=&C_{T_1}^{(B1)}(\tau,E)-
C_{V}^{(B1)}(\tau,E)\,R_\perp(E).
\label{cb1phys}
\end{eqnarray}
Choosing $\lambda_B=0.35\,$GeV, $x=0.85$ (corresponding to 
light meson energy $E=x m_B/2 =2.24\,$GeV or momentum transfer 
$q^2=4.18\,$GeV), and asymptotic distribution amplitudes, the 
curly bracket in (\ref{master2}) evaluates to 
\begin{eqnarray}
X=0+:&& \quad 1+0.283\mbox{ [C]} + 0.371 \mbox{ [jet]} + 
0.059 \mbox{ [log, $\alpha_s$]} - 0.073 \mbox{ [all logs]},  
\nonumber\\[0.2cm]
X=T+:&& \quad 1+0.213\mbox{ [C]} + 0.371 \mbox{ [jet]} + 
0.059 \mbox{ [log, $\alpha_s$]} - 0.073 \mbox{ [all logs]}, 
\nonumber\\[0.2cm]
X=T_1V:&& \quad 1+0.209\mbox{ [C]} + 0.268 \mbox{ [jet]} + 
0.169 \mbox{ [log, $\alpha_s$]} - 0.147 \mbox{ [all logs]},  
\phantom{--}
\end{eqnarray}
where the five terms correspond to the five terms on the right-hand 
side of (\ref{master2}). We observe that the hard correction [C] and 
the jet-function correction [jet] are of similar size, while 
the sum of higher-order logarithms (the sum of the last two terms) 
is at least a factor of 10 
smaller. The total correction to the tree result amounts to an  
enhancement of (50-70)\% of the spectator-scattering effect. 
These features are independent of the value of $E$ 
as can be seen from Figure~\ref{fig:full}, which displays the 
weak energy-dependence of the spectator-scattering correction normalized 
to the tree result.

The dependence of these results on the hadronic input parameters 
$\lambda_B$, $a_1^M$, $a_2^M$ is roughly as follows. $\lambda_B$ 
enters the relative correction through the moments 
(\ref{bmoms}) and therefore affects the jet-function terms only. 
Choosing $\lambda_B=0.5\,$GeV ($0.25\,$GeV) changes the number 
0.371 to 0.295 (0.452), and 0.268 to 0.195 (0.346), an uncertainty 
characteristic for all energy fractions $x$. Furthermore, there is 
an uncertainty due to the model for the shape of the distribution 
amplitude that correlates the logarithmic moments with $\lambda_B$, 
which we do not attempt to quantify. There is a larger 
dependence of the tree result $H_X^{(0)}(E)$ on $\lambda_B$, 
since it is inversely proportional to $\lambda_B$. Positive 
Gegenbauer moments increase the tree result and the relative 
next-to-leading order correction. This can be seen  from (\ref{iperp}) 
for the jet-function correction. For $a_2^M=0.2$, the total relative 
next-to-leading order correction increases from 64\% for $X=+0$ 
(50\% for $X=T_1V$) to 75\% (62\%). Finally, there is a dependence 
on the renormalization scale $\mu$, which we fixed to $1.5\,$GeV. 
In order to estimate this dependence, one must fix the hadronic 
input parameters at some $\mu_0$ and evolve them to $\mu$ using 
(\ref{mrun}), (\ref{brun}). Since the scale-dependence of 
the hadronic parameters is within their uncertainty, we do 
not perform this estimate here.


\section{B decay phenomenology}
\label{sec:symmbr}

In this section we discuss three applications of our results 
to $B$ decays. We restrict ourselves to decays to pions or 
$\rho$ mesons, since the results for kaons are qualitatively very 
similar. 

We use the following parameters: 
the $b$-quark pole mass $m_b=4.8\,$GeV; the renormalization scale 
of the QCD tensor current $\nu=m_b$; the initial scale for the 
renormalization group evolution $\mu_h=m_b$; the renormalization 
scale and final scale of the 
renormalization group evolution $\mu=1.5\,$GeV. This is also the 
(hard-collinear) scale at which all other scale-dependent 
quantities such as meson light-cone distribution amplitudes 
and the scale-dependent decays constants $\hat{f}_B$, $f_{M\perp}$ 
are evaluated. The strong coupling is obtained from 
$\alpha_s(m_b)=0.215$ by employing 2-loop running 
($\Lambda^{(n_f=4)}_{\overline{\rm MS}}=323.6\,$MeV), which gives 
$\alpha_s(1.5\,\mbox{GeV})=0.359$. The pion and $\rho$ meson 
parameters are $f_\pi=130.7\,\mbox{MeV}$, 
$f_{\rho\parallel}=209\,\mbox{MeV}$, $f_{\rho\perp}=150\,\mbox{MeV}$, 
and the second Gegenbauer moment is assumed to be 
$a_2^M=0.1$ for the pion and the distribution amplitudes of both, 
the longitudinal and transverse $\rho$ meson. The $B$ meson mass 
is $m_B=5.28\,$GeV and the decay constant 
$\hat{f}_B=f_B/K(1.5\,\mbox{GeV})=200\,$MeV. We assume the model 
(\ref{bmodel}) for the $B$ meson distribution amplitude and 
$\lambda_B=0.35\,$GeV. This is somewhat 
smaller than the value $0.46\,$GeV suggested by QCD sum rule 
calculations~\cite{Braun:2003wx}. Allowing $\lambda_B$ to vary from 
$0.25\,$GeV to $0.5\,$GeV implies that the value of $\lambda_B$ 
is the single most important uncertainty in the final numerical 
calculation. The \SCETI{} form factors $\xi_a(E)$ 
are defined in the physical form factor 
scheme through full QCD form factors according to (\ref{ffscheme}). 
The full QCD form factors needed for this definition are taken 
from the light-cone QCD sum rule calculations \cite{ff} including 
the parameterization of their $q^2$ dependence. On the basis 
of this input we can compute the remaining seven pion and $\rho$ meson
form factors using (\ref{physscet1}). We relate hadronic to partonic 
variables by first eliminating $E$ through $E=x m_b/2$ in the
coefficient functions. The energy fraction $x$ is then interpreted as 
$x=1-q^2/m_B^2=2 E/m_B$, when we plot hadronic form factors as 
functions of $q^2$ or hadronic energy $E$.

\subsection{Symmetry-breaking corrections to form factor ratios}

In the absence of radiative and power corrections, 
the factorization formula (\ref{eq:factorization}) implies 
parameter-free relations between form factors \cite{Charles:1998dr}, 
since only $\xi_a(E)$ appears on the right-hand side, which cancels 
in ratios. These relations receive corrections, which are 
calculable at leading power in the $1/m_b$ expansion given 
the above-mentioned input parameters~\cite{Beneke:2000wa}. 
The seven relations between the total of ten  pion and 
$\rho$ meson form factors are obtained from the 
two relations (\ref{exactrel}), which do not receive any 
perturbative corrections, and the five relations 
that follow from (\ref{physscet1}) by dividing through 
the appropriate $\xi_a^{\rm FF}$. For instance, the 
second and third equations of (\ref{physscet1}) imply 
\begin{eqnarray}
{\cal R}_1(E) &\equiv& \frac{m_B}{m_B+m_P}\frac{f_T(E)}{f_+(E)}
= R_T(E) +\int_0^1 d\tau \,C_{T+}^{(B1)}(\tau,E)\,
\frac{\Xi_P(\tau,E)}{f_+(E)},
\nonumber\\
{\cal R}_2(E) &\equiv& \frac{m_B+m_V}{m_B}\frac{T_1(E)}{V(E)}= 
R_\perp(E)+ \frac{m_B+m_V}{m_B}\int_0^1 d\tau\,
C_{T_1V}^{(B1)}(\tau,E)\,\frac{\Xi_\perp(\tau,E)}{V(E)},\qquad
\label{ffratios}
\end{eqnarray}
with $C_{T+}^{(B1)}(\tau,E)$, $C_{T_1V}^{(B1)}(\tau,E)$ 
the combinations of coefficient functions defined 
in (\ref{cb1phys}). Similar relations follow for the other 
form factors. The second term on the right-hand side equals  
the hard spectator-scattering term (\ref{nextf}) divided 
by the appropriate $\xi_a^{\rm FF}(E)$. Putting together (\ref{rfactors}), 
(\ref{htree}) and (\ref{master2}) we obtain the form factor 
ratios including the new next-to-leading order (and resummed) 
correction to the spectator-scattering term.

\begin{figure}[p]
\vspace*{0.5cm}
\hspace*{-2.8cm}
\epsfig{file=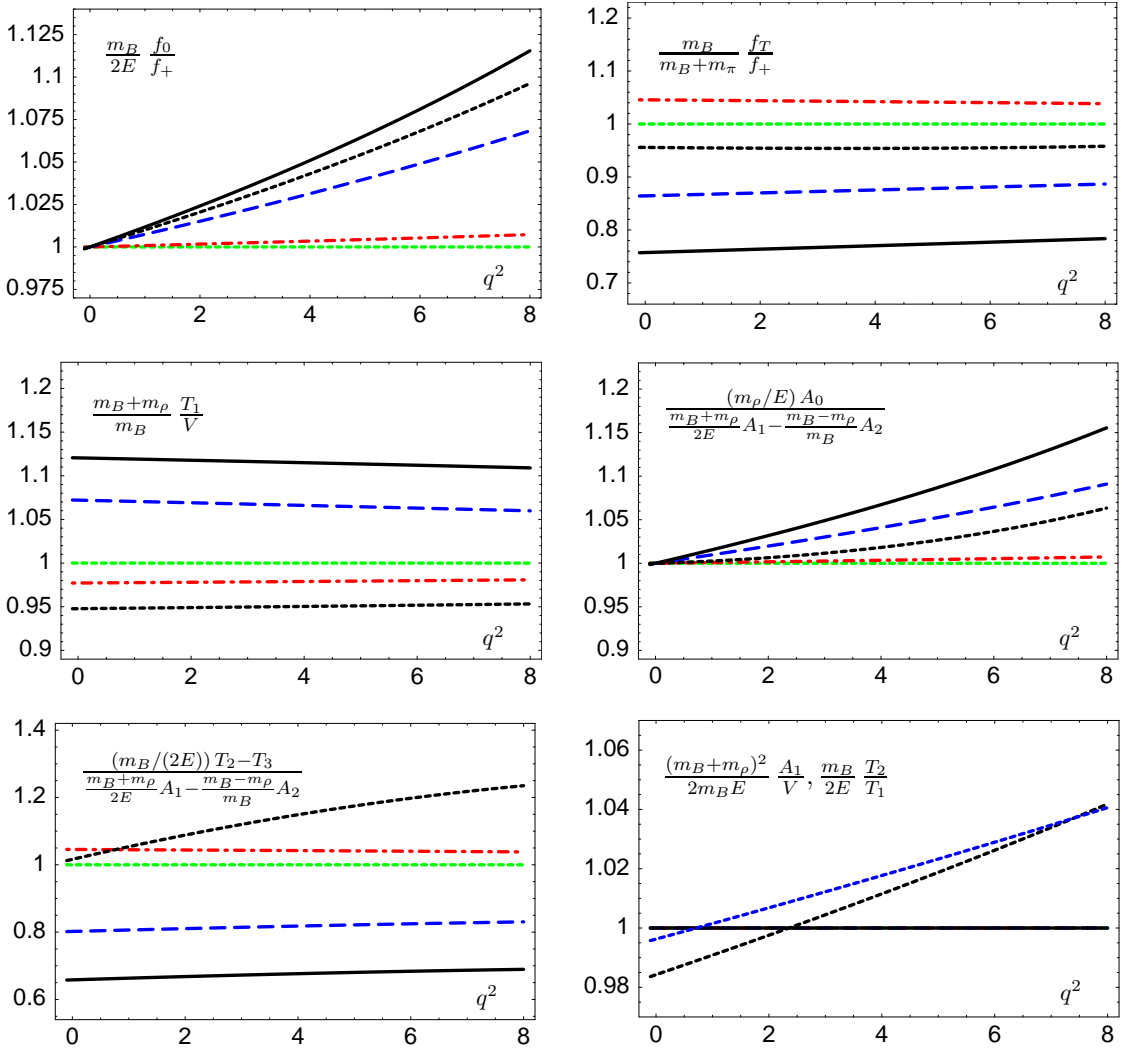,width=18cm}
\vskip0.6cm
\caption{Corrections to the $B\to\pi$ and $B\to\rho$ 
form factor ratios as a function of 
$q^2$. The ratios equal 1 in the absence of radiative 
corrections. Solid (black) line: full result, including NLO and 
resummed leading-logarithmic correction to spectator-scattering; 
Long-dashed (blue): without NLO+LL correction to spectator-scattering;
Dash-dotted (red): without any spectator-scattering term. 
Dashed (black): QCD sum rule calculation. The lower right panel 
shows the two form factor ratios that equal 1 at leading power. 
For comparison, the QCD sum rule results for these two 
ratios are shown (upper (blue) line refers to $A_1/V$, lower (black) 
line to $T_2/T_1$.)
\label{fig:sr}}
\end{figure}

The result of this computation is shown in Figure~\ref{fig:sr} 
for the various form factor ratios. The ratios are normalized 
such that in absence of any radiative corrections they equal 
1 for all $q^2$. Our final result, 
which includes $R$ to order $\alpha_s$ and
the spectator-scattering term to order $\alpha_s^2$ as well as 
the summation of leading logarithms to all orders is 
shown as the solid (black) curves. To display the size of the 
various contributions to the complete result, 
we also show the result following from neglecting the 
spectator-scattering term (dash-dotted (red) curves), 
and from evaluating the spectator-scattering contribution 
in leading-order (long-dashed (blue) curves), which corresponds 
to the previous results~\cite{Beneke:2000wa}. As has already been 
discussed in Section~\ref{sec:numerics} the new NLO correction 
always enhances the symmetry-breaking effect. The correction from  
$R(E)$ in (\ref{ffratios}) is always smaller than the 
spectator-scattering contribution. In fact, it is even smaller 
than the NLO spectator-scattering term, despite the fact that 
the latter is formally of order $\alpha_s^2$. Overall, the 
deviations from the symmetry-limit range up to 40\%, which is 
significant but not anomalously large given that the typical scales 
involved are in the range of $1.5\,$GeV. The theoretical uncertainties
in the relative NLO spectator-scattering term have been discussed 
before in Section~\ref{sec:numerics}. The more important unknown 
factor resides in the normalization of the tree contribution 
(\ref{htree}), which involves the product 
\begin{equation}
\frac{\hat f_B f_M \langle \bar v^{-1}\rangle_M}{\lambda_B \xi_a(E)}
\end{equation}
of hadronic parameters. We estimate the theoretical errors of 
the factors to be around 15\% ($\hat{f}_B$), 15\% ($f_{\rho\perp}$), 
10\% ($\langle \bar v^{-1}\rangle_M$), 30\% ($\lambda_B$) and 
15\% ($\xi_a(E)$), so it is clear that the curves in the figure 
are affected by a significant normalization uncertainty. In
particular, adopting the QCD sum-rule result $\lambda_B=0.46\,$GeV 
rather than $0.35\,$GeV decreases the deviations of the form factor
ratios from unity by about 30\%.

It is instructive to compare this result for the form factor ratios 
with the QCD sum rule calculations. The corresponding sum rule ratios 
are shown as dashed (black; black and blue in the lower right panel) 
curves in Figure~\ref{fig:sr}. One notices that the 
sum rule calculation satisfies the symmetry relations remarkably 
well -- the ratios are in general closer to 1 than predicted on 
the basis of the heavy-quark limit corrected by radiative and 
spectator-scattering effects. There are also significant differences 
concerning the sign of the correction similar to those observed 
already in~\cite{Beneke:2000wa}. It is unclear whether the differences
between the sum rule calculations and those based on the heavy-quark 
limit are due to $1/m_b$ power corrections or ununderstood 
systematics of the sum rule calculation (see the discussion in 
\cite{DeFazio:2005dx}). For instance, the sum-rule result for 
the ratio involving $m_B/(2 E)\,T_2-T_3$ has changed from about 0.7 
to almost 1.2 with the update \cite{ff} of the form factor
calculations. This may not be surprising, since the ratio involves 
cancellations between form factors and may be particularly sensitive 
to the uncertainties of sum rule calculations. In such cases the SCET
calculation of form factor relations is probably more reliable than 
the QCD sum rule method. In general, the comparison of the two methods
leads to the conclusion that the theoretical calculations of form 
factors with QCD sum rules are affected by considerable uncertainties 
until the systematics of and discrepancies with the heavy-quark limit
are better understood.

\subsection{Radiative vs. semi-leptonic decay}

Factorization calculations of radiative and hadronic two-body $B$ decays 
involving only light mesons (and leptons) make use of the 
form factors at maximal recoil. It is therefore of interest 
to investigate the short-distance corrections at $x=1$, i.e. 
$E=m_B/2$ or $q^2=0$. In addition to the exact relations
(\ref{exactrel}), the first and fourth relations of (\ref{physscet1}) 
also degenerate to 
\begin{equation}
\frac{m_B}{2 E}\,f_0(E) = \xi_P^{\rm FF}, 
\qquad \frac{m_V}{E}\,A_0(E) = \xi_\parallel^{\rm FF}
\end{equation}
as a consequence of the equations of motion. This leaves only 
two interesting ratios, namely ${\cal R}_1$ and ${\cal R}_2$ 
defined in (\ref{ffratios}). The last ratio in (\ref{physscet1}) involving 
$T_2$ and $T_3$ can be obtained from ${\cal R}_1$ replacing 
$\xi_P^{\rm FF}=f_+$ by $\xi_\parallel^{\rm FF}$. 

At $x=1$ we obtain the analytic expressions (assuming the 
asymptotic form $\phi_M(v)=6 v\bar v$ for the light-meson 
distribution amplitude)
\begin{eqnarray}
{\cal R}_1 &=& 1 + \frac{\alpha_s}{4\pi}
\left[-\frac{8}{3}\ln\frac{\nu}{m_b}+\frac{8}{3}\right] 
- \frac{8\pi\alpha_s(\mu)}{3}\,\frac{\hat{f}_B f_M}
{M_B\lambda_B \xi_P^{\rm FF}} \,\Bigg\{1+
\frac{\alpha_s(\mu)}{4\pi}
\nonumber\\
&&\times \,\Bigg[
-\frac{8}{3}\ln\frac{\nu}{m_b}
-\frac{8}{3}\ln^2\frac{m_b}{\mu}
+\frac{4}{3}\left\langle\ln^2\frac{m_b\omega}{\mu^2}\right\rangle 
+\left(\frac{8}{3}-\frac{2\pi^2}{9}\right)\ln\frac{m_b}{\mu} 
\nonumber\\
&&\hspace*{0.7cm} +\,\left(-9+\frac{\pi^2}{9}+\frac{2n_f}{3}\right) 
 \left\langle\ln\frac{m_b\omega}{\mu^2}\right\rangle 
+\frac{103}{6}+\frac{5\pi^2}{9}-\frac{19n_f}{9}
+ \delta_{\rm log}^\parallel
\Bigg]\Bigg\}
\nonumber\\[0.2cm]
&=& 1 + 0.046\,(R_T) - 0.165\,\Big\{1+0.540\,(\mbox{NLO spec.}) 
- 0.01 \,(\delta_{\rm log}^\parallel)\Big\} 
\nonumber\\[0.2cm]
&=& 0.794,
\nonumber\\[0.4cm]
{\cal R}_2 &=& 1 + \frac{\alpha_s}{4\pi}
\left[-\frac{8}{3}\ln\frac{\nu}{m_b}-\frac{4}{3}\right] 
+ \frac{4\pi\alpha_s(\mu)}{3}\,\frac{\hat{f}_B f_{M\perp}}
{M_B\lambda_B \xi_\perp^{\rm FF}} \,\Bigg\{1+
\frac{\alpha_s(\mu)}{4\pi}
\nonumber\\
&&\times \,\Bigg[
-\frac{8}{3}\ln\frac{\nu}{m_b}
-\frac{8}{3}\ln^2\frac{m_b}{\mu}
+\frac{4}{3}\left\langle\ln^2\frac{m_b\omega}{\mu^2}\right\rangle 
+\left(-\frac{2}{3}-\frac{2\pi^2}{9}\right)\ln\frac{m_b}{\mu} 
\nonumber\\
&&\hspace*{0.7cm} 
+\,\left(-\frac{26}{3}+\frac{\pi^2}{9}+\frac{2n_f}{3}\right) 
 \left\langle\ln\frac{m_b\omega}{\mu^2}\right\rangle 
+\frac{27}{2}+\pi^2-\frac{2\zeta(3)}{3}-\frac{19n_f}{9}
+ \delta_{\rm log}^\perp
\Bigg]\Bigg\}
\nonumber\\[0.2cm]
&=& 1 - 0.023\,(R_T) + 0.086\,\Big\{1+0.418\,(\mbox{NLO spec.}) 
+ 0.03 \,(\delta_{\rm log}^\parallel)\Big\}
\nonumber\\[0.2cm]
&=& 1.102,
\end{eqnarray} 
where $\delta^a_{\rm log}$ denotes the small effect form the
renormalization-group summation. The numerical results refer to 
the pion (${\cal R}_1$) and $\rho$ meson (${\cal R}_2$) 
with the parameters as specified 
above.  The ratio ${\cal R}_1$ with  
$\xi_P^{\rm FF}=f_+$ replaced by $\xi_\parallel^{\rm FF}$ 
and pion parameters replaced by $\rho$ meson parameters 
gives 0.707 instead of 0.794. For comparison the QCD sum rule 
calculation~\cite{ff} gives ${\cal R}_1=0.96$ (1.02 for $\rho$ 
and the relation involving $T_2$, $T_3$) 
and ${\cal R}_2=0.95$.

The factorization approach allows us to make predictions 
for the exclusive radiative $B$ decays 
$B\to M\gamma$ \cite{Beneke:2001at,Chay:2003kb} and 
$B\to M l^+ l^-$ \cite{bb}. The decays $B\to\rho\gamma$ 
together with  $B\to K^*\gamma$ are particularly interesting, 
because they may lead to a determination of the CKM matrix element 
$|V_{td}|$ or constrain flavour non-universality in penguin 
transitions. The main limitation turns out to be the poor knowledge of
SU(3) flavour symmetry breaking in the ratio of tensor 
form factors $T_1^\rho(0)/T_1^{K^*}(0)$
\cite{Bosch:2004nd,Ali:2004hn}. In \cite{Bosch:2004nd} 
it was therefore suggested to take the ratio of the 
$B\to\rho\gamma$ to the differential semi-leptonic 
$B\to\rho l \nu$ branching fraction, which avoids the problem 
of SU(3)-breaking, but introduces the ratio  $T_1/V$ of $\rho$-meson 
form factors at $q^2=0$. The method relies on normalizing 
the $B\to\rho\gamma$ rate to the differential decay rate 
\begin{equation}
\frac{d^2\Gamma(B\to\rho l\nu)}{dq^2 d\cos\theta} 
\propto (1+\cos\theta)^2\,H_-^2 
+ (1-\cos\theta)^2\,(H_+^2+2 H_0^2)
\end{equation}
near $\cos\theta=1$ (with $\theta$ the angle between the 
neutrino momentum and the $B$ meson momentum in the 
$l\nu$ center-of-mass frame) and $q^2=0$. The angle cut 
has the effect of isolating the negative helicity form factor $H_-$, 
which has a simple expression in the heavy-quark limit. 
Neglecting quadratic effects in the light meson mass, 
\begin{eqnarray}
H_-(q^2) &=& \sqrt{\frac{q^2}{m_B^2}} \left(\frac{m_B+m_\rho}{m_B}
\,A_1(q^2) + \frac{m_B^2-q^2}{m_B (m_B+m_\rho)}\,V(q^2)\right) 
\nonumber\\
&=& 2 \,\sqrt{\frac{q^2}{m_B^2}}
\left(1-\frac{q^2}{m_B^2}\right)\,\frac{m_B}{m_B+m_\rho}\,V(q^2), 
\end{eqnarray}
hence the ratio of branching fractions involves
\begin{equation}
\frac{H_-^2}{T_1^2} = \frac{4q^2}{m_B^2}
\left(1-\frac{q^2}{m_B^2}\right)^{\!2}\frac{H_-^2}{{\xi_\perp^{\rm FF}}^2}
\,\,\stackrel{q^2\to 0}{\to}\,\, \frac{4q^2}{m_B^2} \,
\frac{1}{{\cal R}_2^2}.
\end{equation} 
Assigning a 60\% uncertainty to the spectator-scattering contribution 
to ${\cal R}_2$ we obtain $1/{\cal R}_2^2 =0.82\pm 0.12$ 
to be compared with the QCD sum rule value $1/{\cal R}_2^2 = 
1.11 \pm 0.22$, where we assigned a 10\% uncertainty to the  
calculation of \cite{ff}. The disagreement between the two 
numbers is unfortunate and should be resolved. Assuming the 
result of the calculation in the factorization approach, we obtain 
a 10\% uncertainty on $|V_{td}/V_{ub}|$ from
the form factor ratio in the method proposed in \cite{Bosch:2004nd}. 
This does not include an uncertainty from power-suppressed 
effects.

The tensor-to-vector ratio ${\cal R}_2(q^2)$ also appears in the 
forward-backward asymmetry in the electroweak 
penguin decays $B\to M l^+ l^-$. The complete calculation of 
the decay matrix element divides into ``factorizable'' and 
``non-factorizable'' contributions \cite{Beneke:2000wa,bb}. 
In this terminology, ``factorizable'' contributions are  
related to the heavy-to-light form factors, and hence are the relevant
ones here. Inserting 
\begin{equation}
(m_B+m_V)\,\frac{T_1}{V} = (m_B-m_V)\,\frac{T_2}{A_1} = 
m_B {\cal R}_2 
\end{equation}
into Eq.~(75) of \cite{bb}, we obtain the differential forward-backward 
asymmetry \cite{Beneke:2000wa}
\begin{equation}
\frac{dA_{\rm FB}}{dq^2} \propto 
\mbox{Re} \,\Big[C_9+Y(q^2) + \frac{2 m_b M_B}{q^2}\,C_7^{\rm eff}\,
{\cal R}_2(q^2) + \mbox{non-fact. terms}\Big].
\end{equation}
Since the dependence on $q^2$ is mainly through 
${\cal R}_2(q^2)/q^2$ it follows that the increase of ${\cal
  R}_2(q^2)$ by several percent due to the next-to-leading order
spectator-scattering correction implies an increase in the 
position of the asymmetry zero in approximately the same 
proportion. 

\subsection{Hadronic decays}

The jet-function computed in this paper also appears in the 
next-to-leading order correction to spectator-scattering in hadronic 
two-body $B$ decays to light mesons.  We outline this 
effect by the example of $B\to\pi\pi$ decays, keeping the discussion
short, since the NLO correction is not yet completely available. 

The factorization formula reads \cite{Beneke:1999br} 
\begin{eqnarray}
   \langle \pi\pi |Q_i|\bar B\rangle
   &=&  f_+(0)\,\,T_{i}^I * f_{\pi}\phi_{\pi} +  
        T_i^{II} * f_B\phi_{B+} * f_{\pi}\phi_{\pi} *
        f_{\pi}\phi_{\pi},
\end{eqnarray}
where $Q_i$ denotes an operator from the effective weak Hamiltonian, 
and the formula holds up to power corrections. 
The second term describes spectator-scattering. Its short-distance 
coefficient is a convolution $T_i^{II}=C_i^{II}\star J_\parallel$, 
where $C_i^{II}$ is the coefficient function of a generalization of 
a B-type operator that takes into account the second pion. 
Since the pion that does not pick up the spectator anti-quark from the 
$\bar B$ meson decouples at the hard-scale, the physics at the 
hard-collinear scale is exactly the same as in the $B\to\pi$ 
transition. Hence the jet-function in hadronic decays 
equals $J_\parallel$ \cite{Bauer:2004tj}, which has 
been computed above. Note that this implies that 
the strong rescattering phases are all generated at the hard scale 
$m_b$ (at leading order in the heavy-quark expansion), since the 
jet-function is real. 

Spectator-scattering is particularly important 
for decay amplitudes of the  ``colour-suppressed'' final state
$\pi^0\pi^0$, because the colour-suppression is absent for the 
spectator-scattering term. The situation is opposite for the 
colour-allowed final state $\pi^-\pi^+$. Both amplitudes 
are relevant to $\pi^-\pi^0$. In the following we shall 
therefore focus on the coefficient $\alpha_2(\pi\pi)$ 
that describes the colour-suppressed tree amplitude. We emphasize that
a complete NLO calculation of 
spectator-scattering requires the calculation of the hard coefficient 
$C_i^{II}$ as well. The remarks below must therefore be understood as 
preliminary.

Following the notation of \cite{Beneke:2003zv} (Eqs.~(35) and (47)) 
we write 
\begin{eqnarray}
\alpha_2(\pi\pi) &=& C_2+\frac{C_1}{N_c} + 
\frac{C_1}{N_c}\,\frac{\alpha_s(\mu) C_F}{4\pi}\,V_2(\pi) 
\nonumber\\
&& +\,
\frac{C_1}{N_c}\,\frac{\pi \alpha_s(\mu_h)C_F}{N_c}\,
\Big[H_2^{\rm tw2}(\pi\pi)\,I_\parallel+H_2^{\rm tw3}(\pi\pi)\Big], 
\label{a2}
\end{eqnarray}
where now $\mu$ should be chosen of order $m_b$ and $\mu_h$ 
is a hard-collinear scale assumed to be $\mu_h=\sqrt{\Lambda_h 
\mu}$ with $\Lambda_h=0.5\,\mbox{GeV}$. The $C_i$ are Wilson
coefficients from the effective weak-interaction Hamiltonian, $V_2(\pi)$ is 
a vertex correction, and $H_2^{\rm tw2}(\pi\pi)+
H_2^{\rm tw3}(\pi\pi)$ the spectator-scattering term at
tree level, which we separated into a leading-power (``tw2'') and 
a power-suppressed  ``chirally enhanced'' (``tw3'') term. The new
ingredient in this formula is the factor $I_\parallel$, which equals 1 
in the absence of the NLO correction to the jet-function,  
and is given by (\ref{iperp}) including the correction. 
Exactly the same modification applies to the spectator-scattering 
contribution to $\alpha_1(\pi\pi)$ and the leading-power pieces 
of the penguin amplitudes.
Numerically, with parameters defined in \cite{Beneke:2003zv}, 
we obtain 
\begin{eqnarray}
\alpha_2(\pi\pi) &=& 0.17 - [0.17+0.08i\,]_{V_2} + 
\left\{\begin{array}{lc}
[0.11\cdot 1.37]_{H_2^{\rm tw2}\,\cdot I_\parallel} + 
[0.07] _{H_2^{\rm tw3}} & \qquad \mbox{(default)}
\\
{[0.29\cdot 1.57]}_{H_2^{\rm tw2}\,\cdot I_\parallel} + 
{[0.17]} _{H_2^{\rm tw3}} 
& \qquad \mbox{(S4)}
\end{array}
\right.
\nonumber\\[0.2cm]
&=&
\left\{\begin{array}{lc}
0.22 \,(0.18) - 0.08i & \qquad \mbox{(default)}
\\ 
0.64 \,(0.47) - 0.08i & \qquad \mbox{(S4)}
\end{array}
\right.
\end{eqnarray}
The various terms and factors correspond to those in (\ref{a2}) 
and we show the numbers for the default parameter set and the 
set S4 that provides a better overall description of 
hadronic two-body modes. Due to the near cancellation\footnote{
The size of the loop correction is due to the absence of 
colour-suppression, which makes the tree amplitude small, and is
therefore not an indication of failure of the perturbative 
expansion.} of the 
tree term with the vertex correction the colour-suppressed 
tree amplitude comes essentially from spectator-scattering. 
The factors 1.37 and 1.57 show the effect from the NLO correction 
to the jet-function. In the final line the number in brackets 
gives the result from \cite{Beneke:2003zv}, the unbracketed 
number corresponds to including the new jet-function term. 
To illustrate the implications of these results, we show in 
Table~\ref{tab:bpipi} the CP-averaged $B\to\pi\pi$ branching 
fractions corresponding to the four cases (default vs. S4, 
with and without NLO jet-function correction). For simplicity, 
we only consider the NLO jet function correction to the 
tree amplitudes $\alpha_1(\pi\pi)$ and $\alpha_2(\pi\pi)$  
(colour-allowed and colour-suppressed), but not to the penguin 
amplitudes, since this gives the dominant  
effect (and the results are preliminary anyway, see above). 
It is clearly seen that the NLO correction to spectator-scattering 
can have a significant effect. The enhancement of the
colour-suppressed tree amplitude brings the theoretical computation 
in better agreement with data, since the large $\pi^0\pi^0$ 
rate and the small $\pi^+\pi^-$ to $\pi^-\pi^0$ ratio 
favour a large colour-suppressed tree amplitude~\cite{Beneke:2003zv}. 
We do not discuss the direct 
CP asymmetries, since we expect the still missing 
NLO hard correction to spectator-scattering (which includes 
a new source of rescattering phases) to be the more important 
factor. 

\renewcommand{\arraystretch}{1.4}
\begin{table}
\begin{center}
\renewcommand{\baselinestretch}{2}
\begin{tabular}{|c|c|c|c|c|}\hline
Scenario & default, LO jet& default, NLO jet & S4, LO jet & S4, NLO jet \\
\hline
$\alpha_1(\pi\pi)$ & 
$0.99+0.02i$ & $0.98+0.02i$ & $0.88+0.02i$ & $0.81+0.02i$\\
\hline
$\alpha_2(\pi\pi)$ & 
$0.18-0.08i$ & $0.22-0.08i$ & $0.47-0.08i$ & $0.64-0.08i$\\
\hline
$\mbox{Br}(\bar{B}^0\to \pi^+\pi^-)$ & 
$8.86$ & $8.62$ & $5.17$ &  $4.58$\\
\hline
$\mbox{Br}(\bar{B}^0\to \pi^0\pi^0)$ & 
$0.35$ & $0.40$ & $0.70$ & $1.13$ \\
\hline
 $\mbox{Br}(B^-\to \pi^-\pi^0)$& 
$6.03$ & $6.28$ & $5.07$ & $5.87$ \\
\hline
\end{tabular}
\end{center}
\caption{Tree amplitude coefficients $\alpha_1$ and $\alpha_2$, and the 
CP-averaged $\pi\pi$ branching ratios in units of $10^{-6}$ in 
the default and S4 scenario of \cite{Beneke:2003zv} showing the 
effect of the NLO jet function correction. \label{tab:bpipi}}
\end{table}


\section{Conclusion}
\label{sec:conclusion}

Spectator-scattering plays an important role in the theory of
exclusive $B$ decays. It is also rather complicated, because 
several scales, $m_b$ (hard), $\sqrt{m_b\Lambda}$ (hard-collinear), 
and $\Lambda$ (hadronic) are involved. The development of 
QCD factorization and soft-collinear 
effective theory has made it possible to formulate the calculation 
in terms of two separate matching steps, in which the effects 
from the short-distance scales $m_b$ and $\sqrt{m_b\Lambda}$ 
are calculated in perturbation theory. In previous 
work \cite{Beneke:2004rc} we began the calculation of 1-loop 
corrections to spectator-scattering effects in heavy-to-light 
meson form factors in the large-recoil region with the computation 
of the hard coefficient functions. In this paper we have completed 
the second step with the computation of the hard-collinear 
coefficient function, also called jet-function. Since the calculation 
involves the definition of various renormalized operators in QCD, 
$\SCETI$, and $\SCETII$, 
and the treatment of evanescent operators in dimensional 
regularization, we have described the technical aspects of this 
work in some length. Our results provide a check of similar results 
obtained by Becher et al. \cite{Becher:2004kk,Hill:2004if}. 
The jet-function computed 
here is relevant to many different $B$ decays, including 
radiative and hadronic $B$ decays in the QCD factorization approach. 

The results may be summarized as follows: we find significant 
enhancements of spectator-scattering at next-to-leading order, 
which increase the deviation of form factor ratios from the 
asymptotic heavy quark limit, in which perturbative and power 
corrections are neglected. We have also included the summation of 
formally large logarithms $\ln m_b/\Lambda$, but found this effect 
to be negligible compared to the full 1-loop correction. 
Despite the small scale of order $1.5\,$GeV involved, there is no sign
that a perturbative treatment is not applicable. The 1-loop effects 
from the hard scale and the hard-collinear scale are about equally 
important, being on the order of (20-40)\% (depending on parameters), 
at least in the $\overline{\rm MS}$ factorization 
scheme adopted throughout this work. It follows that the 
dominant theoretical uncertainties are related to hadronic 
input parameters such as moments of light-cone distribution
amplitudes and decay constants. In addition to the symmetry-breaking 
corrections to form factor ratios, we also discussed radiative 
and hadronic two-body decays. Although the jet-function constitutes 
only one aspect of the NLO correction to spectator-scattering in 
hadronic decays, we have seen that the NLO enhancement has interesting
implications for final states with significant 
colour-suppressed tree amplitudes.

We would also like to emphasize the theoretical conclusions from 
this calculation, since the form factors are up to now the only 
observables, for which a complete two-step matching in soft-collinear 
effective theory has been explicitly carried out to the 
1-loop level in a case with spectator-scattering. 
The factorization arguments that lead to the 
formula (\ref{eq:factorization}) rely on the demonstration 
that the B-type \SCETI{} currents can be matched to 
$\SCETII{}$ without encountering endpoint-divergent convolution 
integrals, which would violate naive \SCETII{}
factorization \cite{Beneke:2003pa,Lange:2003pk}. The  
calculations performed here and in \cite{Beneke:2004rc} provide  
an explicit verification of these arguments at the 1-loop 
level.

\vskip0.4cm\noindent
{\em Note added:} We have been informed of related 
work by G.~Kirillin, in which he computes the 1-loop correction to 
the jet-function $J_\parallel$, and to the coefficient function 
$C^{(B1)}_{f_+}$.

\subsubsection*{Acknowledgements}
We are grateful to S.~J\"ager for careful reading of the manuscript. 
M.B. would like to thank the INT, Seattle and KITP, Santa 
Barbara for their  generous hospitality during the summer of 2004, 
when most of this work was being done. D.Y. acknowledges support 
from the Alexander-von-Humboldt Stiftung and the
Japan Society for the Promotion of Science.  
The work of M.B. is supported in part by the 
DFG Sonder\-forschungs\-bereich/Trans\-regio~9 
``Com\-puter\-ge\-st\"utz\-te Theoretische Teilchenphysik''. 


\appendix

\section{Short-distance coefficients}
\label{app:Wilson} 

\subsection{Change of basis}
The coefficient functions of the operators defined in 
(\ref{scalar:newbasis}) to (\ref{tensor:newbasis}) are given in terms 
of those defined and calculated in \cite{Beneke:2004rc} 
(denoted with subscript ``old'') as follows:
\begin{eqnarray}
\mbox{scalar}:&&\nonumber\\[0.1cm]
C_S^{(A0)} &=& C_{S\rm {\rm old}}^{(A0)},
\nonumber\\
C_{S}^{(B1)} &=& C_{S{\rm old}}^{(B1)}-\frac{C_{S {\rm
old}}^{(A0)}}{x}\,,\\[0.2cm]
\mbox{vector}:&&\nonumber\\[0.1cm]
C_V^{(A0)1} &=& C_{V\rm {\rm old}}^{(A0)1}\,,\hspace{2cm}
C_V^{(A0)2} \,=\, C_{V\rm {\rm old}}^{(A0)3}\,,\nonumber\\
C_V^{(A0)3} &=& C_{V\rm {\rm old}}^{(A0)2}-C_{V\rm {\rm
old}}^{(A0)1}\,,
\nonumber\\
C_V^{(B1)1} &=& C_{V{\rm old}}^{(B1)3}-\frac{1}{x}
\left(2 C_{V {\rm old}}^{(A0)1}+C_{V {\rm old}}^{(A0)3}\right)
\,,\nonumber\\
C_V^{(B1)2} &=& - C_{V{\rm old}}^{(B1)2}+\frac{1}{x} \left(2 C_{V
{\rm old}}^{(A0)1}-C_{V {\rm old}}^{(A0)2}\right)
\,,\nonumber\\
C_V^{(B1)3} &=& C_{V{\rm old}}^{(B1)1}+\frac{1}{2} \,C_{V{\rm
old}}^{(B1)4} +\frac{C_{V {\rm old}}^{(A0)2}}{x}\,,\nonumber\\
C_V^{(B1)4} &=& \frac{1}{2} \,C_{V{\rm old}}^{(B1)4}-\frac{1}{x}
\left(C_{V {\rm old}}^{(A0)1}-C_{V {\rm old}}^{(A0)2}\right)
\,,\\[0.2cm]
\mbox{tensor}:&&\nonumber\\[0.1cm]
C_T^{(A0)1} &=& -\frac{1}{2}\,C_{T\rm {\rm old}}^{(A0)1}
\,,\hspace{2cm} C_T^{(A0)2} \,=\, C_{T\rm {\rm old}}^{(A0)3}
\,,\nonumber\\
C_T^{(A0)3} &=& C_{T\rm {\rm old}}^{(A0)2}-
 C_{T\rm {\rm old}}^{(A0)1}
\,,\hspace{1cm} C_T^{(A0)4} \,=\, C_{T\rm {\rm
old}}^{(A0)1}+C_{T\rm {\rm old}}^{(A0)3}-
 C_{T\rm {\rm old}}^{(A0)4}
\,,\nonumber\\
C_T^{(B1)1} &=& C_{T{\rm old}}^{(B1)2}-\frac{1}{2}\,C_{T{\rm
old}}^{(B1)6}+ \frac{1}{x} \left(C_{T {\rm old}}^{(A0)1}+C_{T {\rm
old}}^{(A0)4}\right)\,,
\nonumber\\
C_T^{(B1)2} &=& C_{T{\rm old}}^{(B1)1}-C_{T{\rm old}}^{(B1)5}-
\frac{C_{T {\rm old}}^{(A0)1}}{x} \,,
\nonumber\\
C_T^{(B1)3} &=& -\frac{1}{2}\,C_{T{\rm old}}^{(B1)6}- \frac{1}{x}
\left(C_{T {\rm old}}^{(A0)1}+C_{T {\rm old}}^{(A0)3}- C_{T {\rm
old}}^{(A0)4}\right) \,,
\nonumber\\
C_T^{(B1)4} &=& -C_{T{\rm old}}^{(B1)5}+ \frac{1}{x} \left(C_{T
{\rm old}}^{(A0)1}-C_{T {\rm old}}^{(A0)2}\right)\,,
\nonumber\\
C_T^{(B1)5} &=& C_{T{\rm old}}^{(B1)3}- \frac{1}{x} \left(2 C_{T
{\rm old}}^{(A0)2}+2 C_{T {\rm old}}^{(A0)3}- C_{T {\rm
old}}^{(A0)4}\right)\,,\nonumber
\\
C_T^{(B1)6} &=& \frac{1}{2}\,C_{T{\rm old}}^{(B1)4}- \frac{1}{2 x}
\left(C_{T {\rm old}}^{(A0)1}-2 C_{T {\rm old}}^{(A0)2}\right)\,,
\nonumber\\
C_T^{(B1)7} &=& \frac{1}{4}\left(C_{T{\rm old}}^{(B1)7}- C_{T{\rm
old}}^{(B1)4}\right) + \frac{1}{2 x} \left(C_{T {\rm
old}}^{(A0)1}-C_{T {\rm old}}^{(A0)2}\right)\,.
\end{eqnarray}
In the new basis the pseudoscalar coefficients equal the scalar 
coefficients, and the axial-vector coefficients equal the vector
coefficients. Furthermore $x=\nm v\,\np p^\prime/m_b=2E/m_b$ 
with $E$ the energy of the light meson.

\subsection{Coefficients appearing in the form factors}

The five independent A0-coefficients appearing in the \SCETI{}
representation of the form factors (\ref{pscet1}), (\ref{vscet1}) 
are given by
\begin{eqnarray}
C^{(A0)}_{f_+}&=&
C_V^{(A0)1}+\frac{x}{2}C_V^{(A0)2}+C_V^{(A0)3}\nonumber\\
&=&1 +\frac{\alpha_s C_F}{4\pi}
 \bigg [-2 \ln^2 \bigg (\frac{x}{\hat \mu}\bigg ) +5
         \ln\bigg (\frac{x}{\hat \mu}\bigg )-2 {\rm Li}_2
         (1-x)-\frac{\pi^2}{12}- 3 \ln x-6\bigg
         ]\,,\nonumber\\
C^{(A0)}_{f_0}&=&
C_V^{(A0)1}+\bigg(1-\frac{x}{2}\bigg)C_V^{(A0)2}+C_V^{(A0)3}\nonumber\\
&=&1 +\frac{\alpha_s C_F}{4\pi}
 \bigg [ -
        2 \ln^2 \bigg (\frac{x}{\hat \mu}\bigg ) +5
         \ln\bigg (\frac{x}{\hat \mu}\bigg )-2 {\rm Li}_2
         (1-x)-\frac{\pi^2}{12}- \frac{3-5 x}{1-x} \ln x-4\bigg
         ]\,,\nonumber\\
C^{(A0)}_{f_T}&=&
-2 C_T^{(A0)1}+C_T^{(A0)2}-C_T^{(A0)4},\nonumber\\
&=&1 +\frac{\alpha_s C_F}{4\pi}
 \bigg [-2 \ln\hat \nu -
        2 \ln^2 \bigg (\frac{x}{\hat \mu}\bigg ) +5
         \ln\bigg (\frac{x}{\hat \mu}\bigg )-2 {\rm Li}_2
         (1-x)-\frac{\pi^2}{12}\nonumber\\&&- \frac{3-x}{1-x} \ln x-6\bigg
         ]\,\\
C^{(A0)}_{V}&=&C^{(A0)1}_{V}
\nonumber\\
&=&1 +\frac{\alpha_s C_F}{4\pi}
 \bigg [ -
        2 \ln^2 \bigg (\frac{x}{\hat \mu}\bigg ) +5
         \ln\bigg (\frac{x}{\hat \mu}\bigg )-2 {\rm Li}_2
         (1-x)-\frac{\pi^2}{12}- \frac{3-2 x}{1-x} \ln x-6\bigg
         ]
\,,\nonumber\\
C^{(A0)}_{T_1}&=&
-2 C_T^{(A0)1}+\bigg(1-\frac{x}{2}\bigg )C_T^{(A0)2}+C_T^{(A0)3}\nonumber\\
&=&1 +\frac{\alpha_s C_F}{4\pi}
 \bigg [-2 \ln\hat \nu -
        2 \ln^2 \bigg (\frac{x}{\hat \mu}\bigg ) +5
         \ln\bigg (\frac{x}{\hat \mu}\bigg )-2 {\rm Li}_2
         (1-x)-\frac{\pi^2}{12}- 3 \ln x-6\bigg
         ]\,.\nonumber
\end{eqnarray}
The variable $E$ used in (\ref{pscet1}), (\ref{vscet1}) is related to
$x$ through $x=\nm v\,\np p^\prime/m_b=2E/m_b$. We also define
$\alpha_s=\alpha_s(\mu)$, 
$\hat \mu=\mu/m_b$ and $\hat \nu=\nu/m_b$, where $\nu$ is the 
renormalization scale of the QCD tensor current, and $\mu$ is the \SCETI{} 
renormalization scale. The $\mu$ dependence cancels the 
corresponding dependence of the \SCETI{} form factors $\xi_a(E)$. 
The heavy quark mass is renormalized in the pole scheme. 
The five independent B-coefficients are given by 
\begin{eqnarray}
C^{(B1)}_{f_+}&=&\frac{x}{2}C_V^{(B1)1}+C_V^{(B1)2}\nonumber \\
&=&\bigg({-2}+\frac{1}{x}\bigg )\Bigg \{1+ \frac{\alpha_s
C_F}{4\pi} \bigg[-2 \ln^2 \bigg ( \frac{x}{\hat \mu} \bigg )+\ln
\bigg (\frac{x}{\hat \mu}\bigg )-\frac{3}{1-2 x} \ln x -2 {\rm
Li}_2 (1-x)-\frac{\pi^2}{12}\nonumber \\
&&-\frac{2(1-x)}{1-2 x}+\frac{x}{(1-2x)(1-x\xi)}
-\frac{4(1-x)}{(1-2 x)\bar \xi}\ln \xi+\frac{x(2-x\xi)}{(1-2
x)(1-x\xi)^2}\ln (x\xi)\bigg]
\nonumber\\
&& + \, \frac{\alpha_s}{4\pi} \bigg (C_F-\frac{C_A}{2}\bigg
)\,\bigg[\frac{4}{\xi} \ln \bar \xi \ln \hat \mu+\frac{2}{\xi}
F(x,x\bar \xi)+\frac{2}{x(1-2 x)\xi\bar \xi} G
\nonumber\\
&&+\frac{2 }{1-2 x}\bigg (\frac{2(1-x)}{\xi}\ln \bar\xi+\frac{3-2
x}{\bar\xi}\ln
\xi-\frac{x}{1-x\xi}\ln(x\xi)\bigg)\bigg]\Bigg \},\nonumber\\
C^{(B1)}_{f_0}&=&\bigg(1-\frac{x}{2}\bigg)C_V^{(B1)1}+C_V^{(B1)2}\nonumber \\
&=&-\frac{1}{x}\Bigg\{1+ \frac{\alpha_s C_F}{4\pi} \bigg[-2 \ln^2
\bigg ( \frac{x}{\hat \mu} \bigg )+\ln \bigg (\frac{x}{\hat
\mu}\bigg )-3 \ln x -2 {\rm
Li}_2 (1-x)-\frac{\pi^2}{12} \nonumber \\
&&+\frac{2}{\bar \xi}\bigg (\frac{(2-x)\ln x}{1-x}-\frac{(2-x
\xi)\ln (x\xi)}{1-x \xi}\bigg )+\frac{x (2-x\xi) }{(1-x\xi)^2}\ln
(x\xi)+\frac{x}{1-x\xi}\bigg] \nonumber\\
&& + \, \frac{\alpha_s}{4\pi} \bigg (C_F-\frac{C_A}{2}\bigg
)\,\bigg[\frac{4}{\xi} \ln \bar \xi \ln \hat \mu+\frac{2 }{\bar
\xi}\ln \xi-\frac{2 x}{1-x\xi}\ln (x\xi) +\frac{2}{\xi} F(x,x\bar
\xi)-\frac{2}{x\xi\bar \xi}
G\bigg]\Bigg \}\,,\nonumber\\
C^{(B1)}_{f_T}&=&-C^{(B1)5}_T\nonumber \\
&=& \frac{1}{x}\Bigg \{1+\frac{\alpha_s C_F}{4\pi} \bigg[ -2 \ln
\hat \nu-2 \ln^2\bigg (\frac{x}{\hat \mu}\bigg )+ \ln \bigg
(\frac{x}{\hat \mu}\bigg )-3\ln x - 2{\rm
Li}_2(1-x)-\frac{\pi^2}{12}-2\nonumber \\
&& +\frac{2}{\bar \xi} \bigg (\frac{(2-x)\ln
x}{1-x}-\frac{(2-x\xi)\ln(x\xi)}{1-x\xi}\bigg)
-\frac{x(2-x\xi)\ln (x\xi)}{(1-x\xi)^2}-\frac{x}{1-x\xi}\bigg] 
\nonumber\\
&& + \, \frac{\alpha_s}{4 \pi} \bigg (C_F-\frac{C_A}{2}\bigg
)\,\bigg[ \frac{4}{\xi}\ln \bar \xi \ln\hat \mu+\frac{2}{\bar
\xi}\ln\xi+\frac{2 x}{1-x\xi}\ln (x\xi)
+\,\frac{2}{\xi} F(x,x\bar
\xi) -\frac{2}{x\xi\bar \xi} G
\bigg]\Bigg\}\,,\nonumber\\
C^{(B1)}_V&=& C^{(B1)4}_V\nonumber
\\&=&\frac{\alpha_s C_F}{4\pi} \frac{1}{1-x\xi}
\bigg[\frac{x \ln x}{1-x}-\frac{\ln (x\xi)}{1-x\xi}-\frac{\ln
\xi}{\bar \xi}-1\bigg
]\ \nonumber \\
&& + \,\frac{\alpha_s}{4 \pi} \bigg (C_F-\frac{C_A}{2}\bigg
)\,\bigg[-\frac{2}{x\xi} \ln\bar \xi-\frac{2}{x\bar \xi} \ln\xi
  +\frac{2 \ln (x\bar \xi)}{1-x\bar \xi}+\frac{2 \ln (x\xi)}{1-x\xi}
   -\frac{2}{x^2\xi\bar \xi}G \bigg]\,,\nonumber\\
C^{(B1)}_{T_1}&=&\bigg(1-\frac{x}{2}\bigg)
C^{(B1)3}_T+C^{(B1)4}_T\nonumber \\
&=&  (-1)\,\Bigg \{1+ \frac{\alpha_s C_F}{4\pi} \bigg[-2\ln \hat
\nu-2\ln^2\!\bigg ( \frac{x}{\hat \mu}\bigg )+
\ln\bigg ( \frac{x}{\hat \mu}\bigg )+\ln x-
2 {\rm Li}_2(1-x)-\frac{\pi^2}{12}-1\nonumber \\
&& -\frac{4\xi}{\bar \xi}\ln \xi \ln\hat \mu-\frac{2}{\bar \xi}\ln
\xi-
\frac{2\xi}{\bar \xi}F(x,x\xi) \bigg ] \nonumber\\
&& + \,\frac{\alpha_s}{4 \pi} \bigg (C_F-\frac{C_A}{2}\bigg
)\,\bigg[\bigg ( \frac{4(1+\xi)}{\xi}\ln\bar \xi+\frac{4\xi}{\bar
\xi}
\ln \xi\bigg )\ln \hat \mu-2 \ln\bar \xi+2 \ln \xi\nonumber \\
&&-2 F(x\bar \xi,x\xi)+\frac{2}{\xi}F(x,x\bar \xi)+\frac{2}{\bar
\xi} F(x,x\xi)-\frac{2}{x\bar \xi}G \nonumber \\
&&-\,\frac{2}{x\xi} \bigg ( {\rm Li}_2 (1-x)- {\rm Li}_2 (1-x\bar
\xi) \bigg )-2\bigg ]\Bigg\}\,.
\end{eqnarray}
The variables $E$ and $\tau$ used in (\ref{pscet1}), (\ref{vscet1})
are related to $x$ and $\xi$ through $x=2E/m_b$ and $\xi=\tau$. 
Diagrammatically $\xi$ corresponds to $n_+p^\prime_2/n_+p^\prime$, 
the fractional longitudinal momentum carried by the transverse 
collinear gluon in 
the B-type current operator. We also use $\bar \xi\equiv 1-\xi$, and 
introduced the two abbreviations
\begin{eqnarray}
F(y,z)&\equiv& \ln^2 y -\ln y +{\rm Li}_2(1-y)- \ln^2 z + \ln z
-{\rm
Li}_2(1-z),
\nonumber\\
G &\equiv& {\rm Li}_2 (1-x)-{\rm Li}_2(1-x \bar \xi)-{\rm Li}_2
(1- x\xi)+\frac{\pi^2}{6}.
\end{eqnarray}
These results are obtained by taking the appropriate linear
combinations of the coefficients given in \cite{Beneke:2004rc}. 
The variables  $(x_1,x_2)$ used there are related to $(x,\xi)$ 
by $x_1=x\bar \xi$ and $x_2=x\xi$ ($\xi\in [0,1]$).

\section{Integrals of coefficient functions}
\label{app:convolution}

\subsection{Integration of the jet-function}

The integrals 
$\int_0^1 d\tau\,j_a(\tau;v, \omega)$ of the jet-functions 
(\ref{jparfinal}), (\ref{jperpfinal}) are given by 
\begin{eqnarray}
\int_0^1 d\tau\,j_\parallel(\tau;v, \omega) &=& 
C_F L^2 \nonumber\\
&& \hspace*{-2.2cm} 
+ \,\left(C_F\left[-\frac{7}{3}+2\ln\bar v\right]
+ \left(C_F-\frac{C_A}{2}\right)\left[\frac{22}{3}+\frac{2\ln v}{\bar v} 
\right] +\frac{4}{3}\,n_f T_f\right) L 
\nonumber\\
&& \hspace*{-2.2cm} 
+ \, C_F\left[\frac{53}{9}-\frac{\pi^2}{6}+\left(-\frac{4}{3}+\frac{1}{v}
\right)\ln\bar v + \ln^2\bar v\right]
\nonumber\\
&& \hspace*{-2.2cm} 
+\, \left(C_F-\frac{C_A}{2}\right)\Bigg[-\frac{170}{9}+
\left(\frac{22}{3}-\frac{2}{v}
\right)\ln\bar v 
+\,\frac{1}{\bar v}\,\Bigg(\frac{\pi^2}{3}-2\ln v+\ln^2 v  
\nonumber\\
&& \hspace*{-2.2cm} 
+ \,2 (1-2 v)\Big(\mbox{Li}_2(v)-\mbox{Li}_2(\bar v)\Big)
\Bigg)\Bigg] 
+ \left(-\frac{5}{3}+\ln \bar v\right) \frac{4}{3} \,n_f T_f,
\nonumber \\
\int_0^1 d\tau\,j_\perp(\tau;v, \omega) &=& 
C_F L^2 \nonumber\\
&& \hspace*{-2.2cm} 
+ \,\left(C_F\left[-\frac{7}{3}+2\ln\bar v\right]
+ \left(C_F-\frac{C_A}{2}\right)\left[\frac{22}{3}+\frac{2\bar v}{v} 
\ln\bar v+\frac{2 (2-v) \ln v}{\bar v} 
\right] +\frac{4}{3}\,n_f T_f\right) L 
\nonumber\\
&& \hspace*{-2.2cm} 
+ \, C_F\left[\frac{44}{9}-\frac{\pi^2}{6}+\left(-\frac{7}{3}+\frac{2}{v}
\right)\ln\bar v + \ln^2\bar v\right]
\nonumber\\
&& \hspace*{-2.2cm} 
+\, \left(C_F-\frac{C_A}{2}\right)\Bigg[-\frac{152}{9}+
\left(\frac{22}{3}-\frac{2}{v}
\right)\ln\bar v +\frac{\bar v}{v}\ln^2\bar v
+\,\frac{1}{v \bar v}\,\Bigg(\frac{\pi^2}{3}-2v \ln v 
\nonumber\\
&& \hspace*{-2.2cm} 
+\,v (2-v) \ln^2 v + 2 (1-2 v)\Big(\mbox{Li}_2(v)-\mbox{Li}_2(\bar v)\Big)
\Bigg)\Bigg] 
+ \left(-\frac{5}{3}+\ln \bar v\right) \frac{4}{3} \,n_f T_f.
\end{eqnarray}
The analytic expressions of the integrals entering $I_a$ (see 
(\ref{jetintterm})) read 
(setting $n_f=4$, $T_f=1/2$ and $C_F=4/3$ and $C_A=3$)
\begin{eqnarray}
&&  \hspace*{-2cm}\frac{\lambda_B}{3} \int_0^1 \frac{dv}{\bar v}
\,\phi_M(v) \int_0^\infty \frac{d\omega}{\omega}\,
\phi_{B+}(\omega)\int_0^1 d\tau j_\parallel(\tau;v,\omega)  \nonumber \\
&=& \frac{4}{3} \langle L^2\rangle  + 
\left (-\frac{19}{3}+\frac{\pi^2}{9}\right )
\langle L \rangle
+\frac{169}{18}-\frac{2\pi^2}{9}-\frac{8}{3}\zeta(3)\nonumber\\
&& +a_1^M\bigg [
\frac{4}{3} \langle L^2\rangle
+\left (-\frac{110}{9}+\frac{\pi^2}{3}\right ) \langle L \rangle
+\frac{464}{27}+\frac{\pi^2}{9}-8\zeta(3)\bigg ]
\nonumber\\
&& +a_2^M\bigg [ \frac{4}{3} \langle L^2\rangle +\left
(-\frac{157}{9}+\frac{2\pi^2}{3}\right ) \langle L \rangle +
\frac{646}{27}+\frac{8\pi^2}{9}-16\zeta(3)\bigg ],
\nonumber\\
&&  \hspace*{-2cm}\frac{\lambda_B}{3} \int_0^1 \frac{dv}{\bar v}
\,\phi_M(v) \int_0^\infty \frac{d\omega}{\omega}\,
\phi_{B+}(\omega)\int_0^1 d\tau  
j_\perp(\tau;v,\omega)\nonumber \\
&=& \frac{4}{3}  \langle L^2\rangle+ \left (-6+\frac{\pi^2}{9}\right )
 \langle L \rangle 
+\frac{65}{9}-\frac{\pi^2}{3}-\frac{8}{3}\zeta(3)\nonumber\\
&& +a_1^M\bigg [ \frac{4}{3} \langle L^2\rangle+\left
(-\frac{110}{9}+\frac{\pi^2}{3}\right )  \langle L \rangle + 
\frac{413}{27}+\frac{\pi^2}{9}-8\zeta(3)\bigg ]\nonumber\\
&& +a_2^M\bigg [ \frac{4}{3} \langle L^2\rangle+\left
(-\frac{313}{18}+\frac{2\pi^2}{3}\right )  \langle L \rangle+
\frac{4975}{216}+\frac{7\pi^2}{9}-16\zeta(3)\bigg ].
\end{eqnarray}

\subsection{Convolution of $C^{(B1)}_X$ with the light-cone distribution 
amplitude}

Because the expressions are lengthy, we only list the results for 
the convolution of the three combinations of coefficients 
functions in the physical form factor scheme as defined in 
(\ref{physscet1}), and assume that the light-cone distribution
amplitude are given by their asymptotic forms $\phi_M(v)=6 v\bar v$. 
The convolution integrals read 
\begin{eqnarray}
&&\hspace*{-0.5cm}\int_0^1 \frac{dv}{\bar v}\,\phi_M(v)
\left(C^{(B1)}_{f_0}-C^{(B1)}_{f_+} \,R_0\right)(x,\bar v)\nonumber\\
&=&-\frac{6(1-x)}{x}\,\Bigg\{1+\frac{\alpha_s C_F}{4\pi}\bigg
[-2 \ln^2\left(\frac{x}{\hat\mu}\right )+\ln\left
(\frac{x}{\hat\mu}\right)-\left(2-\frac{1}{(1-x)^2}\right )\ln
x\nonumber \\
&&- \left(2+\frac{2}{x}\right ){\rm
Li}_2(1-x)-\frac{\pi^2}{12}\frac{x-4}{x}+6+\frac{x}{1-x}\bigg]\nonumber
\\
&&+\frac{\alpha_s}{4\pi}\left (C_F-\frac{C_A}{2}\right)\bigg
[\left (\frac{4\pi^2}{3}-8\right )\ln \left (\frac{x}{\hat
\mu}\right )-\frac{4x\ln x}{1-x}+4 \ln x \,{\rm Li}_2(x)-4 {\rm
Li}_3(1-x)\nonumber \\
&&-8 {\rm Li}_3(x)+8-\frac{4\pi^2}{3}-4 \zeta(3)\bigg ]\Bigg\},
\nonumber \\
&&\hspace*{-0.5cm}\int_0^1 \frac{dv}{\bar v}\,\phi_M(v)
\left(C^{(B1)}_{f_T}-C^{(B1)}_{f_+}\, R_T\right)(x,\bar v)\nonumber\\
&=&6\,\Bigg\{1+\frac{\alpha_s C_F}{4\pi}\bigg [-2 \ln\hat \nu-2
\ln^2\left(\frac{x}{\hat\mu}\right )+\ln\left
(\frac{x}{\hat\mu}\right)+\left(2+\frac{2}{x}+\frac{1}{1-x}\right
)\ln
x\nonumber \\
&&- \left(2+\frac{2}{x}+\frac{2}{x^2}\right ){\rm
Li}_2(1-x)-\frac{\pi^2}{12}\frac{x^2-4
x-4}{x^2}+3-\frac{2}{x}\bigg]
\nonumber\\
&&+\frac{\alpha_s }{4\pi}\left (C_F-\frac{C_A}{2}\right)\bigg
[\left (\frac{4\pi^2}{3}-8\right )\ln \left (\frac{x}{\hat
\mu}\right )+\frac{4+4x}{x}\ln x+4 \ln x \,{\rm
Li}_2(x)-\frac{4}{x^2} {\rm
Li}_2(1-x)\nonumber \\
&&-\left(4-\frac{4}{x^2}\right ){\rm Li}_3(1-x)-8 {\rm
Li}_3(x)+8-\frac{4}{x}-\frac{2\pi^2}{3}\left
(2-\frac{1}{x}-\frac{1}{x^2}\right )-\left(4+\frac{4}{x^2}\right
) \zeta(3)\bigg ]\Bigg\},
\nonumber\\
&&\hspace*{-0.5cm} 
\int_0^1\frac{dv}{\bar v}\,\phi_M(v)
\left(C^{(B1)}_{T_1}-C^{(B1)}_{V} \,R_\perp\right)(x,\bar v)\nonumber\\
&=&-3\,\Bigg\{1+\frac{\alpha_s C_F}{4\pi}\bigg [-2 \ln\hat \nu-2
\ln^2\left(\frac{x}{\hat\mu}\right )-\ln\left
(\frac{x}{\hat\mu}\right)+\left(2+\frac{4}{x}+\frac{2}{1-x}\right
)\ln
x\nonumber \\
&&- \left(2+\frac{2}{x}+\frac{4}{x^2}\right ){\rm
Li}_2(1-x)-\frac{\pi^2}{12}\frac{x^2-4
x-8}{x^2}+\frac{9}{2}-\frac{4}{x}\bigg]
\\
&&+\frac{\alpha_s}{4\pi}\left (C_F-\frac{C_A}{2}\right)\bigg
[\left (\frac{4\pi^2}{3}-4\right )\ln \left (\frac{x}{\hat
\mu}\right )+2\ln x-\frac{4(1-x)}{x}\ln x \,{\rm Li}_2(x)+\frac{4}{x} {\rm
Li}_2(1-x)\nonumber \\
&&-\left(4-\frac{4}{x}-\frac{4}{x^2}\right ){\rm
Li}_3(1-x)+\frac{8(1-x)}{x} {\rm
Li}_3(x)+1-\frac{2\pi^2}{3}-4\left
(1+\frac{1}{x}+\frac{1}{x^2}\right ) \zeta(3)\bigg
]\Bigg\}\nonumber
\end{eqnarray}
with $x=2 E/m_b$, $\hat\mu=\mu/m_b$, $\hat\nu=\nu/m_b$, and 
$\alpha_s=\alpha_s(\mu)$.



\end{document}